\documentclass[sn-basic]{sn-jnl}% Basic Springer Nature Reference Style/Chemistry Reference Style

\jyear{2021}%

\theoremstyle{thmstyleone}%
\theoremstyle{thmstyletwo}%

\theoremstyle{thmstylethree}%

\usepackage{framed}
\usepackage{hyperref}
\usepackage{subcaption}

\begin{document}

\title[Ergo, SMIRK is Safe]{Ergo, SMIRK is Safe: A Safety Case for a Machine Learning Component in a Pedestrian Automatic Emergency Brake System}


\author*[1,2]{\fnm{Markus} \sur{Borg}}\email{markus.borg@cs.lth.se}

\author[3]{\fnm{Jens} \sur{Henriksson}}\email{jens.henriksson@semcon.com}

\author[1,2]{\fnm{Kasper} \sur{Socha}}\email{kasper.socha@ri.se}

\author[4]{\fnm{Olof} \sur{Lennartsson}}\email{olof.lennartsson@infotiv.se}

\author[4]{\fnm{Elias} \sur{Sonnsj\"o L\"onegren}}\email{elias.sonnsjo@infotiv.se}

\author[1]{\fnm{Thanh} \sur{Bui}}\email{thanh.bui@ri.se}

\author[1]{\fnm{Piotr} \sur{Tomaszewski}}\email{piotr.tomaszewski@ri.se}

\author[5]{\fnm{Sankar Raman} \sur{Sathyamoorthy}}\email{sankar.sathyamoorthy@qrtech.se}

\author[6]{\fnm{Sebastian} \sur{Brink}}\email{sebastian.brink@combitech.com}

\author[1]{\fnm{Mahshid} \sur{Helali Moghadam}}\email{mahshid.helali.moghadam@ri.se}

\affil*[1]{\orgname{RISE Research Institutes of Sweden}, \orgaddress{\city{Lund}, \country{Sweden}}}

\affil[2]{\orgdiv{Dept. of Computer Science}, \orgname{Lund University}, \orgaddress{\city{Lund}, \country{Sweden}}}

\affil[3]{\orgname{Semcon AB}, \orgaddress{\city{Gothenburg}, \country{Sweden}}}

\affil[4]{\orgname{Infotiv AB}, \orgaddress{\city{Gothenburg}, \country{Sweden}}}

\affil[5]{\orgname{QRTECH AB}, \orgaddress{\city{Gothenburg}, \country{Sweden}}}

\affil[6]{\orgname{Combitech AB}, \orgaddress{\city{Gothenburg}, \country{Sweden}}}

\abstract{Integration of Machine Learning (ML) components in critical applications introduces novel challenges for software certification and verification. New safety standards and technical guidelines are under development to support the safety of ML-based systems, e.g., ISO~21448 SOTIF for the automotive domain and the Assurance of Machine Learning for use in Autonomous Systems (AMLAS) framework. SOTIF and AMLAS provide high-level guidance but the details must be chiseled out for each specific case. We initiated a research project with the goal to demonstrate a complete safety case for an ML component in an open automotive system. This paper reports results from an industry-academia collaboration on safety assurance of SMIRK, an ML-based pedestrian automatic emergency braking demonstrator running in an industry-grade simulator. We demonstrate an application of AMLAS on SMIRK for a minimalistic operational design domain, i.e., we share a complete safety case for its integrated ML-based component. Finally, we report lessons learned and provide both SMIRK and the safety case under an open-source license for the research community to reuse.}




\maketitle

\section{Introduction} \label{sec:intro}
Machine Learning (ML) is increasingly used in critical applications, e.g., supervised learning using Deep Neural Networks (DNN) to support automotive perception. Software systems developed for safety-critical applications must undergo assessments to demonstrate compliance with functional safety standards. However, as the conventional safety standards are not fully applicable for ML-enabled systems \citep{salay2018analysis,tambon2021certify}, several domain-specific initiatives aim to complement them, e.g., organized by the EU Aviation Safety Agency, the ITU-WHO Focus Group on AI for Health, and the International Organization for Standardization.

In the automotive industry, several standardization initiatives are ongoing to allow safe use of ML in road vehicles. It is evident that the established functional safety as defined in ISO~26262 Functional Safety (FuSa) is no longer sufficient for the next generation of Advanced Driver-Assistance Systems (ADAS) and Autonomous Driving (AD). One complementary standard under development is ISO~21448 Safety of the Intended Functionality (SOTIF). SOTIF aims for absence of unreasonable risk due to hazards resulting from \textit{functional insufficiencies}, incl. those originating in ML components.

Standards such as SOTIF mandate high-level requirements on what a development organization must provide in a safety case for an ML-based system. However, how to actually collect the evidence -- and argue that it is sufficient -- is up to the specific organization. Assurance of Machine Learning for use in Autonomous Systems (AMLAS) is one framework that supports the development of corresponding safety cases \citep{amlas2021}. Still, when applying AMLAS on a specific case, there are numerous details that must be analyzed, specified, and validated. The research community lacks demonstrator systems that can be used to explore such details. 

To address the lack of ML-based demonstrator systems with accompanying safety cases, we embarked on an \textit{engineering research} endeavor. Engineering research is described in the evolving ACM SIGSOFT Empirical Standards~\citep{ralph2020empirical} and we consider it synonymous with \textit{design science research}~\citep{wieringa2014design,runeson2020design}. Our engineering research was guided by the following design problem:

\begin{quote}
How to demonstrate and share a complete ML safety case for an open ADAS?
\end{quote}

We report results from an industry-academia collaboration on safety assurance of SMIRK, an ML-based Open-Source Software (OSS) ADAS that provides Pedestrian Automatic Emergency Braking (PAEB) in an industry-grade simulator. SMIRK is an ``original software publication''~\citep{socha2022smirk} available on GitHub~\citep{github_smirk}. In this paper, our main contribution is the carefully described application of AMLAS in conjunction with the development of an ML component. While parts of the framework have been demonstrated before \citep{gauerhof2020assuring}, we believe this is the first comprehensive use of AMLAS conducted independently from its authors. Moreover, we believe this paper constitutes a pioneering safety case for an ML-based component that is OSS and completely transparent. Thus, our contribution can be used as a starting point for studies on safety engineering aspects such as Operational Design Domain (ODD) extension, dynamic safety cases, and reuse of safety evidence.

Our results show that even an ML component in an ADAS designed for a minimalistic ODD results in a large safety case. Furthermore, we consider three lessons learned to be particularly important for the community. First, using a simulator to create synthetic data sets for ML training particularly limits the validity of the negative examples. Second, evaluation of object detection is non-intuitive and necessitates internal training. Third, the fitness function used for model selection encodes essential tradeoff decisions, thus the project team must be aligned. 

%Figure~\ref{fig:paper_overview} presents the structure of this long article. We are aware that we include a large piece of work in a single publication unit. Still, we find that a largely self-contained paper presenting a comprehensive safety case is missing in the research community. As opposed to the ``salami publication'' anti-pattern in academic publishing, we choose to present both SMIRK and its complete safety case in the same article. The figure shows the four main parts of this paper.

The paper is organized into three main parts:

\begin{itemize}
    \item[Part I] Section~\ref{sec:intro} contains this introduction. Section~\ref{sec:bg} contains a background section describing SOTIF, AMLAS, and object detection using YOLO. In Section~\ref{sec:rw} we share an overview of related work. Finally, Section~\ref{sec:method} describes the method used in our R\&D project.
    \item[Part II] Sections~\ref{sec:smirk}--\ref{sec:testres} describe the intertwined development and safety assurance of SMIRK. We present an overall system description (Section~\ref{sec:smirk}), system requirements (Section~\ref{sec:reqts}), system architecture (Section~\ref{sec:arch}), data management strategy (Section~\ref{sec:dms}), ML-based pedestrian recognition component (Section~\ref{sec:mlspec}), test design (Section~\ref{sec:test}), and test results (Section~\ref{sec:testres}), respectively.
     \item[Part III] Section~\ref{sec:lessons} reports lessons learned from our R\&D project. Section~\ref{sec:threats} discusses the main threats to validity before Section~\ref{sec:conc} concludes the paper and outlines future work.
\end{itemize}

Finally, to ensure a self-contained paper, Appendix~\ref{app:amlas} presents the complete AMLAS safety argumentation for the use of ML in SMIRK.
    
%\begin{figure}
%\centering
%\includegraphics[width=0.9\textwidth]{paper_org.png}
%\caption{Paper organization.}
%\label{fig:paper_overview}
%\end{figure}

\section{Background} \label{sec:bg}
This section briefly presents SOTIF and AMLAS, respectively. Also, we present details of object detection and recognition using YOLO, which is fundamental to understand the subsequent safety argumentation.

\subsection{Safety of the Intended Functionality (SOTIF)}
ISO~21448 Safety of the Intended Functionality (SOTIF) is a candidate standard under development to complement the established automotive standard ISO~26262 Functional Safety (FuSa). While FuSa covers hazards caused by \textit{malfunctioning behavior}, SOTIF addresses hazardous behavior caused by the intended functionality. Note that SOTIF covers ``reasonably foreseeable misuse'' but explicitly excludes antagonistic attacks, thus we do not discuss any security concerns in this paper. A system that meets FuSa can still be hazardous due to insufficient environmental perception or inadequate robustness within the ODD. The SOTIF process provides guidance on how to systematically ensure the absence of unreasonable risk due to \textit{functional insufficiencies}. The goal of the SOTIF process is to perform a risk acceptance evaluation and then reduce the probability of 1) known and 2) unknown scenarios causing hazardous behavior. 

\begin{figure}
\centering
\includegraphics[width=\textwidth]{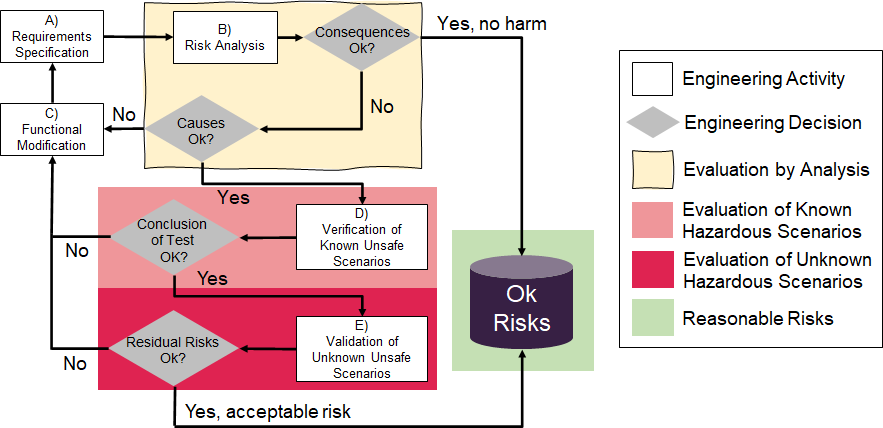}
\caption{A simplified overview of the SOTIF process. Adapted from ISO 21488.}
\label{fig:sotif}
\end{figure}

Figure~\ref{fig:sotif} shows a simplified version of the SOTIF process. The process starts in the upper left with A) Requirements specification. Based on the requirements, a B) Risk Analysis is done. For each identified risk, its potential \textit{Consequences} are analyzed. If the risk of harm is reasonable, it is recorded as an acceptable risk. If not, the activity continues with an analysis of \textit{Causes}, i.e., an identification and evaluation of triggering conditions. If the expected system response to triggering conditions is acceptable, the SOTIF process continues with V\&V activities. If not, the remaining risk forces a C) Functional Modification with a corresponding requirements update.

The lower part of Figure~\ref{fig:sotif} shows the V\&V activities in the SOTIF process, assuming that they are based on various levels of testing. For each risk, the development organization conducts D) Verification to ensure that the system satisfies the requirements for the known hazardous scenarios. If the F) Conclusion of Verification Tests are satisfactory, the V\&V activities continues with validation. It not, the remaining risk requires a C) Functional Modification. In the E) Validation, the development organization explores the presence of unknown hazardous scenarios -- if any are identified, they turn into known hazardous scenarios. The H) Conclusion of Validation Tests estimates the likelihood of encountering unknown scenarios that lead to hazardous behavior. If the residual risk is sufficiently small, it is recorded as an acceptable risk. If not, the remaining risk again necessitates a C) Functional Modification.

\subsection{Safety Assurance Using the AMLAS Process}
The methodology for the Assurance of Machine Learning for use in Autonomous Systems (AMLAS) was developed by the Assuring Autonomy International Programme, University of York~\citep{amlas2021}. AMLAS provides a process that results in 34 safety evidence artifacts. Moreover, AMLAS provides a set of recurring safety case patterns for ML components presented using the graphical format Goal Structuring Notation (GSN) \citep{gsn2021}.

\begin{figure}
\centering
\includegraphics[width=\textwidth]{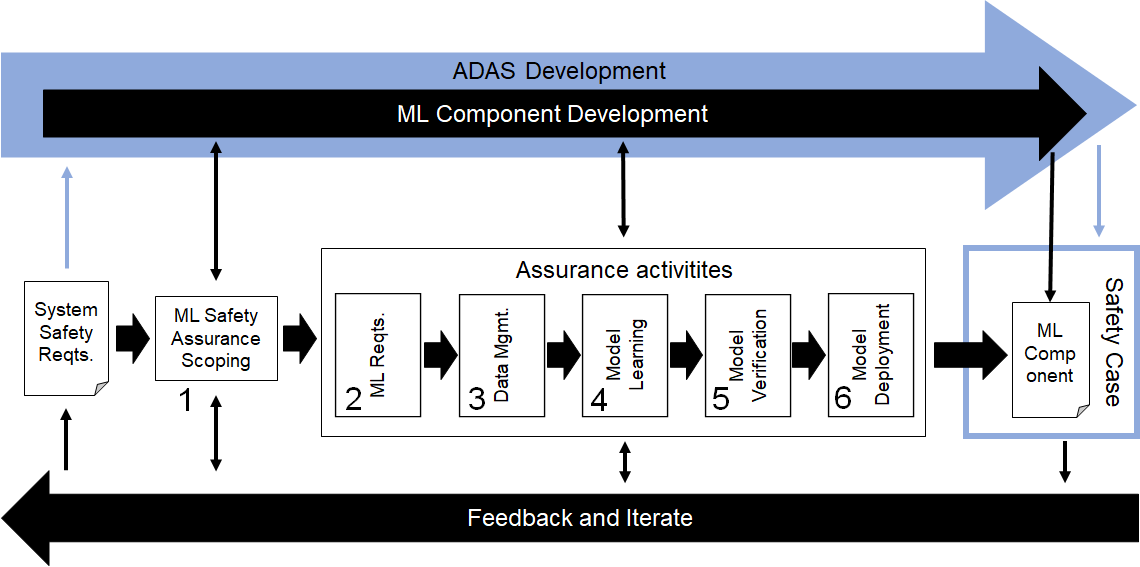}
\caption{An overview of the AMLAS process, adapted from \citet{amlas2021}. Blue color denotes systems engineering, whereas black color relates specifically to the ML component. Numbers refer to the AMLAS stages, not sections in this paper.}
\label{fig:amlas}
\end{figure}

Figure~\ref{fig:amlas} shows an overview of the six stages of the AMLAS process. The upper part stresses that the development of an ML component and its corresponding safety case is done in the context of larger systems development, indicated by the blue arrow. Analogous to SOTIF, AMLAS is an iterative process as highlighted by the black arrow in the bottom.

Starting from the System Safety Requirements from the left, Stage~1 is \textbf{ML Safety Assurance Scoping}. This stage operates on a systems engineering level and defines the scope of the safety assurance process for the ML component as well as the scope of its corresponding safety case -- the interplay with the non-ML safety engineering is fundamental. The next five stages of AMLAS all focus on assurance activities for different constituents of ML development and operations. Each of these stages conclude with an assurance argument that when combined, and complemented by evidence through artifacts, compose the overall ML safety case.

\begin{itemize}
    \item[Stage~2] \textbf{ML Safety Requirements Assurance}. Requirements engineering is used to elicit, analyze, specify, and validate ML safety requirements in relation to the software architecture and the ODD.
    \item[Stage~3] \textbf{Data Management Assurance}. Requirements engineering is first used to develop data requirements that match the ML safety requirements. Subsequently, data sets are generated (development data, internal test data, and verification data) accompanied by quality assurance activities.
    \item[Stage~4] \textbf{Model Learning Assurance}. The ML model is trained using the development data. The fulfilment of the ML safety requirements is assessed using the internal test data.
    \item[Stage~5] \textbf{Model Verification Assurance}. Different levels of testing or formal verification to assure that the ML model meets the ML safety requirements. Most importantly, the ML model shall be tested on verification data that has not influenced the training in any way.
    \item[Stage~6] \textbf{Model Deployment Assurance}. Integrate the ML model in the overall system and verify that the system safety requirements are satisfied. Conduct integration testing in the specified ODD.
\end{itemize}

The rightmost part of Figure~\ref{fig:amlas} shows the overall safety case for the system under development with the argumentation for the ML component as an essential part, i.e., the target of the AMLAS process. %All semantics used in the figures in Section~\ref{app:amlas} are defined in this open standard.

\subsection{Object Detection and Recognition Using YOLO} \label{sec:yolo}
YOLO is an established real-time object detection and recognition algorithm that was originally released by \citet{redmon2016you}. The first version of YOLO introduced a novel object detection process that uses a single DNN to perform both prediction of bounding boxes around objects and classification at once. YOLO was heavily optimized for fast inference to support real-time applications. A fundamental concept of YOLO is that the algorithm considers each image only once, hence its name ``You Only Look Once.'' Thus, YOLO is referred to as a single-stage object detector. %While there are several versions of YOLO, the fundamental ideas remain the same across versions.% - including YOLOv5 used in SMIRK.

Single-stage object detectors consist of three core parts: 1) the model backbone, 2) the model neck, and 3) the model head. The model backbone extracts important features from input images. The model neck generates so called ``feature pyramids'' using PANet \citep{liu2018path} that support generalization to different sizes and scales. The model head performs the detection task, i.e., it generates the final output vectors with bounding boxes and class probabilities.

In a nutshell, YOLO segments input images into smaller images. Each input image is split into a square grid of individual cells. Each cell predicts bounding boxes capturing potential objects and provides confidence scores for each box. Furthermore, YOLO does a class prediction for objects in the bounding boxes. Relying on the Intersection over Union (IoU) method for evaluating bounding boxes, YOLO eliminates redundant bounding boxes. The final output from YOLO consists of unique bounding boxes with class predictions. There are several versions of YOLO and each version provides different model architectures that balance the tradeoff between inference speed and accuracy differently -- additional layers of neurons provide better accuracy at the expense of computation time.

%Figure~\ref{fig:yolov5_benchmarking} shows the speed/accuracy tradeoffs for different YOLOv5 architectures.  On the y-axis, COCO~AP~val denotes the mAP@0.5:0.95 metric measured on the 5,000-image COCO~val2017 data set over various inference sizes from 256 to 1,536. On the x-axis, GPU Speed measures average inference time per image on the COCO~val2017~data set using an AWS p3.2xlarge V100 instance at batch-size 32. The curve EfficientDet illustrates results from Google AutoML at batch size 8.

%\begin{figure}
%\centering
%\includegraphics[width=1\textwidth]{yolov5_benchmarking.png}
%\caption{Speed/accuracy tradeoffs for different YOLOv5 architectures. (Image source: Ultralytics under GPLv3)}
%\label{fig:yolov5_benchmarking}
%\end{figure}

\section{Related Work} \label{sec:rw}
Many researchers argue that software and systems engineering practices must evolve as ML enters the picture. As proposed by \citet{bosch2021engineering}, we refer to this emerging area as AI engineering.
AI engineering increases the importance of several system qualities. Quality characteristics discussed in the area of ``explainable AI'' are particularly important to safety argumentation. In our paper, we adhere to the following definitions provided by \citet{arrieta2020explainable}: ``Given an audience, an \textit{explainable AI} is one that produces details or reasons to make its functioning clear or easy to understand.'' In the same vein, we refer to \textit{interpretability} as ``a passive characteristic referring to the level at which a given model makes sense for a human observer.'' In our work on SMIRK, the human would be an assessor of the safety argumentation. In contrast, we regard \textit{explainability} as an active characteristic of a model that clarifies its internal functions. The current ML models in SMIRK do not actively explain their outputs. For a review of DNN explainability techniques in automotive perception, e.g., post-hoc saliency methods, counterfactual explanations, and model translations, we refer to a recent review by \citet{zablocki2022explainability}.

In light of AI engineering, the rest of this section presents related work on safety argumentation and testing of automotive ML, respectively.

%In this section, we focus on aspects of AI engineering for safety-critical ML-based automotive systems, i.e., safety argumentation in general and DNNs used for automotive perception in particular. %Moreover, we stress that the remainder of the paper contains several references to related work as we discuss various design choices.
\subsection{Safety Argumentation for Machine Learning}
Many publications address the issue of safety argumentation for systems with ML-based components and highlight corresponding safety gaps. A solid argumentation is required to enable safety certification, for example to demonstrate compliance with future standards such as SOTIF and ISO~8800 Road Vehicles -- Safety and Artificial Intelligence. While there are several established safety patterns for non-AI systems (e.g., simplicity, substitution, sanity check, condition monitoring, comparison, diverse redundancy, replication redundancy, repair, degradation, voting, override, barrier, and heartbeat \citep{wu2004safety,preschern2015building}), considerable research is now directed at understanding to what extent existing approaches apply to AI engineering. For example, \citet{picardi2020assurance} presented a set of patterns that later turned into AMLAS, i.e., the process that guides most of our work in this paper.

\citet{mohseni2020practical} reviewed safety-related ML research and mapped ML techniques to to safety engineering strategies. Their mapping identified three main categories toward safe ML: 1) \textit{inherently safe design}, 2) \textit{enhanced robustness}, a.k.a. ``safety margin'', and 3) \textit{run-time error detection}, a.k.a. ``safe fail''. The authors further split each category into sub-categories. Based on their proposed categories, the safety argumentation we present for SMIRK used an inherently safe design as the starting point through a careful ``design specification'' and ``implementation transparency.'' Moreover, from the third category, SMIRK relies on ``OOD error detection'' as will be presented in Section~\ref{sec:safetycage}.

\citet{schwalbe2020survey} presented a survey on specific considerations for safety argumentation targeting DNNs, organized into four development phases. First, they state that \textit{requirements engineering} must focus on data representativity and entail scenario coverage, robustness, fault tolerance, and novel safety-related metrics. Second, \textit{development} shall seek safety by design by, e.g., acknowledging uncertainty and enhancing robustness (in line with recommendations by \citet{mohseni2020practical}). Third, the authors discuss \textit{verification} primarily through formal model checks using solvers. Fourth, they consider \textit{validation} as the task of identifying missing requirements and test cases, for which they recommend data validation followed by both qualitative and quantitative analyses.

\citet{tambon2021certify} conducted a systematic literature review on certification of safety-critical ML-based systems. Based on 217 primary studies, the authors identified five overall categories in the literature: 1) \textit{robustness}, 2) \textit{uncertainty}, 3) \textit{explainability}, 4) \textit{verification}, and 5) \textit{safe reinforcment learning}. Moreover, the paper concludes by calling for deeper industry-academia collaborations related to certification. Our work on SMIRK responds to this call and explicitly discusses the identified categories on an operational level (except reinforcement learning since this type of ML does not apply to SMIRK). By conducting hands-on development of an ADAS and its corresponding safety case, we have identified numerous design decisions that have not been discussed in prior work. As the devil is in the detail, we recommend other researchers to transparently report from similar endeavors in the future.

%\citet{kochanthara2021functional} propose a safety assessment method on the systems-of-systems level, i.e., for cooperative driving systems. While the method does not target ML specifically, it discusses the context of platooning with a manually driven lead vehicle is followed by autonomous vehicles -- a solution that most likely requires ML-based perception. 
\citet{schwalbe2020structuring} systematically established and broke down safety requirements to argue the sufficient absence of risk arising from SOTIF-style functional insufficiencies. The authors stress the importance of diverse evidence for a safety argument involving DNNs. Moreover, they provide a generic approach and template to thoroughly report DNN specifics within a safety argumentation structure. Finally, the authors show its applicability for an example use case based on pedestrian detection. In our work, 34 artifacts of different types constitute the safety evidence. Furthermore, pedestrian detection is a core feature of SMIRK.

Several research projects selected ML-based pedestrian detection systems to illustrate different aspects of safety argumentation. \citet{wozniak2020safety} provided a safety case pattern for ML-based systems and showcase its applicability for pedestrian detection. The pattern is integrated within an overall encompassing approach for safety case generation. On a similar note, \citet{willers2020safety} discussed safety concerns for DNN-based automotive perception, including technical root causes and mitigation strategies. The authors argue that it remains an open question how to conclude whether a specific concern is sufficiently covered by the safety case -- and stress that safety cannot be determined analytically through ML accuracy metrics. In our work on SMIRK, we provide safety evidence that goes beyond the level of the ML model. Using AMLAS, we also claim that we demonstrate sufficient evidence for ML in SMIRK. Finally, related to pedestrian detection, we find that the work by \citet{gauerhof2020assuring} is the closest to this study, and the reader will find that we repeatedly refer to it in relation to SMIRK requirements in Section~\ref{sec:reqts}.

In the current paper, we present a holistic ML safety case building on previous work for a demonstrator system available under an OSS license. Furthermore, in contrast to a discussion restricted to pedestrian detection, we discuss an ADAS that subsequently commences emergency braking in a simulated environment. This addition responds to a call from \citet{haq2021automatic} to go from offline to online testing, as many safety violations cannot be detected on the level of the ML model. While online testing is not novel, research papers on ML testing have largely considered standalone image data sets disconnected from concerns of running systems.

\subsection{Testing of Machine Learning in Automated Vehicles}
According to AMLAS, the two primary means to generate safety evidence for ML-based systems through V\&V are \textit{testing} and \textit{verification} \citep{amlas2021}. As test-based verification is substantially more mature for DNNs such as used in SMIRK, we restrict ourselves to testing. In the context of an ML-based system, this can be split into: 1) data set testing, 2) model testing, 3) unit testing, 4) integration testing, 5) system testing, and 6) acceptance testing \citep{song2022exploring}. The related work focuses on model and system testing.

Model testing shall be performed on the verification data set that must not have been used in the training process. Depending on the test subject and the test level, the inputs for corresponding ADAS testing might be images such as in DeepRoad \citep{zhang2018deeproad} and DeepTest \citep{tian2018deeptest}, or test scenario configurations as used by \citet{ebadi2021efficient}.

Many research projects investigated efficient design of effective test cases for ADAS. Thus, several approaches to test case generation in simulated environments have been proposed. \textit{Pseudo-test oracle differential testing} focuses on detecting divergence between programs’ behaviors when provided the same input data. For example, DeepXplore \citep{pei2017deepxplore} changes the test input---like solving an optimization problem---to find the inputs triggering different behaviors between similar autonomous driving DNN models, while trying to increase the neuron coverage. \textit{Metamorphic testing} works based on detecting violations of metamorphic relations to identify erroneous behaviors. For example, DeepTest \citep{tian2018deeptest} applies different transformations to a set of seed images and utilizes metamorphic relations to detect faulty behaviors of different Udacity DNN models for self-driving cars, while aiming for increasing neuron coverage as well. \textit{Gradient-based test input generation} regards the test input generation as an optimization problem and generates a high number of failure-revealing or difference-inducing test inputs while maximizing the test adequacy criteria, e.g., neuron coverage. DeepXplore \citep{pei2017deepxplore} utilizes gradient ascent to generate inputs provoking different behaviors among similar DNN models. \textit{Generative Adversarial Network (GAN)-based test input generation} is intended to generate realistic test input, which can not be easily distinguished from original input. DeepRoad \citep{zhang2018deeproad} utilizes a GAN-based metamorphic testing approach to generate test images for testing Udacity DNN models for autonomous driving.

System testing entails ensuring that system safety standards are met following the integration of the model into the system. As for model testing, many research projects proposed techniques to generate test cases. A commonly proposed approach is to use search-based techniques to identify failure-revealing or collision-provoking scenarios. Many papers argue that simulated test scenarios are effective complements to field testing -- which tends to be costly and sometimes dangerous.

\citet{ben_abdessalem_testing_2016} proposed the multi-objective search algorithm NSGA-II along with surrogate models to find critical test scenarios with few simulations for the pedestrian detection system PeVi. PeVi constitutes the reference architecture for SMIRK, as described in Section~\ref{sec:arch}, for which we obtained deep insights during a replication study \citep{borg2021digital}. In a subsequent study, \citet{abdessalem2018testing} used MOSA developed by \citet{panichella2015reformulating}---a \textit{many-objective optimization search algorithm}--- along with objectives based on branch coverage and failure-based heuristics to detect undesired and failure-revealing feature interaction scenarios for integration testing in an automated vehicle. Furthermore, in a third related study, \citet{abdessalem2018testinglearnable} combined NSGA-II and decision tree classifier models---which they referred to as a \textit{learnable evolutionary algorithm}---to guide the search for critical test scenarios. Moreover, the proposed approach characterizes the failure-prone regions of the test input space. Inspired by previous work, we have used various search-techniques, including NSGA-II, to generate test scenarios for pedestrian detection and emergency braking of the autonomous driving platform Baidu Apollo in the SVL simulator \citep{ebadi2021efficient}.

The test cases providing safety evidence for SMIRK correspond to systematic grid searches rather than any metaheuristic search. We think this is a necessary starting point for a safety argumentation. On the other hand, as argued in several related papers, we believe that search techniques could be a useful complement during testing of ML-based systems to pinpoint weaknesses and guide functional modifications -- for example as part of the SOTIF process. In the SMIRK context, we plan to investigate this further as part of future work.

\section{Method: Engineering Research} \label{sec:method} \label{sec:smile}
The overall frame of our work is the \textit{engineering research}\footnote{As noted in the empirical standard for engineering research, some methodological guidelines refer to the same approach as \textit{design science research.}} standard as defined in the community endeavor ACM SIGSOFT Empirical Standards \citep{ralph2020empirical}. Engineering research is an appropriate standard when evaluating technological artifacts, e.g., methods, systems, and tools -- in our case SMIRK and its safety case.  To support the validity of our research, we consulted the essential attributes of the corresponding checklist provided in the standard. We provide three clarifications for readers cross-checking our work against these attributes: 1) empirical evaluations of SMIRK are done using simulation in ESI Pro-SiVIC, 2) empirical evaluation of the safety case has been done through workshops and peer-review, and 3) we compare the SMIRK safety case against the state-of-the-art as we build on previous work.

Engineering research aims to develop general design knowledge in a specific field to help practitioners create solutions to their problems. As discussed by \citet{aken2004management}, \textit{technological rules} can be used to express design knowledge by capturing general knowledge about the mappings between problems and proposed solutions. Technological rules can normally be expressed as ``To achieve $<$Effect$>$ in $<$Context$>$ apply $<$Intervention$>$''. Starting from a technological rule, researchers can extend the rule's scope of validity by applying the intervention to new contexts or develop rules from the general to more specific contexts~\citep{runeson2020design}, i.e., new studies can lead to both generalization and specialization.

Our work aims to specialize a technological rule for AMLAS. Starting from:

\begin{quote}
To \textit{develop a safety case} for \textit{ML in autonomous systems} apply \textit{AMLAS}.
\end{quote}

we seek general design knowledge for the more specific rule:

\begin{quote}
To \textit{develop a safety case} for \textit{ML-based perception in ADAS} apply \textit{AMLAS}.
\end{quote}

\begin{figure}
\centering
\includegraphics[width=\textwidth]{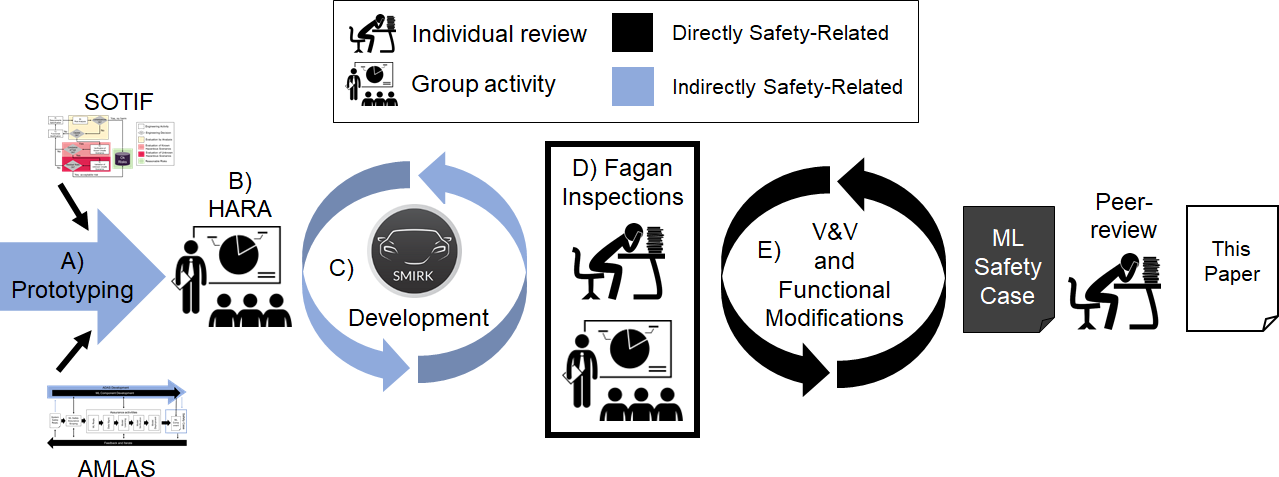}
\caption{An overview of the SMIRK development in the SMILE~III project.}
\label{fig:smile}
\end{figure}

Figure~\ref{fig:smile} shows an overview of the two-year R\&D project (SMILE~III\footnote{\url{https://tinyurl.com/smileiii}}) that resulted in the SMIRK MVP (Minimum Viable Product) and the safety case for its ML component. Note that collaborations in the preceding SMILE projects started already in 2016 \citep{borg2019safely}, i.e., we report from research in the context of prolonged industry involvement. Starting from the left, we relied on A) Prototyping to get an initial understanding of the problem and solution domain~\citep{kapyaho2015agile}. As our pre-understanding during prototyping grew, SOTIF and AMLAS were introduced as fundamental development processes and we established a first System Requirements Specification (SRS). The AMLAS process starts in the System Safety Requirements, which in our case come from following the SOTIF process.

Based on the SRS, we organized a B) Hazard Analysis and Risk Assessment (HARA) workshop (cf. ISO~262626) with all author affiliations represented. Then, the iterative C) SMIRK development phase commenced, encompassing both software development, ML development, and a substantial amount of documentation. When meeting our definition of done, i.e., an MVP implementation and stable requirements specifications, we conducted D) Fagan Inspections as described in Section~\ref{sec:fagan}. After corresponding updates, we baselined the SRS and the Data Management Specification (DMS). Note that due to the Covid-19 pandemic, all group activities were conducted in virtual settings.

Subsequently, the development project turned to E) V\&V and Functional Modifications as limitations were identified. In line with the SOTIF process (cf. Figure~\ref{fig:sotif}), also this phase of the project was iterative. The various V\&V activities generated a significant part of the evidence that supports our safety argumentation. The rightmost part of Figure~\ref{fig:smile} depicts the safety case for the ML component in SMIRK, which is peer-reviewed as part of the submission process of this paper.

\subsection{Safety Evidence from Fagan Inspections} \label{sec:fagan} \label{amlas:j} \label{amlas:m}
We conducted two formal Fagan inspections \citep{fagan1976design} during the SMILE~III project with representatives from the organizations listed as co-authors of this paper. All reviewers are active in automotive R\&D. The inspections targeted the Software Requirements Specification and the Data Management Specification, respectively. The two formal inspections constitute essential activities in the AMLAS safety assurance and result in ML Safety Requirements Validation Results \textbf{[J]} and a Data Requirements Justification Report \textbf{[M]}. A Fagan inspection consists of the steps 1) Planning, 2) Overview, 3) Preparation, 4) Inspection meeting, 5) Rework, and 6) Follow-up.

\begin{enumerate}
    \item Planning: The authors prepared the document and invited the required reviewers to an inspection meeting.
    \item Overview: During one of the regular project meetings, the lead authors explained the fundamental structure of the document to the reviewers, and introduced an inspection \href{checklist}{https://github.com/RI-SE/smirk/blob/main/docs/support/ETSA03_SRS_Review_Checklist_v1.0.pdf}. Reviewers were assigned particular inspection perspectives based on their individual expertise. All information was repeated in an email, as not all reviewers were present at the meeting.
    \item Preparation: All reviewers conducted an individual inspection of the document, noting any questions, issues, and required improvements.
    \item Inspection meeting: Two weeks after the individual inspections were initiated, the lead authors and all reviewers met for a virtual meeting. The entire document was discussed, and the findings from the independent inspections were compared. All issues were compiled in inspection \href{protocols}{https://github.com/RI-SE/smirk/tree/main/docs/protocols}.
    \item Rework: The lead authors updated the SRS according to the inspection protocol. The independent inspection results were used as input to capture-recapture techniques to estimate the remaining amount of work \citep{petersson2004capture}. All changes are traceable through individual GitHub commits.
    \item Follow-up: Reviewers who previously found issues verified that those had been correctly resolved.
\end{enumerate}

\subsection{Presentation Structure for the Safety Evidence}
In the remainder of this paper, the AMLAS stages and the resulting artifacts act as the backbone in the presentation. Table~\ref{tab:amlas_index} provides an overview of how those artifacts relate to the stages of AMLAS and where in this paper they are described for SMIRK. Throughout the paper, the notation \textbf{[A]}--\textbf{[HH]}, in bold font, refers to the 34 individual artifacts prescribed by AMLAS. Finally, in Appendix~\ref{app:amlas}, the same 34 artifacts are used to present a complete safety case for the ML component in SMIRK.

\begin{table}[]
\caption{SMIRK Safety Assurance Table. Numbers in the Input/Output columns refer to the AMLAS stages in Figure~\ref{fig:amlas}. GitHub repository: \url{https://github.com/RI-SE/smirk}. AI~Sweden hosts our annotated data set (185~GB) at: % and provides it under a CC-BY-NC 4.0 license:
\url{https://www.ai.se/en/data-factory/datasets/data-factory-datasets/smirk-dataset}}
\label{tab:amlas_index}
\begin{tabular}{|c|p{4cm}|p{1cm}|p{1cm}|l|}
\hline
\textbf{ID} & \multicolumn{1}{c|}{\textbf{AMLAS Artifact Title}}            & \textbf{Input to} & \textbf{Output from} & \multicolumn{1}{c|}{\textbf{Where?}} \\ \hline
{[}A{]}     & System Safety Requirements                     & 1, 6              &                      & Sec.~\ref{amlas:a}                   \\ \hline
{[}B{]}     & Description of Operating Environment of System & 1, 6              &                      & Sec.~\ref{amlas:b}                    \\ \hline
{[}C{]}     & System Description                             & 1, 6              &                      & Sec.~\ref{amlas:c}                     \\ \hline
{[}D{]}     & ML Component Description                       & 1                 &                      & Sec.~\ref{amlas:d}                     \\ \hline
{[}E{]}     & Safety Requirements Allocated to ML Component  & 2                 & 1                    & Sec.~\ref{amlas:e}                  \\ \hline
{[}F{]}     & ML Assurance Scoping Argument Pattern          & 1                 &                      & Fig.~\ref{amlas:f}                     \\ \hline
{[}G{]}     & ML Safety Assurance Scoping Argument           &                   & 1                    & Sec.~\ref{amlas:g}                     \\ \hline
{[}H{]}     & ML Safety Requirements                         & 3, 4, 5           & 2                    & Sec.~\ref{amlas:h}                   \\ \hline
{[}I{]}     & ML Safety Requirements Argument Pattern        & 2                 &                      & Fig.~\ref{amlas:i}                     \\ \hline
{[}J{]}     & ML Safety Requirements Validation Results      &                   & 2                    & Sec.~\ref{amlas:j}                      \\ \hline
{[}K{]}     & ML Safety Requirements Argument                &                   & 2                    & Sec.~\ref{amlas:k}                     \\ \hline
{[}L{]}     & Data Requirements                              &                   & 3                    & Sec.~\ref{amlas:l}                       \\ \hline
{[}M{]}     & Data Requirements Justification Report         &                   & 3                    & Sec.~\ref{amlas:m}                     \\ \hline
{[}N{]}     & Development Data                               &                   & 3                    & AI Sweden                              \\ \hline
{[}O{]}     & Internal Test Data                             &                   & 3                    & AI Sweden                                  \\ \hline
{[}P{]}     & Verification Data                              &                   & 3                    & AI Sweden                                  \\ \hline
{[}Q{]}     & Data Generation Log                            &                   & 3                    & Sec.~\ref{amlas:q}                      \\ \hline
{[}R{]}     & ML Data Argument Pattern                       & 3                 &                      & Fig.~\ref{amlas:r}                       \\ \hline
{[}S{]}     & ML Data Validation Results                     &                   & 3                    & Sec.~\ref{amlas:s}                      \\ \hline
{[}T{]}     & ML Data Argument                               &                   & 3                    & Sec.~\ref{amlas:t}                     \\ \hline
{[}U{]}     & Model Development Log                          &                   & 4                    & Sec.~\ref{amlas:u}                    \\ \hline
{[}V{]}     & ML Model                                       & 5, 6              & 4                    & GitHub repo                                  \\ \hline
{[}W{]}     & ML Learning Argument Pattern                   & 4                 &                      & Fig.~\ref{amlas:w}                  \\ \hline
{[}X{]}     & Internal Test Results                          &                   & 4                    & Sec.~\ref{amlas:x}                      \\ \hline
{[}Y{]}     & ML Learning Argument                           &                   & 4                    & Sec.~\ref{amlas:y}                     \\ \hline
{[}Z{]}     & ML Verification Results                        &                   & 5                    & Sec.~\ref{amlas:z}                      \\ \hline
{[}AA{]}    & Verification Log                               &                   & 5                    & Sec.~\ref{amlas:aa}                     \\ \hline
{[}BB{]}    & ML Verification Argument Pattern               & 5                 &                      & Fig.~\ref{amlas:bb}                     \\ \hline
{[}CC{]}    & ML Verification Argument                       &                   & 5                    & Sec.~\ref{amlas:cc}                      \\ \hline
{[}DD{]}    & Erroneous Behaviour Log                        &                   & 6                    & Sec.~\ref{amlas:dd}                    \\ \hline
{[}EE{]}    & Operational scenarios                          & 6                 &                      & Sec.~\ref{amlas:ee}                   \\ \hline
{[}FF{]}    & Integration Testing Results                    &                   & 6                    & Sec.~\ref{amlas:ff}                      \\ \hline
{[}GG{]}    & ML Deployment Argument Pattern                 & 6                 &                      & Fig.~\ref{amlas:gg}                       \\ \hline
{[}HH{]}    & ML Deployment Argument                         &                   & 6                    & Sec.~\ref{amlas:hh}                     \\ \hline
\end{tabular}
\end{table}

\section{SMIRK System Description \textbf{[C]}} \label{sec:smirk} \label{amlas:c}
SMIRK is a PAEB system that relies on ML. As an example of an ADAS, SMIRK is intended to act as one of several systems supporting the driver in the dynamic driving task, i.e., all the real-time operational and tactical functions required to operate a vehicle in on-road traffic. SMIRK, including the accompanying safety case, is developed with full transparency under an OSS license. We develop SMIRK as a demonstrator in a simulated environment provided by ESI Pro-SiVIC\footnote{We stress that SMIRK shall never be used in a real vehicle and the authors take no responsibility in any such endeavors.}.

The SMIRK product goal is to assist the driver on country roads in rural areas by performing emergency braking in the case of an imminent collision with a pedestrian. The level of automation offered by SMIRK corresponds to SAE Level 1: Driver Assistance, i.e., ``the driving mode-specific execution by a driver assistance system of either steering or acceleration/deceleration'' --- in our case only braking. The first release of SMIRK is an MVP, i.e., an implementation limited to a highly restricted ODD, but, future versions might include steering and thus comply with SAE Level 2.

%However, SMIRK is developed with evolvability in mind, thus future versions might include steering and thus comply with SAE Level 2. The first release of SMIRK is an MVP, i.e., an implementation limited to a highly restricted ODD.

Sections~\ref{sec:smirk} and~\ref{sec:reqts} presents the core parts of the SMIRK SRS. The SRS, as well as this section, largely follows the structure proposed in IEEE 830-1998: IEEE Recommended Practice for Software Requirements Specifications~\citep{ieee_srs} and the template provided by~\citet{wiegers_srs}. %While the standard has been replaced by ISO/IEC/IEEE 29148:2011, the old standard serves the SMIRK development well.
This section presents a SMIRK overview whereas Section~\ref{sec:reqts} specifies the system requirements.  

\subsection{Product Scope}
\begin{figure}
\centering
\includegraphics[width=0.7\textwidth]{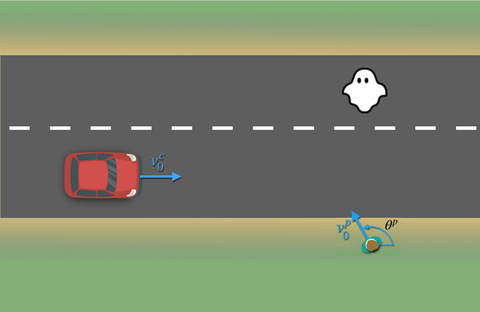}
\caption{Example scenario with a pedestrian crossing the road at an arbitrary angle. A false positive is also presented, i.e., a ghost.}
\label{fig:arbitrary_scenario}
\end{figure}

Figure~\ref{fig:arbitrary_scenario} illustrates the overall function provided by SMIRK. SMIRK sends a brake signal when a collision with a pedestrian is imminent. Pedestrians are expected to cross the road at arbitrary angels, including perpendicular movement and moving toward or away from the car. Furthermore, a stationary pedestrian on the road must also trigger emergency braking, i.e., a scenario known to be difficult for some pedestrian detection systems. Finally, Figure~\ref{fig:arbitrary_scenario} stresses that SMIRK must be robust against false positives, also know as ``braking for ghosts.'' In this work, this refers to the ML-based component recognizing a pedestrian although another type of object is on collision course (e.g., an animal or a traffic cone) rather than radar noise. Trajectories are illustrated with blue arrows accompanied by a speed ($v$) and possibly an angle ($\theta$). In the superscript, $c$ and $p$ denote car and pedestrian, respectively, and $0$ in the subscript indicates initial speed.

Note that the sole purpose of SMIRK is PAEB. The design of SMIRK assumes deployment in a vehicle with complementary ADAS, e.g., large animal detection, lane keeping assistance, and various types of collision avoidance. We also expect that sensors and actuators will be shared between ADAS. For the SMIRK MVP, however, we do not elaborate further on ADAS co-existence and we do not adhere to any particular higher-level automotive architecture. In the same vein, we do not assume a central perception system that fuses various types of sensor input. SMIRK uses a standalone ML model trained for pedestrian detection and recognition. In the SMIRK terminology, to mitigate confusion, the radar detects objects and the ML-based pedestrian recognition component identifies potential pedestrians in the camera input.

The SMIRK development necessitated quality trade-offs. The software product quality model defined in the ISO/IEC~25010 standard consists of eight characteristics. Furthermore, as recommend in requirements engineering research~\citep{horkoff2019non}, we add the two novel quality characteristics interpretability and fairness. For each characteristic, we share its importance for SMIRK by assigning it a low \textbf{[L]}, medium \textbf{[M]} or high \textbf{[H]} priority. The priorities influence architectural decisions in SMIRK and support elicitation of architecturally significant requirements~\citep{chen2012characterizing}.

\begin{itemize}
    \item \textbf{Functional suitability}. No matter how functionally restricted the SMIRK MVP is, stated and implied needs of a prototype ADAS must be met. \textbf{[H]}
    \item \textbf{Performance efficiency}. SMIRK must be able to process input, do ML inference, and commence emergency braking in realistic driving scenarios. Identifying when performance efficiency reached excessive levels is vital in the requirements engineering process. \textbf{[M]}
    \item \textbf{Compatibility}. SMIRK shall be compatible with other ADAS, but, this is an ambition beyond the MVP development. \textbf{[L]}
    \item \textbf{Usability}. SMIRK is an ADAS that operates in the background without a user interface for direct driver interaction. \textbf{[L]}
    \item \textbf{Reliability}. A top priority in the SMIRK development that motivates the application of AMLAS. \textbf{[H]}%Note, however, that safety is not covered in the ISO/IEC 25010 product quality model but in its complementary quality-in-use model. 
    \item \textbf{Security}. Not included in the SOTIF scope, thus not prioritized. \textbf{[L]}
    \item \textbf{Maintainability}. Evolvability from the SMIRK MVP is a key concern for future projects, thus maintainability is important. \textbf{[M]}
    \item \textbf{Portability}. We plan to port SMIRK to other simulators and physical demonstrators in the  future. Initially, it is not a primary concern. \textbf{[L]}
    \item \textbf{Interpretability}. While interpretability is vital for any cyber-physical system, SMIRK's ML exacerbates the need further. \textbf{[M]}
    \item \textbf{Fairness}. A vital quality that primarily impacts the data requirements specified in the Data Management Specification~\citep{borg2021exploring}. \textbf{[H]}
\end{itemize}

\subsection{Product Functions} \label{sec:prod_func}
SMIRK comprises implementations of four algorithms and uses external vehicle functions. In line with SOTIF, we organize all constituents into the categories sensors, algorithms, and actuators.

\begin{itemize}
    \item Sensors
    \begin{itemize}
        \item Radar detection and tracking of objects in front of the vehicle (see Section~\ref{sec:log_view}).
        \item A forward-facing mono-camera (see Section~\ref{sec:log_view}).
    \end{itemize}
    \item Algorithms
    \begin{itemize}
        \item Time-to-collision (TTC) calculation for objects on collision course.
        \item Pedestrian detection and recognition based on the camera input where the radar detected an object (see Section~\ref{sec:pedrec}).
        \item Out-Of-distribution (OOD) detection of never-seen-before input (part of the safety cage mechanism, see Section~\ref{sec:safetycage}).
        \item A braking module that commences emergency braking. In the SMIRK MVP, maximum braking power is always used.
    \end{itemize}
    \item Actuators
    \begin{itemize}
        \item Brakes (provided by ESI Pro-SiVIC, not elaborated further).
    \end{itemize}
\end{itemize}

\begin{figure}
\centering
\includegraphics[width=0.5\textwidth]{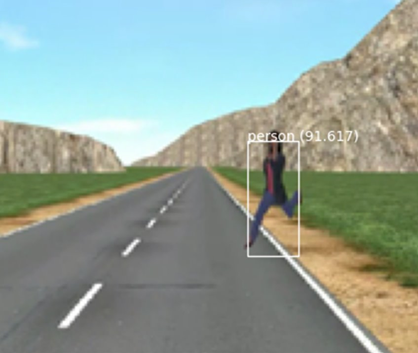}
\caption{Illustrative example of pedestrian detection that shall trigger emergency braking.}
\label{fig:example}
\end{figure}

Figure~\ref{fig:example} illustrates detection of a pedestrian on a collision course, i.e., PAEB shall be commenced. The ML-based functionality of pedestrian detection and recognition, including the corresponding OOD detection, is embedded in the \textbf{Pedestrian Recognition Component} (defined in Section~\ref{sec:log_view}).

\section{SMIRK System Requirements} \label{sec:reqts}
This section specifies the SMIRK system requirements, organized into system safety requirements and ML safety requirements. ML safety requirements are further refined into performance requirements and robustness requirements. The requirements are largely re-purposed from the system for pedestrian detection at crossings described by \citet{gauerhof2020assuring} to our PAEB ADAS, thus allowing for comparisons to previous work within the research community.

\subsection{System Safety Requirements \textbf{[A]}} \label{amlas:a}
\begin{itemize}
    \item \textbf{SYS-SAF-REQ1} SMIRK shall commence automatic emergency braking if and only if collision with a pedestrian on collision course is imminent.
\end{itemize}

Rationale: This is the main purpose of SMIRK. If possible, ego car will stop and avoid a collision. If a collision is inevitable, ego car will reduce speed to decrease the impact severity. Hazards introduced from false positives, i.e., braking for ghosts, are mitigated under ML Safety Requirements.

\subsection{Safety Requirements Allocated to ML Component \textbf{[E]}} \label{amlas:e}
Based on the HARA (see Section~\ref{sec:smile}), two categories of hazards were identified. First, SMIRK might miss pedestrians and fail to commence emergency braking -- we refer to this as a \textit{missed pedestrian}. Second, SMIRK might commence emergency braking when it should not -- we refer to this as an instance of \textit{ghost braking}. 

\begin{itemize}
    \item Missed pedestrian hazard: The severity of the hazard is very high (high risk of fatality). Controllability is high since the driver can brake ego car.
    \item Ghost braking hazard: The severity of the hazard is high (can be fatal). Controllability is very low since the driver would have no chance to counteract the braking.
\end{itemize}

We concluded that the two hazards shall be mitigated by ML safety requirements.

\subsection{Machine Learning Safety Requirements \textbf{[H]}} \label{amlas:h}
This section refines \textbf{SYS-SAF-REQ1} into two separate requirements corresponding to missed pedestrians and ghost braking, respectively.

\begin{itemize}
    \item \textbf{SYS-ML-REQ1}. The pedestrian recognition component shall identify pedestrians in all valid scenarios when the radar tracking component returns a $TTC < 4s$ for the corresponding object.
    \item \textbf{SYS-ML-REQ2} The pedestrian recognition component shall reject false positive input that does not resemble the training data.
\end{itemize}
    
Rationale: \textbf{SYS-SAF-REQ1} is interpreted in light of missed pedestrians and ghost braking and then broken down into the separate ML safety requirements \textbf{SYS-ML-REQ1} and \textbf{SYS-ML-REQ2}. The former requirement deals with the ``if'' aspect of \textbf{SYS-SAF-REQ1} whereas its ``and only if'' aspect is targeted by \textbf{SYS-ML-REQ2}. SMIRK follows the reference architecture from ~\citet{ben_abdessalem_testing_2016} and \textbf{SYS-ML-REQ1} uses the same TTC threshold (4 seconds, confirmed during a research stay in Luxembourg). Moreover, we have validated that this TTC threshold is valid for SMIRK based on calculating braking distances for different car speeds. \textbf{SYS-ML-REQ2} motivates the primary contribution of the SMILE~III project, i.e., an OOD detection mechanism that we refer to as a \textit{safety cage}.

\subsubsection{Performance Requirements}
The performance requirements are specified with a focus on quantitative targets for the pedestrian recognition component. All requirements below are restricted to pedestrians on or close to the road.

For objects detected by the radar tracking component with a TTC $\lt$ 4s, the following requirements must be fulfilled:

\begin{itemize}
    \item \textbf{SYS-PER-REQ1} The pedestrian recognition component shall identify pedestrians with an accuracy of 93\% when they are within 80 m.
    \item \textbf{SYS-PER-REQ2} The false negative rate of the pedestrian recognition component shall not exceed 7\% within 50 m.
    \item \textbf{SYS-PER-REQ3} The false positives per image of the pedestrian recognition component shall not exceed 0.1\% within 80 m.
    \item \textbf{SYS-PER-REQ4} In 99\% of sequences of 5 consecutive images from a 10 FPS video feed, no pedestrian within 80 m shall be missed in more than 20\% of the frames.
    \item \textbf{SYS-PER-REQ5} For pedestrians within 80 m, the pedestrian recognition component shall determine the position of pedestrians within 50 cm of their actual position.
    \item \textbf{SYS-PER-REQ6} The pedestrian recognition component shall allow an inference speed of at least 10 FPS in the ESI Pro-SiVIC simulation.
\end{itemize}

Rationale: SMIRK adapts the performance requirements specified by~\citet{gauerhof2020assuring} for the SMIRK ODD. \textbf{SYS-PER-REQ1} reuses the accuracy threshold from Example 7 in AMLAS, which we empirically validated for SMIRK -- initially in a feasibility analysis, subsequently during system testing. \textbf{SYS-PER-REQ2} and \textbf{SYS-PER-REQ3} are two additional requirements inspired by~\citet{henriksson2019towards}. Note that \textbf{SYS-PER-REQ3} relies on the metric false positive per image rather than false positive rate as true negatives do not exist for object detection (further explained in Section~\ref{amlas:aa} and discussed in Section~\ref{sec:lessons}). \textbf{SYS-PER-REQ6} means that any further improvements to reaction time have a negligible impact on the total brake distance.

\subsubsection{Robustness Requirements}
Robustness requirements are specified to ensure that SMIRK performs adequately despite expected variations in input. For pedestrians present within 50 m of Ego, captured in the field of view of the camera:

\begin{itemize}
    \item \textbf{SYS-ROB-REQ1} The pedestrian recognition component shall perform as required in all situations Ego may encounter within the defined ODD.
    \item \textbf{SYS-ROB-REQ2} The pedestrian recognition component shall identify pedestrians irrespective of their upright pose with respect to the camera.
    \item \textbf{SYS-ROB-REQ3} The pedestrian recognition component shall identify pedestrians irrespective of their size with respect to the camera.
    \item \textbf{SYS-ROB-REQ4} The pedestrian recognition component shall identify pedestrians irrespective of their appearance with respect to the camera.
\end{itemize}

Rationale: SMIRK reuses robustness requirements for pedestrian detection from previous work. \textbf{SYS-ROB-REQ1} is specified in~\citet{gauerhof2020assuring}. \textbf{SYS-ROB-REQ2} is presented as Example 7 in AMLAS, which has been limited to upright poses, i.e., SMIRK is not designed to work for pedestrians sitting or lying on the road. \textbf{SYS-ROB-REQ3} and \textbf{SYS-ROB-REQ4} are additions identified during the Fagan inspection of the System Requirements Specification (see Section~\ref{sec:fagan}).

\subsection{Operational Design Domain \textbf{[B]}} \label{amlas:b}
This section briefly describes the SMIRK ODD. As the complete ODD specification, based on the taxonomy developed by NHTSA \citep{thorn2018framework}, is lengthy, we only present the fundamental aspects in this section. We refer interested readers to the GitHub repository. Note that we deliberately specified a minimalistic ODD, i.e., ideal conditions, to allow the development a complete safety case for the SMIRK MVP.

\begin{itemize}
    \item \textbf{Physical infrastructure} Asphalt single-lane roadways with clear lane markings and a gravel shoulder. Rural settings with open green fields.
    \item \textbf{Operational constraints} Maximum speed of 70~km/h and no surrounding traffic.
    \item \textbf{Objects} No objects except 0-1 pedestrians, either stationary or moving with a constant speed ($\lt$ 15~ km/h) and direction.
    \item \textbf{Environmental Conditions} Clear daytime weather with overhead sun. Headlights turned off. No particulate matter leading to dirt on the sensors.
\end{itemize}

\section{SMIRK System Architecture} \label{sec:arch}
SMIRK is a pedestrian emergency braking ADAS that demonstrates safety-critical ML-based driving automation on SAE Level 1. The system uses input from two sensors (camera and radar/LiDAR) and implements a DNN trained for pedestrian detection and recognition. If the radar detects an imminent collision between the ego car and an object, SMIRK will evaluate if the object is a pedestrian. If SMIRK is confident that the object is a pedestrian, it will apply emergency braking. To minimize hazardous false positives, SMIRK implements a SMILE safety cage to reject input that is OOD. To ensure industrial relevance, SMIRK builds on the reference architecture from PeVi, an ADAS studied in previous work by~\cite{ben_abdessalem_testing_2016}.

%Based on a stakeholder analysis in the SMILE~III project, this architecture description considers the following stakeholders:

%\begin{itemize}
%    \item Researchers who want to study the design of SMIRK.
%    \item Safety assessors who want to investigate the general design in the light of the safety case.
%    \item Software developers building or evolving SMIRK.
%    \item ML developers designing and tuning the ML perception model.
%    \item Hardware developers interested in the SMIRK sensors, incl. replacing them or adding sensor fusion.
%    \item Simulator developers looking for ways to port SMIRK to their virtual prototyping environments.
%    \item Testers developing test plans for SMIRK.
%    \item System integrators who are about to include SMIRK in other systems, incl. co-existence with other ADAS.
%\end{itemize}

Explicitly defined architecture viewpoints support effective communication of certain aspects and layers of a system architecture. The different viewpoints of the identified stakeholders are covered by the established 4+1 view of architecture by~\citet{kruchten19954+}. For SMIRK, we describe the logical view using a simple illustration with limited embedded semantics complemented by textual explanations. The process view is presented through a bulleted list, whereas the interested reader can find the remaining parts in the GitHub repository~\citep{github_smirk}. Scenarios are illustrated with figures and explanatory text.

\subsection{Logical View} \label{sec:log_view}
\begin{figure}
\centering
\includegraphics[width=1\textwidth]{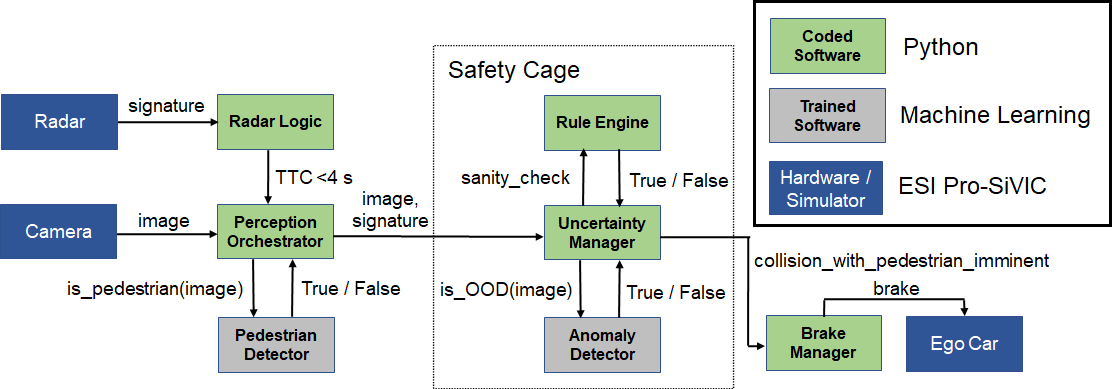}
\caption{SMIRK logical view.}
\label{fig:logical_view}
\end{figure}

The SMIRK logical view is constituted by a description of the entities that realize the PAEB. Figure~\ref{fig:logical_view} provides a graphical description.

SMIRK interacts with three external resources, i.e., hardware sensors and actuators in ESI Pro-SiVIC: A) Mono Camera (752$\times$480 (WVGA), sensor dimension 3.13 cm $\times$ 2.00 cm, focal length 3.73 cm, angle of view 45 degrees), B) Radar unit (providing object tracking and relative lateral and longitudinal speeds), and C) Ego Car (An Audi A4 for which we are mostly concerned with the brake system). SMIRK consists of the following constituents. We refer to E), F), G), I), and J) as the \textbf{Pedestrian Recognition Component}, i.e., the ML-based component for which this study presents a safety case.

\begin{itemize}
    \item Software components implemented in Python:
    \begin{itemize}
        \item D) Radar Logic (calculating TTC based on relative speeds)
        \item E) Perception Orchestrator (the overall perception logic)
        \item F) Rule Engine (part of the safety cage, implementing heuristics such as pedestrians do not fly in the air)
        \item G) Uncertainty Manager (main part of the safety cage, implementing logic to avoid false positives)
        \item H) Brake Manager (calculating and sending brake signals to the ego car)
    \end{itemize}
    \item Trained Machine Learning models:
    \begin{itemize}
        \item I) Pedestrian Detector (a YOLOv5 model trained using PyTorch\footnote{\url{https://pytorch.org/}}
        \item J) Anomaly Detector (an autoencoder provided by Seldon\footnote{\url{https://www.seldon.io/}})
    \end{itemize}
\end{itemize}

\subsection{Process View}
The process view deals with the dynamic aspects of SMIRK including an overview of the run time behavior of the system. The overall SMIRK flow is as follows:

\begin{enumerate}
    \item The Radar detects an object and sends the signature to the Radar Logic class.
    \item The Radar Logic class calculates the TTC. If a collision between the ego car and the object is imminent, i.e., TTC is less than 4 seconds assuming constant motion vectors, the Perception Orchestrator is notified.
    \item The Perception Orchestrator forwards the most recent image from the Camera to the Pedestrian Detector to evaluate if the detected object is a pedestrian.
    \item The Pedestrian Detector performs a pedestrian detection in the image and returns the verdict (True/False) to the Pedestrian Orchestrator.
    \item If there appears to be a pedestrian on a collision course, the Pedestrian Orchestrator forwards the image and the radar signature to the Uncertainty Manager in the safety cage.
    \item The Uncertainty Manager sends the image to the Anomaly Detector and requests an analysis of whether the camera input is OOD or not.
    \item The Anomaly Detector analyzes the image in the light of the training data and returns its verdict (True/False).
    \item If there indeed appears to be an imminent collision with a pedestrian, the Uncertainty Manager forwards all available information to the Rule Engine for a sanity check.
    \item The Rule Engine does a sanity check based on heuristics, e.g., in relation to laws of physics, and returns a verdict (True/False).
    \item The Uncertainty Manager aggregates all information and, if the confidence is above a threshold, notifies the Brake Manager that collision with a pedestrian is imminent.
    \item The Brake Manager calculates a safe brake level and sends the signal to Ego Car to commence PAEB.
\end{enumerate}

\section{SMIRK Data Management Specification} \label{sec:dms}
This section describes the overall approach to data management for SMIRK and the explicit data requirements. SMIRK is a demonstrator for a simulated environment. Thus, as an alternative to longitudinal traffic observations and consideration of accident statistics, we have analyzed the SMIRK ODD through the ESI Pro-SiVIC ``Object Catalog.'' We conclude that the demographics of pedestrians in the ODD is constituted of the following: adult males and females in either casual, business casual, or business casual clothes, young boys wearing jeans and a sweatshirt, and male road workers. As other traffic is not within the ODD (e.g., cars, motorcycles, and bicycles), we consider the following basic shapes from the object catalog to as examples of OOD objects (that still can appear in the ODD) for SMIRK to handle in operation: boxes, cones, pyramids, spheres, and cylinders.

\subsection{Data Requirements \textbf{[L]}} \label{amlas:l}
This section specifies requirements on the data used to train and test the pedestrian recognition component. The data requirements are specified to comply with the ML Safety Requirements in the SRS. All data requirements are organized according to the assurance-related desiderata proposed by~\citet{ashmore2021assuring}, i.e., the key assurance requirements that ensure that the data set is relevant, complete, balanced, and accurate.

Table~\ref{tab:data_reqts_matrix} shows a requirements traceability matrix between ML safety requirements and data requirements. The matrix presents an overview of how individual data requirements contribute to the satisfaction of ML Safety Requirements. Entries in individual cells denote that the ML safety requirement is addressed, at least partly, by the corresponding data requirement. \textbf{SYS-PER-REQ4} and \textbf{SYS-PER-REQ6} are not related to the data requirements.

\begin{table}
\centering
\caption{Requirements-Data traceability matrix.}
\label{tab:data_reqts_matrix}
\includegraphics[width=1\textwidth]{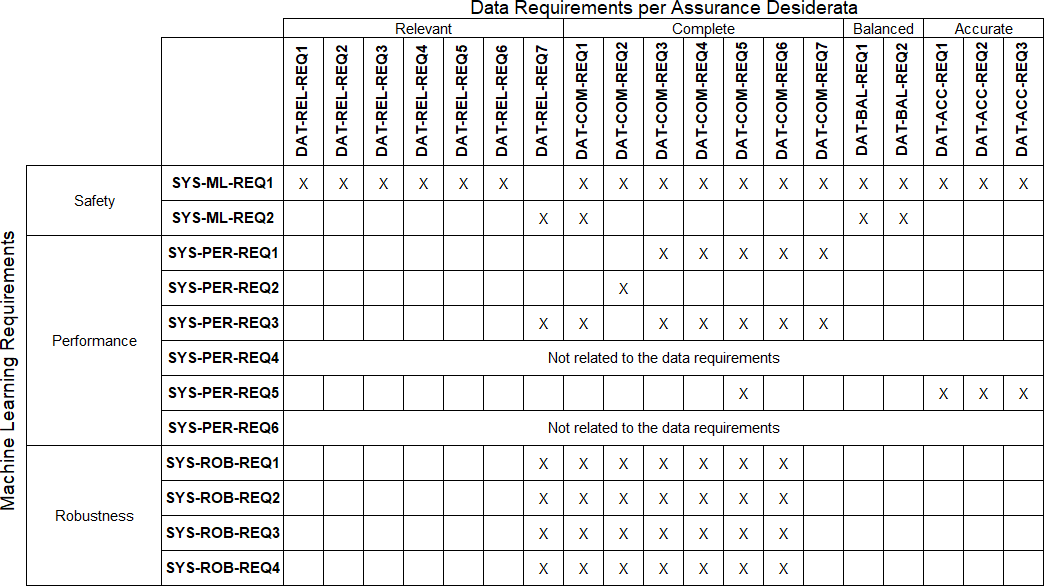}
\end{table}

\subsubsection{Desideratum: Relevant}
This desideratum considers the intersection between the data set and the supported dynamic driving task in the ODD. The SMIRK training data will not cover operational environments that are outside of the ODD, e.g., images collected in heavy snowfall.

\begin{itemize}
    \item \textbf{DAT-REL-REQ1} All data samples shall represent images of a road from the perspective of a vehicle.
    \item \textbf{DAT-REL-REQ2} The format of each data sample shall be representative of that which is captured using sensors deployed on the ego car.
    \item \textbf{DAT-REL-REQ3} Each data sample shall assume sensor positioning representative of the positioning used on the ego car.
    \item \textbf{DAT-REL-REQ4} All data samples shall represent images of a road environment that corresponds to the ODD.
    \item \textbf{DAT-REL-REQ5} All data samples containing pedestrians shall include one single pedestrian.
    \item \textbf{DAT-REL-REQ6} Pedestrians included in data samples shall be of a type that may appear in the ODD.
    \item \textbf{DAT-REL-REQ7} All data samples representing non-pedestrian OOD objects shall be of a type that may appear in the ODD.
\end{itemize}
    
Rationale: SMIRK adapts the requirements from the Relevant desiderata specified by \citet{gauerhof2020assuring} for the SMIRK ODD. \textbf{DAT-REL-REQ5} is added based on the corresponding fundamental restriction of the ODD of the SMIRK MVP. \textbf{DAT-REL-REQ7} restricts data samples providing OOD examples for testing.

\subsubsection{Desideratum: Complete}
This desideratum considers the sampling strategy across the input domain and its subspaces. Suitable distributions and combinations of features are particularly important. \citet{ashmore2021assuring} refer to this as the external perspective on the data.

\begin{itemize}
    \item \textbf{DAT-COM-REQ1} The data samples shall include the complete range of environmental factors within the scope of the ODD.
    \item \textbf{DAT-COM-REQ2} The data samples shall include images representing all types of pedestrians according to the demographics of the ODD.
    \item \textbf{DAT-COM-REQ3} The data samples shall include images representing pedestrians paces from standing still up to running at $15 km/h$.
    \item \textbf{DAT-COM-REQ4} The data samples shall include images representing all angles an upright pedestrian can be captured by the given sensors on the ego car.
    \item \textbf{DAT-COM-REQ5} The data samples shall include images representing all distances to crossing pedestrians from 10 m up to 100 m away from ego car.
    \item \textbf{DAT-COM-REQ6} The data samples shall include examples with different levels of occlusion giving partial views of pedestrians crossing the road.
    \item \textbf{DAT-COM-REQ7} The data samples shall include a range of examples reflecting the effects of identified system failure modes.
\end{itemize}
    
Rationale: SMIRK adapts the requirements from the Complete desiderata specified by \citet{gauerhof2020assuring} for the SMIRK ODD. We deliberately replaced the original adjective ``sufficient'' to make the data requirements more specific. Furthermore, we add \textbf{DAT-COM-REQ3} to cover different poses related to the pace of the pedestrian and \textbf{DAT-COM-REQ4} to cover different observation angles.

\subsubsection{Desideratum: Balanced}
This desideratum considers the distribution of features in the data set, e.g., the balance between the number of samples in each class. \citet{ashmore2021assuring} refer to this as an internal perspective on the data.

\begin{itemize}
    \item \textbf{DAT-BAL-REQ1} The data set shall have a representation of samples for each relevant class and feature that ensures AI fairness with respect to gender.
    \item \textbf{DAT-BAL-REQ2} The data set shall have a representation of samples for each relevant class and feature that ensures AI fairness with respect to age.
    \item \textbf{DAT-BAL-REQ3} The data set shall contain both positive and negative examples.
\end{itemize}
    
Rationale: SMIRK adapts the requirements from the Balanced desiderata specified by \citet{gauerhof2020assuring} for the SMIRK ODD. The concept of AI fairness is to be interpreted in the light of the Ethics guidelines for trustworthy AI published by the European Commission \citep{eu2019}. Note that the number of ethical dimensions that can be explored in through the ESI Pro-SiVIC object catalog is limited to gender (\textbf{DAT-BAL-REQ1}) and age (\textbf{DAT-BAL-REQ2}). Moreover, the object catalog does only contain male road workers and all children are boys. Furthermore, \textbf{DAT-BAL-REQ3} is primarily included to align with \citet{gauerhof2020assuring} and to preempt related questions by safety assessors. In practice, the concept of negative examples when training object detection models are typically satisfied implicitly as the parts of the images that do not belong to the annotated class are true negatives (further explained in Section~\ref{amlas:aa}).

\subsubsection{Desideratum: Accurate}
This desideratum considers how measurement issues can affect the way that samples reflect the intended ODD, e.g., sensor accuracy and labeling errors.

\begin{itemize}
    \item \textbf{DAT-ACC-REQ1}: All bounding boxes produced shall include the entirety of the pedestrian.
    \item \textbf{DAT-ACC-REQ2}: All bounding boxes produced shall be no more than 10\% larger in any dimension than the minimum sized box capable of including the entirety of the pedestrian.
    \item \textbf{DAT-ACC-REQ3}: All pedestrians present in the data samples shall be correctly labeled.
\end{itemize}
    
Rationale: SMIRK reuses the requirements from the Accurate desiderata specified by \citet{gauerhof2020assuring}.

\subsection{Data Generation Log \textbf{[Q]}} \label{amlas:q}
This section describes how the data used for training the ML model in the pedestrian recognition component was generated. Based on the data requirements, we generate data using ESI Pro-SIVIC. The data are split into three sets in accordance with AMLAS.

\begin{itemize}
    \item Development data: Covering both training and validation data used by developers to create models during ML development.
    \item Internal test data: Used by developers to test the model.
    \item Verification data: Used in the independent test activity when the model is ready for release.
\end{itemize}

The SMIRK data collection campaign focuses on generation of annotated data in ESI Pro-SiVIC. All data generation is script-based and fully reproducible. Section~\ref{sec:positive_examples} describes positive examples [PX], i.e., humans that shall be classified as pedestrians. Section~\ref{sec:ood_examples} describes examples that represent OOD shapes [NX], i.e., objects that shall not initiate PAEB in case of an imminent collision. These images, referred to as OOD examples, shall either not be recognized as a pedestrian or be rejected by the safety cage (see Section~\ref{sec:safetycage}).

In the data collection scripts, ego car is always stationary whereas pedestrians and objects move according to specific configurations. The parameters and values were selected to systematically cover the ODD. Finally, images are sampled from the camera at 10 frames per second with a resolution of $752 \times 480$ pixels. For each image, we add a separate image file containing the ground truth pixel-level annotation of the position of the pedestrian. In total, we generate data representing $8 \times 616 = 4,928$ execution scenarios with positive examples and $5 \times 20 = 100$ execution scenarios with OOD examples. In total, the data collection campaign generates roughly 185~GB of image data, annotations, and meta-data (including bounding boxes).

\subsubsection{Positive Examples} \label{sec:positive_examples}
We generate positive examples from humans with eight visual appearances (see the upper part of Figure~\ref{fig:example_objects}) available in the ESI Pro-SiVIC object catalog:

\begin{itemize}
    \item[\textbf{P1}] Casual female pedestrian
    \item[\textbf{P2}] Casual male pedestrian
    \item[\textbf{P3}] Business casual female pedestrian
    \item[\textbf{P4}] Business casual male pedestrian
    \item[\textbf{P5}] Business female pedestrian
    \item[\textbf{P6}] Business male pedestrian
    \item[\textbf{P7}] Child
    \item[\textbf{P8}] Male construction worker
\end{itemize}
    
For each of the eight visual appearances, we specify the execution of 616 scenarios in ESI Pro-SiVIC organized into four groups (A-D). The pedestrians always follow rectilinear motion (a straight line) at a constant speed during scenario execution. Groups A and B describe pedestrians crossing the road, either from the left (Group A) or from the right (Group B). There are three variation points, i.e., 1) the speed of the pedestrian, 2) the angle at which the pedestrian crosses the road, and 3) the longitudinal distance between ego car and the pedestrian's starting point. In all scenarios, the distance between the starting point of the pedestrian and the edge of the road is 5~m.

\begin{itemize}
    \item A. Crossing the road from left to right (4 $\times$ 7 $\times$ 10 = 280 scenarios)
    \begin{itemize}
        \item Speed (m/s): [1, 2, 3, 4]
        \item Angle (degree): [30, 50, 70, 90, 110, 130, 150]
        \item Longitudinal distance (m): [10, 20, 30, 40, 50, 60, 70, 80, 90, 100]
    \end{itemize}
    \item B. Crossing the road from right to left (4 $\times$ 7 $\times$ 10 = 280 scenarios)
    \begin{itemize}
        \item Speed (m/s): [1, 2, 3, 4]
        \item Angle (degree): [30, 50, 70, 90, 110, 130, 150]
        \item Longitudinal distance (m): [10, 20, 30, 40, 50, 60, 70, 80, 90, 100]
    \end{itemize}
\end{itemize}

Groups C and D describe pedestrians moving parallel to the road, either toward ego car (Group C) or away (Group D). There are two variation points, i.e., 1) the speed of the pedestrian and 2) an offset from the road center. The pedestrian always moves 90~m, with a longitudinal distance between ego car and the pedestrian's starting point of 100~m for Group C (towards) and 10~ m for Group D (away).

\begin{itemize}
    \item C. Movement parallel to the road toward ego car (4 $\times$ 7= 28 scenarios)
    \begin{itemize}
        \item Speed (m/s): [1, 2, 3, 4]
        \item Lateral offset (m): [-3, -2, -1, 0, 1, 2, 3]
    \end{itemize}
    \item D. Movement parallel to the road away from ego car (4 $\times$ 7= 28 scenarios)
    \begin{itemize}
        \item Speed (m/s): [1, 2, 3, 4]
        \item Lateral offset (m): [-3, -2, -1, 0, 1, 2, 3]
    \end{itemize}
\end{itemize}

\subsubsection{Out-of-Distribution Examples} \label{sec:ood_examples}
\begin{figure}
\centering
\includegraphics[width=1\textwidth]{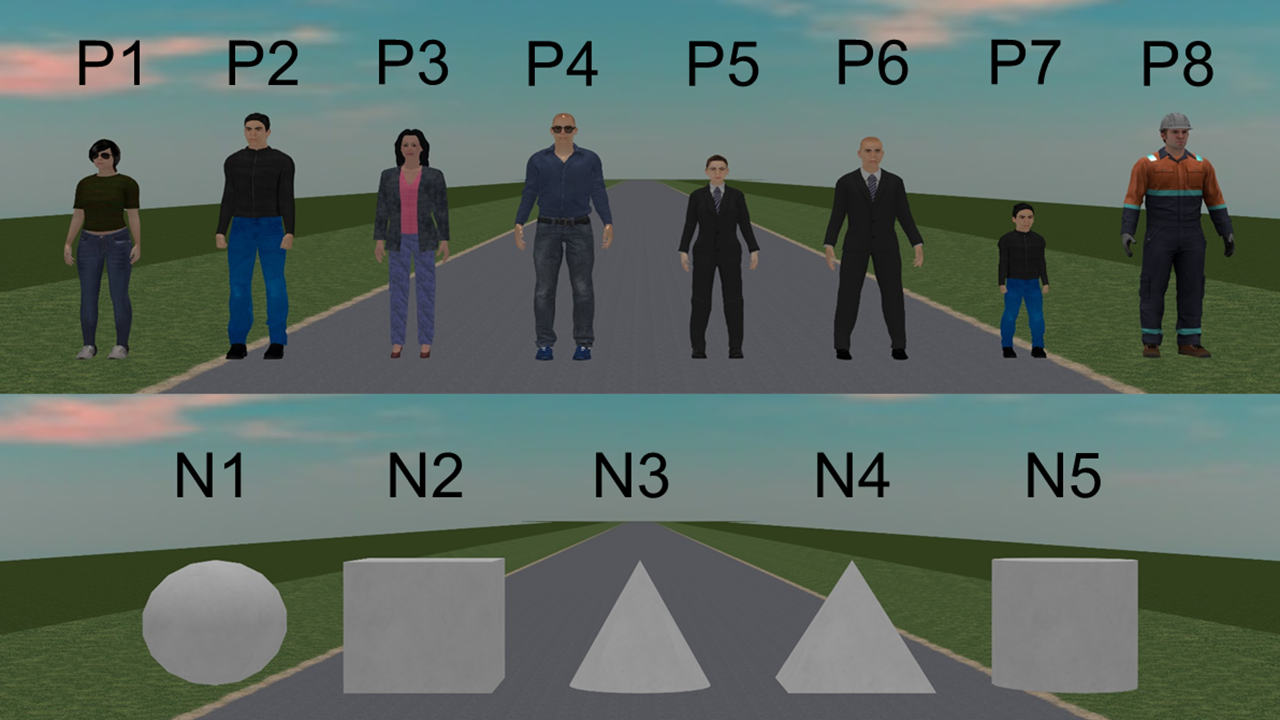}
\caption{Visual appearance of pedestrians (P1--P8) and basic shapes (N1--N5).}
\label{fig:example_objects}
\end{figure}

We generate OOD examples using five basic shapes (see the lower part of Figure~\ref{fig:example_objects}) available in the ESI Pro-SiVIC object catalog:

\begin{itemize}
    \item[\textbf{N1}] Sphere
    \item[\textbf{N2}] Cube
    \item[\textbf{N3}] Cone
    \item[\textbf{N4}] Pyramid
    \item[\textbf{N5}] Cylinder
\end{itemize}

For each of the five basic shapes, we specify the execution of 20 scenarios in ESI Pro-SiVIC. The scenarios represent a basic shape crossing the road from the left or right at an angle perpendicular to the road. Since basic shapes are not animated, we fix the speed at 4~m/s. Moreover, as lateral offsets and different angles make little to no difference in front of the camera, we disregard these variation points. In all scenarios, the distance between the starting point of the basic shape and the edge of the road is 5~m. The only variation points are the crossing direction and the longitudinal distance between ego car and the objects' starting point. As for pedestrians, the objects always follow a rectilinear motion at a constant speed during scenario execution.

\begin{itemize}
    \item Crossing direction: [left, right]
    \item Longitudinal distance (m): [10, 20, 30, 40, 50, 60, 70, 80, 90, 100]
\end{itemize}

\subsubsection{Preprocessing and Data Splitting} \label{sec:data_split}
As the SMIRK data collection campaign relies on data generation in ESI Pro-SiVIC, the need for pre-processing differs from counterparts using naturalistic data. The script-based data generation ensures that the crossing pedestrians and objects appear at the right distance with specified conditions and with controlled levels of occlusion. All output images share the same characteristics, thus no normalization is needed. SMIRK includes a script to generate bounding boxes for training the object detection model. ESI Pro-SiVIC generates ground truth image segmentation on a pixel-level. The script is used to convert the output to the appropriate input format for model training.

The development data contains images with no pedestrians, in line with the description of ``background images'' in the YOLOv5 training tips provided by Ultralytics\footnote{\url{https://github.com/ultralytics/yolov5/wiki/Tips-for-Best-Training-Results}}. Background images have no objects for the model to detect, and are added to reduce FPs. Ultralytics recommends 0-10\% background images to help reduce FPs and reports that the fraction of background images in the well-known COCO data set is 1\% \citep{lin2014microsoft}. In our case, we add background images with cylinders (N5) to the development data. In total, the SMIRK development data contains 1.98\% background images, i.e., 1.75\% images without any objects and 0.23\% with a cylinder.

The generated data are used in three sequestered (separated) data sets:

\begin{itemize}
    \item Development data: \textbf{P2}, \textbf{P3}, \textbf{P6}, and \textbf{N5}
    \item Internal test data: \textbf{P1}, \textbf{P4}, \textbf{N1}, and \textbf{N3}
    \item Verification data: \textbf{P5}, \textbf{P7}, \textbf{P8}, \textbf{N2}, and \textbf{N4}
\end{itemize}

Note that we deliberately avoid mixing pedestrian models from the ESI Pro-SiVIC object catalog in the data sets due to the limited diversity in the images within the ODD.

\section{Machine Learning Component Specification} \label{sec:mlspec} \label{amlas:d}
The pedestrian recognition component consists of two ML-based constituents: a pedestrian detector and an anomaly detector (see Figure~\ref{fig:logical_view}).

\subsection{Pedestrian Detection Using YOLOv5s} \label{sec:pedrec}
SMIRK implements its pedestrian recognition component using the third-party OSS framework YOLOv5 by Ultralytics. Based on Ultralytics' publicly reported experiments on real-time characteristics of different YOLOv5 architectures\footnote{\url{https://github.com/ultralytics/yolov5}}, we found that YOLOv5s stroke the best balance between inference time and accuracy for SMIRK. After validating the feasibility in our experimental setup, we proceeded with this ML architecture selection.

The pedestrian recognition component uses the YOLOv5 architecture without any modifications. YOLOv5s has 191 layers and $\approx$7.5 million parameters. We use the default configurations proposed in YOLOv5s regarding activation, optimization, and cost functions. As activation functions, YOLOv5s uses Leaky ReLU in the hidden layers and the sigmoid function in the final layer. We use the default optimization function in YOLOv5s, i.e., stochastic gradient descent. The default cost function in YOLOv5s is binary cross-entropy with logits loss as provided in PyTorch, which we also use. We refer the interested reader to further details provided by \citet{rajput2020} and Ultralytics' GitHub repository. 

%Figure~\ref{fig:yolov5_benchmarking} shows the speed/accuracy tradeoffs for different YOLOv5 architectures with YOLOv5s depicted in orange. The results are provided by Ultralytics including instructions for reproduction. On the y-axis, COCO AP val denotes the mAP@0.5:0.95 metric measured on the 5,000-image COCO val2017 data set over various inference sizes from 256 to 1,536. On the x-axis, GPU Speed measures average inference time per image on the COCO val2017 data set using an AWS p3.2xlarge V100 instance at batch-size 32. The curve EfficientDet illustrates results from Google AutoML at batch size 8.

%\begin{figure}
%\centering
%\includegraphics[width=1\textwidth]{yolov5_benchmarking.png}
%\caption{Speed/accuracy tradeoffs for different YOLOv5 architectures. (Image source: Ultralytics under GPLv3)}
%\label{fig:yolov5_benchmarking}
%\end{figure}

\subsection{Model Development Log \textbf{[U]}} \label{amlas:u}
This section describes how the YOLOv5s model has been trained for the SMIRK MVP. We followed the general process presented by Ultralytics for training on custom data.

First, we manually prepared two SMIRK data sets to match the input format of YOLOv5. In this step, we also divided the development data \textbf{[N]} into two parts. The first part containing approximately 80\% of development data, was used for training. The second part, consisting of the remaining data, was used for validation. Camera frames from the same video sequence were kept together in the same partition to avoid having almost identical images in the training and validation sets. Additionally, we kept the distribution of objects and scenario types consistent in both partitions. The internal test data \textbf{[O]} was used as a test set. We then prepared these three data sets, training, validation, and test, according to Ultralytic's instructions and trained YOLOv5 for a single class, i.e., pedestrians. The data sets were already annotated using ESI Pro-SiVIC, thus we only needed to export the labels to the YOLO format with one txt-file per image. Finally, we organize the individual files (images and labels) according to the YOLOv5 instructions. More specifically, each label file contains the following information:

\begin{itemize}
    \item One row per object.
    \item Each row contains class, x\_center, y\_center, width, and height.
    \item Box coordinates are stored in normalized xywh format (from 0 -- 1).
    \item Class numbers are zero-indexed, i.e., they start from 0.
\end{itemize}

Second, we trained a YOLO model using the YOLOv5s architecture with the development data without any pre-trained weights. The model was trained for 10 epochs with a batch-size of 8. The results from the validation subset (27,843 images in total) of the development data guide the selection of the confidence threshold for the ML model. We select a threshold to meet \textbf{SYS-PER-REQ3} with a safety margin for the development data, i.e., an FPPI of 0.1\%. This yields a confidence threshold for the ML model to classify an object as a pedestrian that equals 0.448. The final pedestrian detection model, i.e., the ML model \textbf{[V]}, has a size of $\approx$ 14~MB.

\subsection{OOD Detection for the Safety Cage Architecture} \label{sec:safetycage}
SMIRK detects OOD input images as part of its safety cage architecture. The OOD detection relies on the OSS third-party library Alibi Detect\footnote{\url{https://github.com/SeldonIO/alibi-detect}} from Seldon. Alibi Detect is a Python library that provides several algorithms for outlier, adversarial, and drift detection for various types of data~\citep{klaise2020monitoring}. For SMIRK, we trained Alibi Detect's autoencoder for outlier detection, with three convolutional and deconvolutional layers for the encoder and decoder respectively. 

\begin{figure}
\centering
\includegraphics[width=0.8\textwidth]{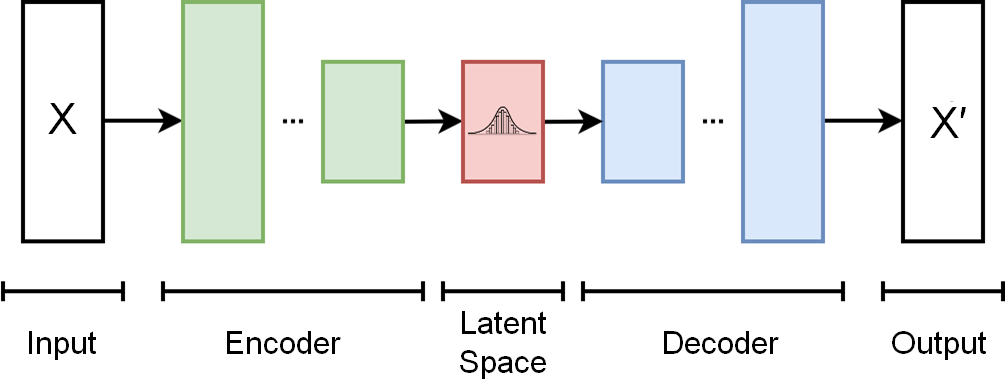}
\caption{Overview architecture of an autoencoder. Adapted from WikiUser:EugenioTL (CC BY-SA 4.0)}
\label{fig:vae}
\end{figure}

Figure~\ref{fig:vae} shows an overview of the DNN architecture of an autoencoder. An encoder and a decoder are trained jointly in two steps to minimize a \textit{reconstruction error}. First, the autoencoder receives input data $X$ and encodes it into a latent space of fewer dimensions. Second, the decoder tries to reconstruct the original data and produces output $X'$. \citet{an2015variational} proposed using the reconstruction error from a autoencoder to identify input that differs from the training data. Intuitively, if \textit{inlier} data is processed by the autoencoder, the difference between $X$ and $X'$ will be smaller than for \textit{outlier} data, i.e., OOD data will stand out. By carefully selecting a tolerance threshold, this approach can be used for OOD detection.

For SMIRK, we trained Alibi Detect's autoencoder for OOD detection on the training data subset of the development data. The encoder part is designed with three convolutional layers followed by a dense layer resulting in a bottleneck that compresses the input by 96.66\%. The latent dimension is limited to 1,024 variables to limit requirements on processing VRAM of the GPU. The reconstruction error from the autoencoder is measured as the mean squared error between the input and the reconstructed instance. The mean squared error is used for OOD detection by computing the reconstruction error and considering an input image as an outlier if the error surpasses a threshold $\theta$. The threshold used for OOD detection in SMIRK is 0.004, roughly corresponding to the threshold that rejects a number of samples that equals the amount of outliers in the validation set. As explained in Section~\ref{amlas:dd}, the OOD detection is only active for objects at least 10 m away from ego car as the results for close-up images are highly unreliable. Furthermore, as the constrained SMIRK ODD ensures that only one single object appears in each scenario, the safety cage architecture applies the policy ``once an anomaly, always an anomaly'' -- objects that get rejected once will remain anomalous no matter what subsequent frames might contain.

\section{SMIRK System Test Specification} \label{sec:test}
This section describes the overall SMIRK test strategy. The ML-based pedestrian recognition component is tested on multiple levels. We focus on four aspects of the ML testing scope facet proposed by \citet{song2022exploring}:

\begin{itemize}
    \item Data set testing: This level refers to automatic checks that verify that specific properties of the data set are satisfied. As described in the ML Data Validation Results, the data validation, presented in Section~\ref{amlas:s}, includes automated testing of the Balance desiderata. Since the SMIRK MVP relies on synthetic data, the distribution of pedestrians is already ensured by the scripts. However, other distributions such as distances to objects and bounding box sizes are important targets for data set testing. %\citet{zhang2022machine} refer to data set testing as input testing.
    \item Model testing: Testing that the ML model provides the expected output. This is the primary focus of academic research on ML testing, and includes white-box, black-box, and data-box access levels during testing \citep{riccio2020testing}. SMIRK model testing is done independently from model development and results in ML Verification Results \textbf{[Z]} as described in Section~\ref{amlas:z}.
    \item Unit testing: Conventional unit testing on the level of Python classes. A test suite developed for the pytest framework is maintained by the developers and not elaborated further in this paper.
    \item System testing: System-level testing of SMIRK based on a set of Operational Scenarios \textbf{[EE]}. All test cases are designed for execution in ESI Pro-SiVIC. The system testing targets the requirements in the System Requirements Specification. This level of testing results in Integration Testing Results \textbf{[FF]} presented in Section~\ref{amlas:ff}.
\end{itemize}
    
\subsection{ML Model Testing \textbf{[AA]}} \label{amlas:aa}
This section corresponds to the Verification Log \textbf{[AA]} in AMLAS Step 5, i.e., Model Verification Assurance. Here we explicitly document the ML Model testing strategy, i.e., the range of tests undertaken and bounds and test parameters motivated by the SMIRK system requirements.

The testing of the ML model is based on assessing the object detection accuracy for the sequestered verification data set. A fundamental aspect of the verification argument is that this data set was never used in any way during the development of the ML model. To further ensure the independence of the ML verification, engineers from Infotiv, part of the SMILE~III research consortium, led the corresponding verification and validation work package and were not in any way involved in the development of the ML model. As described in the Machine Learning Component Specification (see Section~\ref{sec:mlspec}), the ML development was led by Semcon with support from RISE Research Institutes of Sweden.

The ML model test cases provide results for both 1) the entire verification data set and 2) eight slices of the data set that are deemed particularly important. The selection of slices was motivated by either an analysis of the available technology or ethical considerations, especially from the perspective of AI fairness \citep{borg2021exploring}. Consequently, we measure the performance for the following slices of data. Identifiers in parentheses show direct connections to requirements.

\begin{itemize}
    \item[S1] The entire verification data set
    \item[S2] Pedestrians close to the ego car (longitudinal distance $\lt$ 50~m) (\textbf{SYS-PER-REQ1, SYS-PER-REQ2})
    \item[S3] Pedestrians far from the ego car (longitudinal distance $\ge$ 50~m)
    \item[S4] Running pedestrians (speed $\ge$ 3~m/s) (\textbf{SYS-ROB-REQ2})
    \item[S5] Walking pedestrians (speed $\gt$ 0~m/s but $\lt$ 3~m/s) (\textbf{SYS-ROB-REQ2})
    \item[S6] Occluded pedestrians (entering or leaving the field of view, defined as bounding box in contact with any edge of image) (\textbf{DAT-COM-REQ4})
    \item[S7] Male pedestrians (\textbf{DAT-COM-REQ2})
    \item[S8] Female pedestrians (\textbf{DAT-COM-REQ2})
    \item[S9] Children (\textbf{DAT-COM-REQ2})
\end{itemize}

Evaluating the output from an object detection model in computer vision is non-trivial. We rely on the established IoU metric to evaluate the accuracy of the YOLOv5 model. After discussions in the development team, supported by visualizations\footnote{\url{https://zapirfe.com/docs/visualizing-ml/iou.html}}, we set the target at 0.5. We recognize that there are alternative measures tailored for pedestrian detection, such as the log-average miss rate \citep{dollar2011pedestrian} but we find such metrics to be unnecessarily complex for the restricted SMIRK ODD with a single pedestrian. There are also entire toolboxes that can be used to assess object detection \citep{bolya2020tide}.
Whereas more complex metrics could be used, we decide to use IoU in SMIRK's safety argumentation. Relying on a simpler metric supports interpretability, which is a vital to any safety case, especially if ML is involved~\citep{jia2022role}.
%In our safety argumentation, however, we argue that the higher interpretability of a simpler -- but valid -- evaluation metric outweighs the potential benefits of a customized metric for a more complex ODD.  

Even using the standard IoU metric to assess how accurate SMIRK’s ML model is, the evaluation results are not necessarily intuitive to non-experts. Each image in the SMIRK data set either has a ground truth bounding box containing the pedestrian or no bounding box at all. Similarly, when performing inference on an image, the ML model will either predict a bounding box containing a potential pedestrian or no bounding box at all. IoU is the intersection over the union of the two bounding boxes. An IoU of 1 implies a perfect overlap. For the ML model in SMIRK, we evaluate pedestrian detection at IoU = 0.5, which for each image means:

\begin{itemize}
    \item[TP] True positive: IoU $\ge$ 0.5
    \item[FP] False positive: IoU $\lt$ 0.5
    \item[FN] False negative: There is a ground truth bounding box in the image, but no predicted bounding box.
    \item[TN] True negative: All parts of the image with neither a ground truth nor a predicted bounding box. This output carries no meaning in our case.
\end{itemize}

\begin{figure}
\centering
\includegraphics[width=1\textwidth]{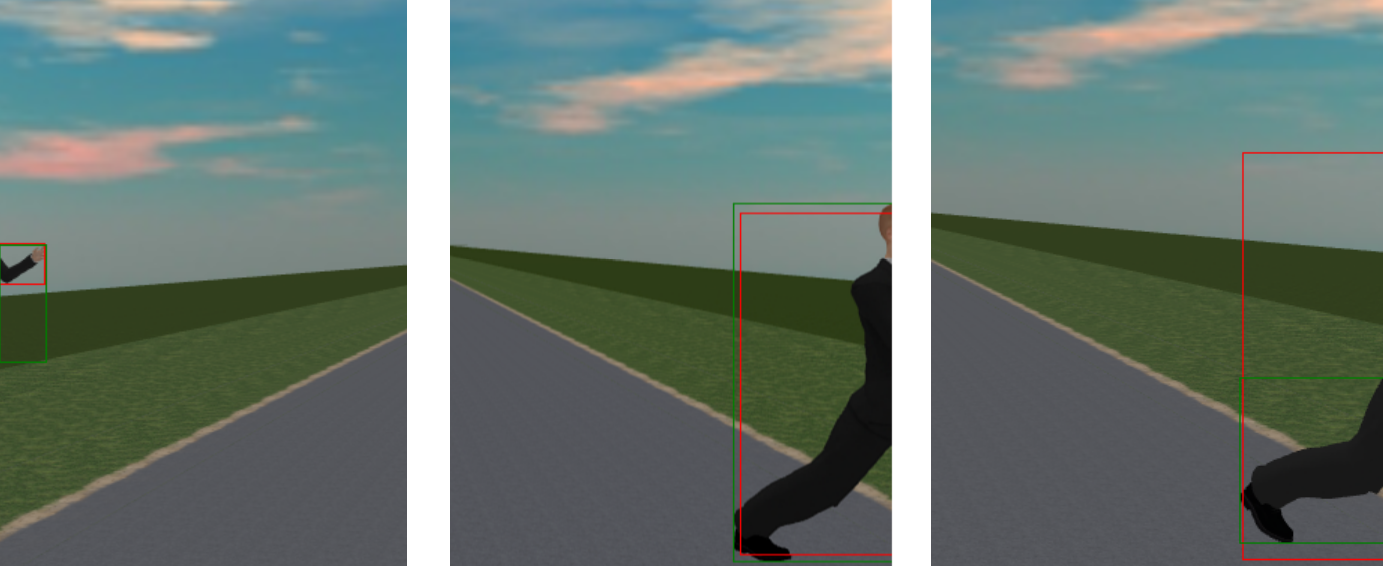}
\caption{Example predictions from the SMIRK ML model. The center image represents a TP, whereas the left and right examples are FPs with IoU scores of 0.3 and 0.4, respectively.}
\label{fig:FPs}
\end{figure}

Figure~\ref{fig:FPs} shows predictions from the the ML model. The green rectangles show the ground truth and the red rectangles show ML model's prediction of where a pedestrian is present. The left example is a FP since IoU=0.3 with a predicted box substantially smaller than the ground truth. On the other hand, the ground truth is questionable, as there probably is only a single pixel containing the pedestrian below the visible arm that drastically increases the size of the green box. The center example is a TP with IoU=0.9, i.e., the overlap between the boxes is very large. The right example is another FP with IoU=0.4 where the predicted box is much larger than the ground truth. These examples show that FPs during model testing do not directly translate to FPs on the system level as discussed in the HARA (Safety Requirements Allocated to ML Component \textbf{[E]}). If any of the objects within the red bounding boxes were on a collision course with the ego car, commencing PAEB would indeed be the right action for SMIRK and thus not violate \textbf{SYS-SAF-REQ1}. This observation corroborates the position by \citep{haq2021can}, i.e., system level testing that goes beyond model testing on single frames is critically needed.

All results from running ML model testing, i.e., ML Verification Results \textbf{[Z]}, are documented in the Protocols folder.

\subsection{System Level Testing}
System-level testing of SMIRK involves integrating the ML model into the pedestrian recognition component and the complete PAEB ADAS. We do this by defining a set of Operational Scenarios \textbf{[EE]} for which we assess the satisfaction of the ML Safety Requirements. The results from the system-level testing, i.e., the Integration Testing Results \textbf{[FF]}, are presented in Section~\ref{amlas:ff}.

\subsubsection{Operational Scenarios} \label{sec:op_scenarios} \label{amlas:ee}
SOTIF defines an operational scenario as ``a description of an imagined sequence of events that includes the interaction of the product or service with its environment and users, as well as interaction among its product or service components''. Consequently, the set of operational scenarios used for testing SMIRK on the system level must represent the diversity of real scenarios that may be encountered when SMIRK is in operation. Furthermore, for testing purposes, it is vital that the set of defined scenarios are meaningful with respect to the verification of SMIRK's safety requirements.

As SMIRK is designed to operate in ESI Pro-SiVIC, the difference between defining operational scenarios in text and implementation scripts to execute the same scenarios in the simulated environment is very small. We will not define any operational scenarios that cannot be scripted for execution in ESI Pro-SiVIC. To identify a meaningful set of operational scenarios, we use equivalence partitioning as proposed by \citet{masuda2017software} as one approach to limit the number of test scenarios to execute in vehicle simulators. Originating in the equivalence classes, we use combinatorial testing to reduce the set of operational scenarios. Using combinatorial testing to create test cases for system testing of a PAEB testing in a vehicle simulator has previously been reported by \citet{tao2019industrial}. We create operational scenarios that provide complete pair-wise testing of SMIRK considering the identified equivalence classes using the AllPairs test combinations generator\footnote{\url{https://github.com/thombashi/allpairspy}}.

Based on an analysis of the ML Safety Requirements and the Data Requirements, we define operational scenarios addressing \textbf{SYS-ML-REQ1} and \textbf{SYS-ML-REQ2} separately. For each subset of operational scenarios, we identify key variation dimensions (i.e., parameters in the test scenario generation) and split dimensions into equivalence classes using explicit ranges. Note that ESI Pro-SiVIC enables limited configurability of basic shapes compared to pedestrians, thus the corresponding number of operational scenarios is lower.
\\\\Operational Scenarios for \textbf{SYS-ML-REQ1}:
    \begin{itemize}
        \item Pedestrian starting point (lateral offset from the road in meters): Left side of the road (-5~m), Center of the road (0~m), Right side of the road (5~m)
        \item Longitudinal distance from ego car (offset in meters): Close distance ($\lt$ 25~m), Medium distance (25--50~m), Far away ($\gt$ 50~m)
        \item Pedestrian appearance: Female casual, Male business casual, Male business, Female business, Child, Male worker
        \item Pedestrian speed (m/s): Stationary (0~m/s), Slow (1~m/s), Fast (3~m/s)
        \item Pedestrian crossing angle (degrees): Toward ego car (0), Diagonal toward (45), Perpendicular (90), Diagonal away (135), Away from car (180)
        \item Ego car speed (m/s): Slow ($\lt$ 10~m/s), Medium (10--15~m/s), Fast (15--20~m/s)
    \end{itemize}

The dimensions and ranges listed above result in 2,430 possible combinations. Using combinatorial testing, we create a set of 25 operational scenarios that provides pair-wise coverage of all equivalence classes. 
\\\\Operational Scenarios for \textbf{SYS-ML-REQ2}:
\begin{itemize}
\item Object starting point (lateral offset from the road in meters): Left side of the road (-5~m), On the road (0~m), Right side of the road (5~m)
\item Longitudinal distance from ego car (offset in meters): Close ($\lt$ 25~m), Medium distance (25--50~m), Far away ($\gt$ 50~m)
\item Object appearance: Sphere, Cube, Cone, Pyramid
\item Object speed (m/s): Stationary (0~m/s), Slow (1~m/s), Fast (3~m/s)
\item Ego car speed (m/s): Slow ($\lt$10~m/s), Medium (10--15~m/s), Fast (15--20~m/s)
\end{itemize}

The dimensions and ranges listed above result in 324 possible combinations. Using combinatorial testing, we create a set of 13 operational scenarios that provides pair-wise coverage of all equivalence classes.

For each operational scenario, two test parameters represent ranges of values, i.e., the longitudinal distance between ego car and the pedestrian and the speed of ego car. For these two test parameters, we identify a combination of values that result in a collision unless SMIRK initiates emergency braking. Table~\ref{tab:operational_scenarios} shows an overview of the 38 operational scenarios, whereas all details are available as executable test scenarios in the GitHub repository.

\begin{table}[]
\caption{Pairwise-testing of equivalence classes constituting 38 operational scenarios.}
\label{tab:operational_scenarios}
\begin{tabular}{|l|c|c|c|c|c|c|}
\hline
\textbf{ID} & \textbf{Object} & \textbf{X} & \textbf{Y} & \textbf{Angle}    & \textbf{Speed} & \textbf{Car\_Speed} \\ \hline
TC-OS-1     & P7                  & close      & left       & diagonal\_away    & slow           & slow                \\ \hline
TC-OS-2     & P7                  & medium     & center     & away              & stationary     & medium              \\ \hline
TC-OS-3     & P7                  & far        & right      & diagonal\_towards & fast           & fast                \\ \hline
TC-OS-4     & P5                  & medium     & right      & perpendicular     & slow           & medium              \\ \hline
TC-OS-5     & P5                  & close      & center     & towards           & fast           & fast                \\ \hline
TC-OS-6     & P1                  & far        & left       & diagonal\_away    & fast           & fast                \\ \hline
TC-OS-7     & P6                  & medium     & left       & perpendicular     & fast           & medium              \\ \hline
TC-OS-8     & P6                  & far        & center     & away              & slow           & slow                \\ \hline
TC-OS-9     & P1                  & close      & right      & diagonal\_towards & slow           & slow                \\ \hline
TC-OS-10    & P4                  & left       & right      & diagonal\_away    & slow           & slow                \\ \hline
TC-OS-11    & P8                  & close      & left       & perpendicular     & fast           & slow                \\ \hline
TC-OS-12    & P8                  & far        & left       & diagonal\_towards & slow           & medium              \\ \hline
TC-OS-13    & P4                  & far        & center     & perpendicular     & stationary     & fast                \\ \hline
TC-OS-14    & P5                  & close      & left       & diagonal\_away    & slow           & slow                \\ \hline
TC-OS-15    & P6                  & close      & right      & diagonal\_away    & slow           & slow                \\ \hline
TC-OS-16    & P4                  & medium     & left       & diagonal\_towards & slow           & slow                \\ \hline
TC-OS-17    & P8                  & close      & right      & diagonal\_away    & slow           & slow                \\ \hline
TC-OS-18    & P1                  & close      & center     & diagonal\_towards & stationary     & medium              \\ \hline
TC-OS-19    & P1                  & medium     & left       & diagonal\_towards & slow           & slow                \\ \hline
TC-OS-20    & P6                  & medium     & left       & diagonal\_towards & slow           & slow                \\ \hline
TC-OS-21    & P1                  & medium     & left       & perpendicular     & slow           & medium              \\ \hline
TC-OS-22    & P5                  & medium     & left       & diagonal\_towards & slow           & slow                \\ \hline
TC-OS-23    & P4                  & medium     & left       & diagonal\_away    & slow           & medium              \\ \hline
TC-OS-24    & P7                  & medium     & left       & perpendicular     & slow           & slow                \\ \hline
TC-OS-25    & P8                  & medium     & left       & diagonal\_towards & slow           & slow                \\ \hline
TC-OS-26    & N2                  & close      & left       & perpendicular     & slow           & slow                \\ \hline
TC-OS-27    & N3                  & medium     & right      & perpendicular     & fast           & medium              \\ \hline
TC-OS-28    & N4                  & far        & right      & perpendicular     & slow           & fast                \\ \hline
TC-OS-29    & N1                  & far        & left       & perpendicular     & slow           & medium              \\ \hline
TC-OS-30    & N1                  & medium     & left       & perpendicular     & fast           & fast                \\ \hline
TC-OS-31    & N1                  & close      & right      & perpendicular     & fast           & slow                \\ \hline
TC-OS-32    & N4                  & close      & left       & perpendicular     & fast           & fast                \\ \hline
TC-OS-33    & N3                  & medium     & left       & perpendicular     & slow           & slow                \\ \hline
TC-OS-34    & N2                  & far        & right      & perpendicular     & slow           & slow                \\ \hline
TC-OS-35    & N3                  & far        & left       & perpendicular     & slow           & fast                \\ \hline
TC-OS-36    & N2                  & medium     & left       & perpendicular     & fast           & medium              \\ \hline
TC-OS-37    & N3                  & close      & left       & perpendicular     & slow           & slow                \\ \hline
TC-OS-38    & N1                  & close      & left       & perpendicular     & slow           & medium              \\ \hline
\end{tabular}
\end{table}

\subsubsection{System Test Cases} \label{sec:system_TCs}
The system test cases are split into three categories. First, each operational scenario identified in Section~\ref{sec:op_scenarios} constitutes one system test case, i.e., Test Cases 1-38. Second, to increase the diversity of the test cases in the simulated environment, we complement the reproducible Test Cases 1-38 with test case counterparts adding random jitter to the parameters. For Test Cases 1-38, we create analogous test cases that randomly add jitter in the range from -10\% to +10\% to all numerical values. Partial random testing has been proposed by \citet{masuda2017software} in the context of test scenarios execution in vehicle simulators. Note that introducing random jitter to the test input does not lead to the test oracle problem \citep{barr2014oracle}, as we can automatically assess whether there is a collision between ego car and the pedestrian without emergency braking in ESI Pro-SiVIC or not (TC-RAND-[1-25]). Furthermore, for the test cases related to provoking ghost braking, we know that emergency braking shall not commence.

The third category is requirements-based testing (RBT). RBT is used to gain confidence that the functionality specified in the ML Safety Requirements has been implemented correctly \citep{hauer2019did}. The top-level safety requirement \textbf{SYS-SAF-REQ1} will be verified by testing of all underlying requirements, i.e., its constituent detailed requirements. The test strategy relies on demonstrating that \textbf{SYS-ML-REQ1} and \textbf{SYS-ML-REQ2} are satisfied when executing TC-OS-[1-38] and TC-RAND-[1-38]. \textbf{SYS-PER-REQ1} -- \textbf{SYS-PER-REQ5} and \textbf{SYS-ROB-REQ1} -- \textbf{SYS-ROB-REQ4} are verified through the model testing described in Section~\ref{amlas:aa}. The remaining performance requirement \textbf{SYS-PER-REQ6} is verified by TC-REQ-3. Table~\ref{tab:systemtests} lists all system test cases, of all three categories, using the Given-When-Then structure as used in behavior-driven development \citep{tsilionis2021unifying}. For the test cases TC-RBT-[1-3], the ``Given'' condition is that all metrics have been collected during execution of TC-OS-[1-38] and TC-RAND-[1-38]. The set includes seven metrics:

\begin{enumerate}
    \item \textbf{MinDist}: Minimum distance between ego car and the pedestrian during a scenario.
    \item \textbf{TimeTrig}: Time when the radar tracking component first returned TTC~$\lt$~4~s for an object.
    \item \textbf{DistTrig}: Distance between ego car and the object when the radar component first returned TTC~$\lt$~4~s for an object.
    \item \textbf{TimeBrake}: Time when emergency braking was commenced.
    \item \textbf{DistBrake}: Distance between ego car and the object when emergency braking commenced.
    \item \textbf{Coll}: Whether a scenario involved a collision between ego car and a pedestrian.
    \item \textbf{CollSpeed}: Speed of ego car at the time of collision.
\end{enumerate}
    
\begin{table}[]
\caption{SMIRK system test cases. VMC means valid metrics were collected during execution of the 38 preceding scenarios.}
\label{tab:systemtests}
\resizebox{\textwidth}{!}{
\begin{tabular}{|p{1cm}|p{1cm}|p{1.3cm}|p{4cm}|p{3.3cm}|}
\hline
\multicolumn{1}{|c|}{\textbf{ID}} & \multicolumn{1}{c|}{\textbf{Type}} & \multicolumn{1}{c|}{\textbf{Given}} & \multicolumn{1}{c|}{\textbf{When}}                                                                     & \multicolumn{1}{c|}{\textbf{Then}}                                                                \\ \hline
TC-OS-{[}1-25{]}                            & Op. Scenario               & Scenario {[}1-25{]}                 & Pedestrian crosses the street and ego car is on collision course                                       & SMIRK commences PAEB                                                                              \\ \hline
TC-OS-{[}26-38{]}                           & Op. Scenario               & Scenario {[}26-38{]}                & Object crosses the street and ego car is on collision course                                           & SMIRK does not commence PAEB                                                                      \\ \hline
TC-RAND-{[}1-25{]}                          & Random Testing                     & TC-OS-{[}1-25{]}+jitter             & Pedestrian crosses the street and ego car is on collision course                                       & SMIRK commences PAEB                                                                              \\ \hline
TC-RAND-{[}19-38{]}                         & Random Testing                     & TC-OS-{[}19-38{]}+jitter            & Object crosses the street and ego car is on collision course                                           & SMIRK does not commence PAEB                                                                      \\ \hline
TC-RBT-1                                    & RBT (SYS-ML-REQ1)                  & VMC                                 & The radar tracking component returns a pedestrian with TTC~$\lt$~4~s                                & The pedestrian recognition component identifies the pedestrian                                    \\ \hline
TC-RBT-2                                    & RBT (SYS-ML-REQ2)                  & VMC                                 & The radar tracking component returns a basic shape with TTC~$\lt$~4~s                               & The pedestrian recognition component does not identify a pedestrian                               \\ \hline
TC-RBT-3                                    & RBT (SYS-PER-REQ6)                 & VMC                                 & The radar tracking component returns a pedestrian with TTC~$\lt$~4~s within 80~m                    & The inference speed of the pedestrian recognition component is at least 10 FPS                                    \\ \hline
\end{tabular}
}
\end{table}

\section{SMIRK Test Results} \label{sec:testres}
This section presents the most important test results from three levels of ML testing, i.e., data testing, model testing, and system testing. Complete test reports are available in the protocols subfolder on GitHub\footnote{\url{https://github.com/RI-SE/smirk/tree/main/docs/protocols}}. Moreover, this section presents the Erroneous Behaviour Log.

\subsection{Results from Data Testing [S]} \label{amlas:s}
This section describes the results from testing the SMIRK data set. The data testing primarily involves a statistical analysis of its distribution and automated data validation using Great Expectations\footnote{\url{https://greatexpectations.io/}}. Together with the outcome of the Fagan inspection of the Data Management Specification (described in Section~\ref{sec:fagan}), this constitutes the ML Data Validation Results in AMLAS. As depicted later in Figure~\ref{amlas:r}, the results entail evidence mapping to the four assurance-related desiderata, i.e., we report a validation of 1) data relevance, 2) data completeness, 3) data balance, and 4) data accuracy. Since we generate synthetic data using ESI Pro-SiVIC, data relevance has been validated through code reviews and data accuracy is implicit as the tool's ground truth is used. For both the relevance and accuracy desiderata, we have manually analyzed a sample of the generated data to verify requirements satisfaction.

We validate the ethical dimension of the data balance by analyzing the gender (\textbf{DAT-BAL-REQ1}) and age (\textbf{DAT-BAL-REQ2}) distributions of the pedestrians in the SMIRK data set. SMIRK evolves as a demonstrator in a Swedish research project, which provides a frame of reference for this analysis. Table~\ref{tab:data_dist} shows how the SMIRK data set compares to Swedish demographics from the perspective of age and gender. The demographics originate in a study on collisions between vehicles and pedestrians by the Swedish Civil Contingencies Agency~\citep{msb2014}. We notice that 1) children are slightly over-represented in accidents but under-represented in deadly accidents, and that 2) adult males account for over half of the deadly accidents in Sweden. The rightmost column shows the distribution of pedestrian types in the entire SMIRK data set. We designed the SMIRK data generation process to result in a data set that resembles the deadly accidents in Sweden, but, motivated by AI fairness, we increased the fraction of female pedestrians to mitigate a potential gender bias. Finally, as discussed in Section~\ref{sec:data_split}, code reviews confirmed that the development data contains roughly 2\% ``background images''.

\begin{table}[]
\caption{Distribution of pedestrian types in Sweden and in the SMIRK data set.}
\label{tab:data_dist}
\begin{tabular}{|l|c|c|c|c|}
\hline
\textbf{Pedestrian types}       & \textbf{Population} & \textbf{Accidents} & \textbf{Deadly accidents} & \textbf{SMIRK} \\ \hline
Children \& young adults (0-19) & 23\%                & 27\%               & 12\%                      & 12.5\%         \\ \hline
Adult males (20+)               & 39\%                & 31\%               & 57\%                      & 50\%           \\ \hline
Adult females (20+)             & 38\%                & 42\%               & 31\%                      & 37.5\%         \\ \hline
\end{tabular}
\end{table}

Automated data testing is performed by defining conditions that shall be fulfilled by the data set. These conditions are checked against the existing data and any new data that is added. Some tests are fixed and simple, such as expecting the dimensions of input images to match the ones produced by the vehicle's camera. Similarly, all bounding boxes are expected to be within the dimensions of the image. Other tests look at the distribution and ranges of values to assure the completeness, accuracy, and balance of the data set and catch human errors. This includes validating enough coverage of pedestrians at different positions of the image, coverage of varying range of pedestrian distances, and bounding box aspect ratios. For values that are hard to define rules for, a known good set of inputs can be used as a starting point and remaining and new inputs can be checked to against these reference inputs. As an example, this can be used to verify that the color distribution and pixel intensity are within expected ranges. This can be used to identify images that are too dark or dissimilar to existing images. 

\begin{figure*}
        \centering
        \begin{subfigure}[b]{0.475\textwidth}
            \centering
            \includegraphics[width=\textwidth]{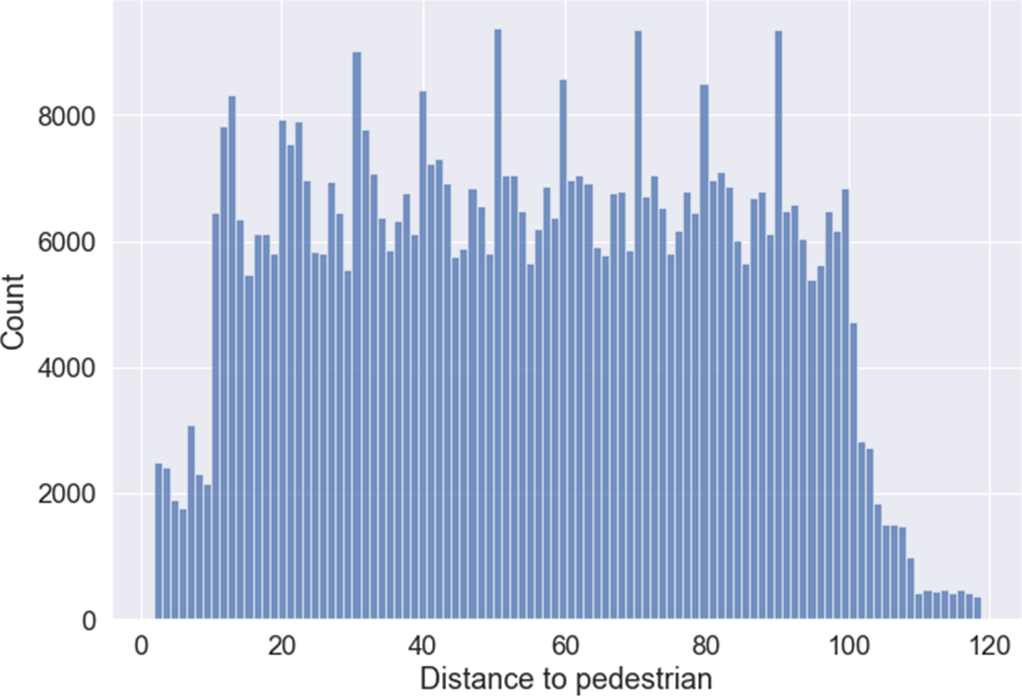}
            \caption[GreatExp A]%
            {{\small Distance distributions between ego car and pedestrians (m).}}    
            \label{fig:great_exp_a}
        \end{subfigure}
        \hfill
        \begin{subfigure}[b]{0.475\textwidth}  
            \centering 
            \includegraphics[width=\textwidth]{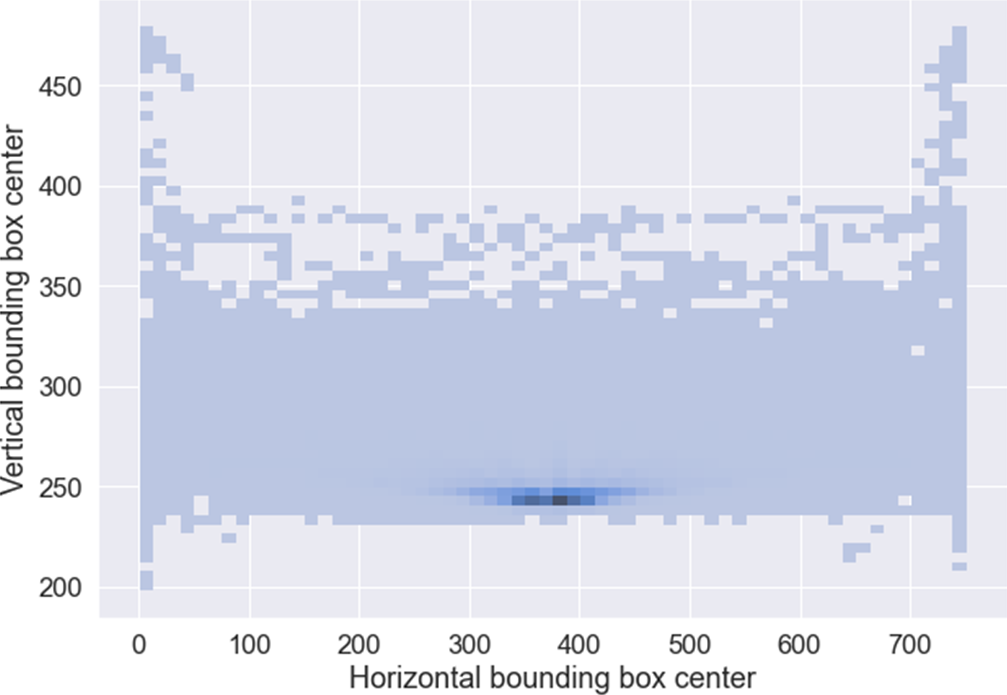}
            \caption[GreatExp B]%
            {{\small Heatmap of center positions of the bounding boxes (WVGA pixel).}}    
            \label{fig:great_exp_b}
        \end{subfigure}
        \vskip\baselineskip
        \begin{subfigure}[b]{0.475\textwidth}   
            \centering 
            \includegraphics[width=\textwidth]{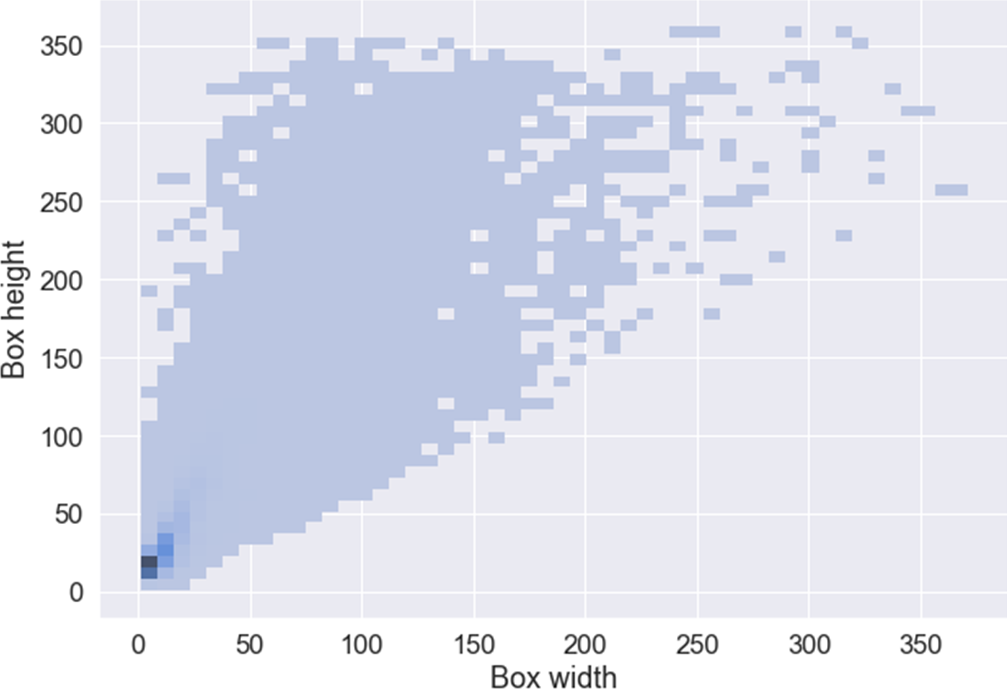}
            \caption[GreatExp C]%
            {{\small Heatmap of bounding box dimensions (pixels).}}    
            \label{fig:great_exp_c}
        \end{subfigure}
        \hfill
        \begin{subfigure}[b]{0.475\textwidth}   
            \centering 
            \includegraphics[width=\textwidth]{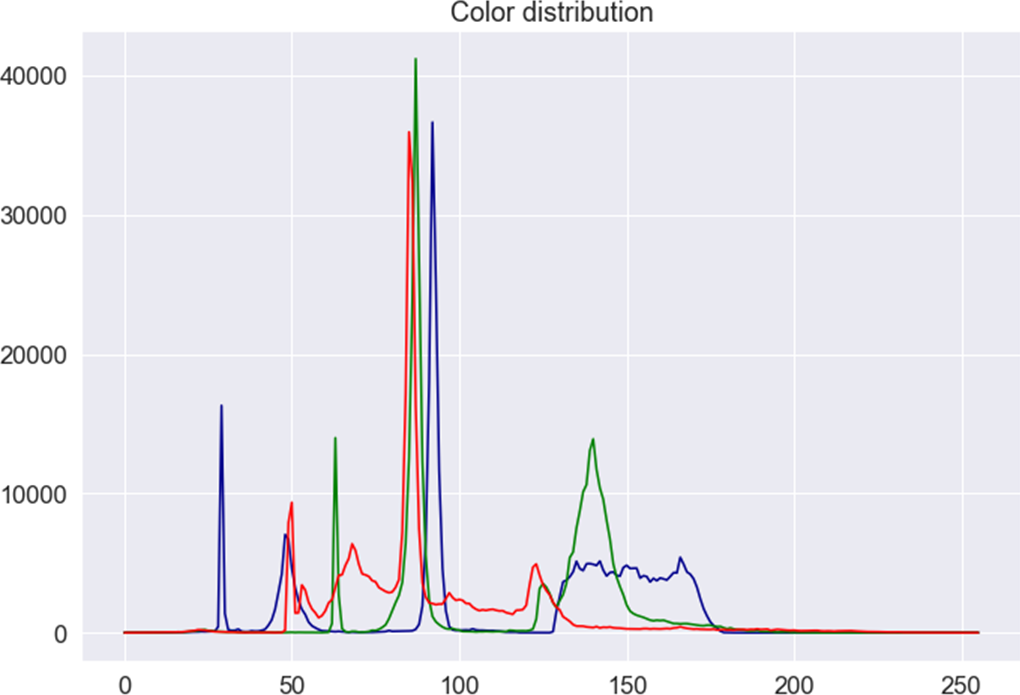}
            \caption[GreatExp D]%
            {{\small The color histogram of the data set.\\}}    
            \label{fig:great_exp_d}
        \end{subfigure}
        \caption[ The average and standard deviation of critical parameters ]
        {\small Four visualizations from the data testing.} 
        \label{fig:great_exp}
\end{figure*}

Figure~\ref{fig:great_exp} shows a selection of summary plots from the data testing that support our claims for data validity, in particular from the perspective of data completeness. Subfigure~\ref{fig:great_exp_a} presents the distance distribution between ego car and pedestrians, verifying that the data set contains pedestrians at distances 10--100~m (\textbf{DAT-COM-REQ5}). Subfigure~\ref{fig:great_exp_b} shows a heatmap of the bounding boxes' centers captures by the 752x480 WVGA camera. We confirm that pedestrians appear from the sides of the field of view and a larger fraction of images contain a pedestrian just in front of ego car. The position distribution supports our claim that \textbf{DAT-COM-REQ4} is satisfied, i.e., the data samples represent different camera angles. Subfigure~\ref{fig:great_exp_c} shows a heatmap of bounding box dimensions, i.e., different aspect ratios. A variety of aspect ratios indicate that pedestrians move with a diversity of arm and leg movements -- indicating walking and running -- and thus support our claim that \textbf{DAT-COM-REQ3} is fulfilled. Finally, Subfigure~\ref{fig:great_exp_d} shows the color histogram of the data set. In the automated data testing, we use this as a reference value when adding new images to ensure that they match the ODD. For example, a sample of nighttime images would have a substantially different color distribution.

\subsection{Results from Model Testing}
This section is split into results from testing done during development and the subsequent independent verification. Throughout this section, the following abbreviations are used for a set of standard evaluation metrics: Precision (P), Recall (R), F1-score (F1), Intersection over Union (IoU), True Positive (TP), False Positive (FP), FPs Per Image (FPPI), False Negative (FN), and Average Precision for IoU at 0.5 (AP@0.5), and Confidence (Conf).
%Mean Average Precision for a range of ten IoU values between 0.5 to 0.95 (AP@0.5:0.95),

\subsubsection{Internal Test Results \textbf{[X]}} \label{amlas:x}
\begin{figure*}
        \centering
        \begin{subfigure}[b]{0.475\textwidth}
            \centering
            \includegraphics[width=\textwidth]{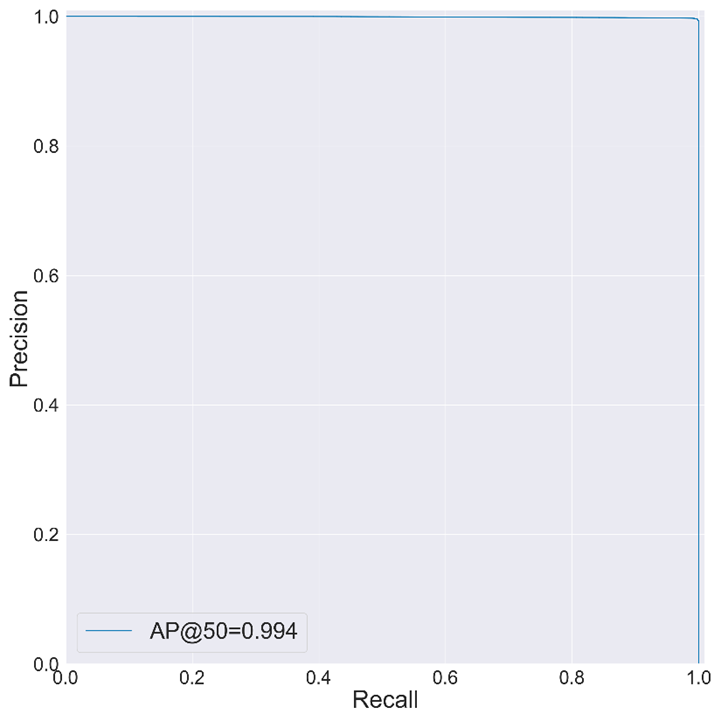}
            \caption[GreatExp A]%
            {{\small P--R curve.}}    
            \label{fig:internal_test_data_a}
        \end{subfigure}
        \hfill
        \begin{subfigure}[b]{0.475\textwidth}  
            \centering 
            \includegraphics[width=\textwidth]{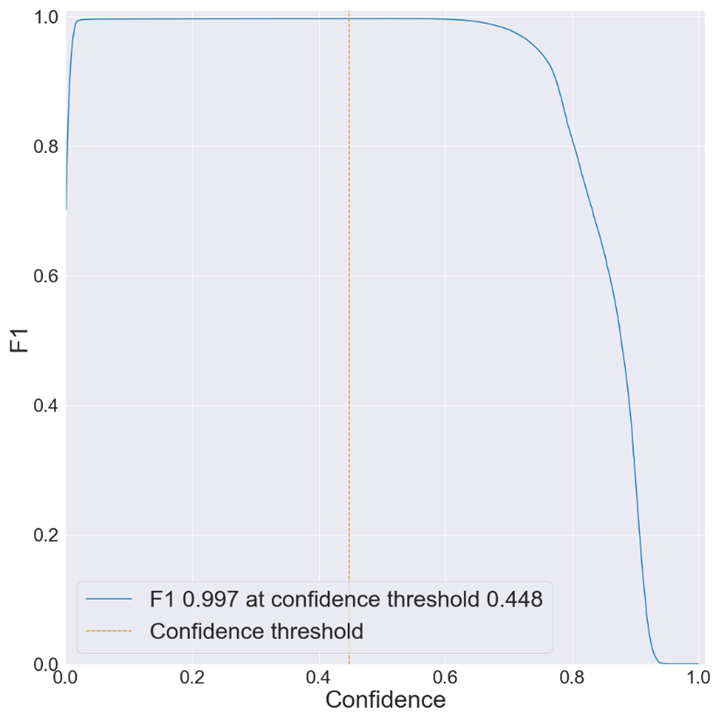}
            \caption[GreatExp B]%
            {{\small F1 vs. Conf.}}    
            \label{fig:internal_test_data_b}
        \end{subfigure}
        \vskip\baselineskip
        \begin{subfigure}[b]{0.475\textwidth}   
            \centering 
            \includegraphics[width=\textwidth]{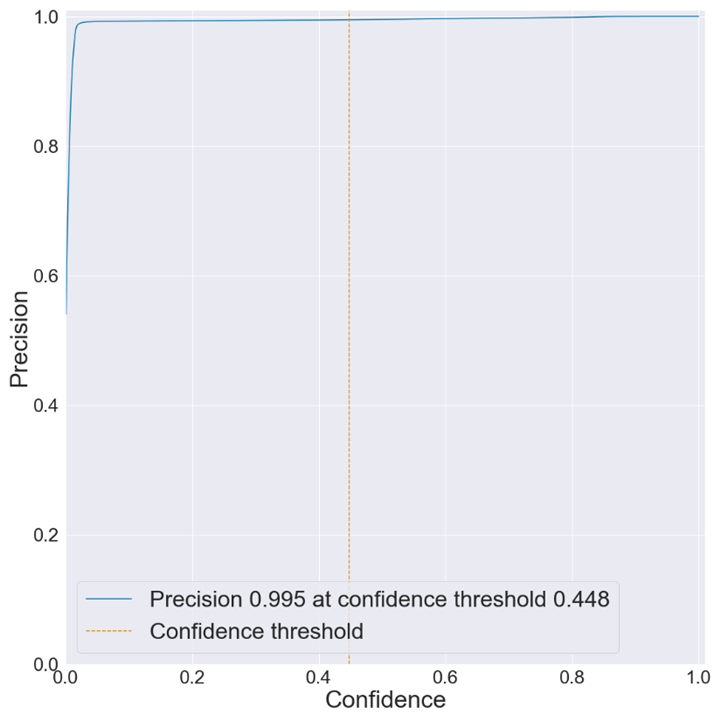}
            \caption[GreatExp C]%
            {{\small P vs. Conf.}}    
            \label{fig:internal_test_data_c}
        \end{subfigure}
        \hfill
        \begin{subfigure}[b]{0.475\textwidth}   
            \centering 
            \includegraphics[width=\textwidth]{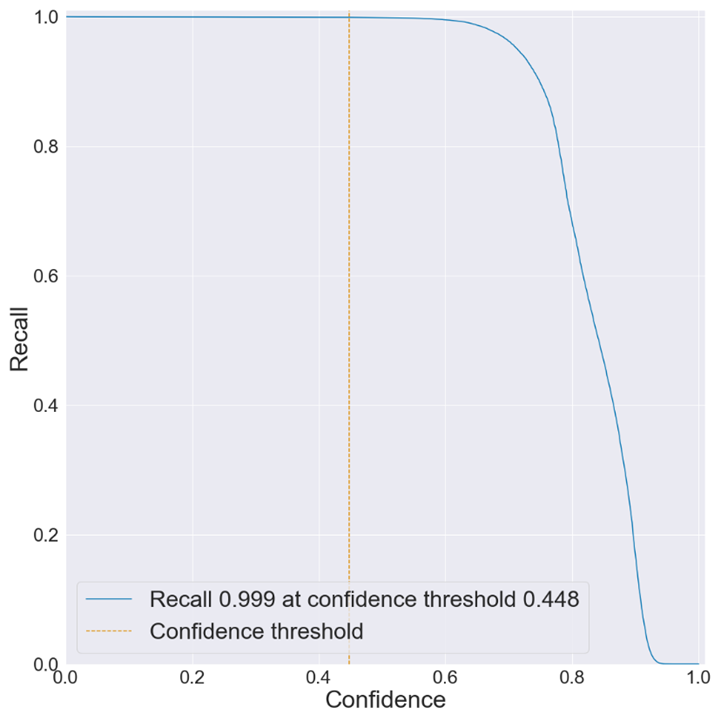}
            \caption[GreatExp D]%
            {{\small R vs. Conf.\\}}    
            \label{fig:internal_test_data_d}
        \end{subfigure}
        \caption[]
        {\small Evaluation of the ML model on the internal test data at IoU=0.5.} 
        \label{fig:internal_test_data}
\end{figure*}

In this section, we present the most important results from the internal testing. These results provide evidence that the ML model satisfies the ML safety requirements (see Section~\ref{amlas:h}) on the internal test data. The total number of images in the internal test data is 139,526 (135,139 pedestrians (96.9\%) and 4,387 non-pedestrians (3.1\%)). As described in Section~\ref{amlas:aa}, Figure~\ref{fig:internal_test_data} depicts four subplots representing IoU = 0.5: A) P vs R, B) F1 vs. Conf, C) P vs. Conf, and D) R vs. Conf. Subfigure~\ref{fig:internal_test_data_a} shows that the ML model is highly accurate, i.e., the unavoidable discrimination-efficiency tradeoff of object detection \citep{wu2008optimizing} is only visible in the upper right corner. Subfigures~\ref{fig:internal_test_data_b}--\ref{fig:internal_test_data_d} show how P, R, and F1 vary with different Conf thresholds. Table~\ref{tab:internal_test} presents further details of the accuracy of the ML model for the selected Conf threshold, organized into 1) all distances from the ego car, 2) within 80 m, and 3) within 50 m, respectively. The table also shows the effect of adding OOD detection using the autoencoder, i.e., a substantially reduced number of FPs.

\begin{table}[]
\caption{ML model accuracy on the internal test data at the Conf threshold 0.448. The rows show results for all distances, within 80 m, and within 50 m, respectively. Every second row show results for the ML model followed by OOD detection using the autoencoder.}
\label{tab:internal_test}
\begin{tabular}{|lc|c|c|c|c|c|c|c|}
\hline
\multicolumn{1}{|l|}{\textbf{Distance}} & \textbf{Total} & \textbf{TP} & \textbf{FP} & \textbf{FN} & \textbf{P} & \textbf{R} & \textbf{F1} & \textbf{AP@0.5} \\ \hline
\multicolumn{1}{|l|}{All}               & 139,526        & 134,948     & 711         & 191         & 0.9948     & 0.9986     & 0.9967      & 0.9942          \\ \hline
\multicolumn{2}{|c|}{+OOD}                               & 134,927     & 20          & 212         & 0.9999     & 0.9984     & 0.9991       & 0.995          \\ \hline \hline
\multicolumn{1}{|l|}{$\le$ 80 m}        & 105,588        & 101,320     & 444         & 173         & 0.9956     & 0.9983     & 0.997       & 0.9948          \\ \hline
\multicolumn{2}{|c|}{+OOD}                               & 101,300     & 13          & 193         & 0.9999     & 0.9981     & 0.999      & 0.995          \\ \hline \hline
\multicolumn{1}{|l|}{$\le$ 50 m}        & 61,845         & 57,877      & 186         & 173         & 0.9968     & 0.9970     & 0.9969      & 0.9944          \\ \hline
\multicolumn{2}{|c|}{+OOD}                               & 57,857      & 13          & 193         & 0.9998     & 0.9967      & 0.9982      & 0.995          \\ \hline 
\end{tabular}
\end{table}

Table~\ref{tab:internal_test_reqts} demonstrates how the ML model satisfies the performance requirements on the internal test data. First, the TP rate (95.9\%) and the FN rate (0.31\%) for the respective distances meet the requirements. The model's FPPI (0.42\%), on the other hand, is too high to meet \textbf{SYS-PER-REQ3} as we observed 444 FPs (cones outnumber spheres by 2:1). This observation reinforces the need to use a safety cage architecture, i.e., OOD detection that can reject input that does not resemble the training data. The rightmost column in Table~\ref{tab:internal_test_reqts} shows how the FPPI decreased to 0.012\% with the autoencoder. All basic shapes were rejected, but 13 images with pedestrians led to FPs within the ODD due to too low IoU scores.

\begin{table}[]
\caption{ML model satisfaction of the performance requirements on the internal test data at the Conf threshold 0.448. R1--R4 = SYS-PER-REQ1--4. The rightmost column show results for the YOLOv5 model followed by OOD detection using the autoencoder.}
\label{tab:internal_test_reqts}
\begin{tabular}{|l|c|c|c|}
\hline
\multicolumn{1}{|c|}{\textbf{Req.}} & \textbf{Expected}                                                                                                           & \textbf{Observed (Model)}          & \textbf{Observed (Model+OOD)}      \\ \hline
R1                                  & \begin{tabular}[c]{@{}c@{}}TP rate $\geq$ 93\%\\ for $\le$ 80 m\end{tabular}                                                & $\frac{101,320}{105,588} = 96.0\%$ & $\frac{101,300}{105,588} = 95.9\%$ \\ \hline
R2                                  & \begin{tabular}[c]{@{}c@{}}FN rate $\leq$ 7\%\\ for $\le$ 50 m\end{tabular}                                                 & $\frac{173}{61,845} = 0.28\%$      & $\frac{193}{61,845} = 0.31\%$      \\ \hline
R3                                  & \begin{tabular}[c]{@{}c@{}}FPPI $\leq$ 0.1\%\\ for $\le$ 80 m\end{tabular}                                                 & $\frac{444}{105,588} = 0.42\%$     & $\frac{13}{105,588} = 0.012\%$     \\ \hline
R4                                  & \begin{tabular}[c]{@{}c@{}}$\leq$ 3\% of rolling windows\\ contain $\ge$ 2 misses in\\ 5 frames for $\le$ 80 m\end{tabular} & $\frac{216}{101,564} = 0.21\%$     & $\frac{239}{101,564} = 0.24\%$     \\ \hline
\end{tabular}
\end{table}

\begin{figure}
\centering
\includegraphics[width=0.7\textwidth]{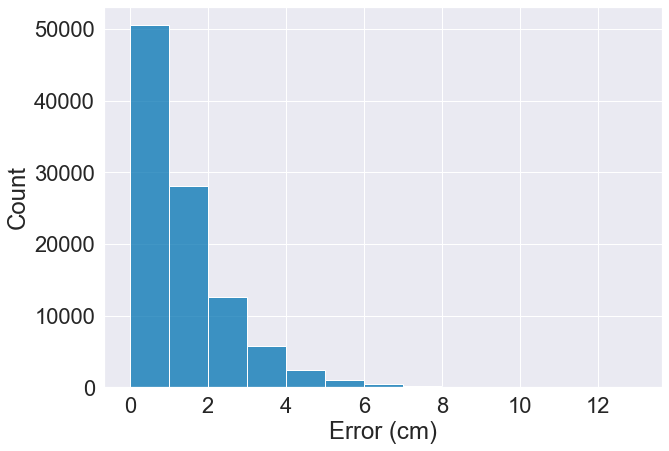}
\caption{Distribution of position errors for the internal test data.}
\label{fig:internal_test_distances}
\end{figure}

\textbf{SYS-PER-REQ4} is met as the fraction of rolling windows with more than a single FN is 0.24\%, i.e., $\leq$ 3\%. Figure~\ref{fig:internal_test_distances} shows the distribution of position errors in the object detection for pedestrians within 80~m of ego car, i.e., the difference between the object detection position and ESI Pro-SiVIC ground truth. The median error is 1.0~cm, the 99\% percentile is 5.6~cm, and the largest observed error is 12.7~cm. Thus, we show that \textbf{SYS-PER-REQ5} is satisfied for the internal test data, i.e., $\leq$ 50~cm position error for pedestrian detection within 80~m. Note that satisfaction of \textbf{SYS-PER-REQ6}, i.e., sufficient inference speed, is demonstrated as part of the system testing reported in Section~\ref{amlas:ff}. The complete test report is available on GitHub.

Table~\ref{tab:internal_slices} presents the output of the ML model on the eight slices of internal test data defined in Section~\ref{amlas:aa}. Note that we saved the children in the ESI Pro-SiVIC object catalog for the verification data, i.e., S9 does not exist in the internal test data. Apart from the S6 slice with occlusion, the model accuracy is consistent across the slices which corroborates satisfaction of the robustness requirements on the internal test data, e.g., in relation to pose (\textbf{SYS-ROB-REQ2}), size (\textbf{SYS-ROB-REQ2}), and appearance (\textbf{SYS-ROB-REQ2}).  

\begin{table}[]
\caption{ML model accuracy on eight slices of the internal test data. S1=All data, S2=close distance, S3=far distance, S4=running pedestrians, S5=walking pedestrians, S6=occluded pedestrians, S7=males, and S8=females. Every second rows show results for the ML model followed by OOD detection using the autoencoder.}
\label{tab:internal_slices}
\begin{tabular}{|lc|c|c|c|c|c|c|c|}
\hline
\multicolumn{1}{|l|}{\textbf{Slice}} & \textbf{Total} & \textbf{TP} & \textbf{FP} & \textbf{FN} & \textbf{P} & \textbf{R} & \textbf{F1} & \textbf{AP@0.5} \\ \hline
\multicolumn{1}{|l|}{S1}             & 139,526        & 134,948     & 711         & 191         & 0.9948     & 0.9986     & 0.9967      & 0.9942          \\ \hline
\multicolumn{2}{|r|}{+OOD}                            & 134,927     & 20          & 212         & 0.9999     & 0.9984     & 0.9991      & 0.995           \\ \hline
\multicolumn{1}{|l|}{S2}             & 61,333         & 57,774      & 16          & 172         & 0.9997     & 0.997      & 0.9984      & 0.995           \\ \hline
\multicolumn{2}{|r|}{+OOD}                            & 57,753      & 13          & 193         & 0.9998     & 0.9967     & 0.9982      & 0.995           \\ \hline
\multicolumn{1}{|l|}{S3}             & 43,547         & 43,547      & 0           & 0           & 1          & 1          & 1           & 0.995           \\ \hline
\multicolumn{2}{|r|}{+OOD}                            & 43,547      & 0           & 0           & 1          & 1          & 1           & 0.995           \\ \hline
\multicolumn{1}{|l|}{S4}             & 38,786         & 37,804      & 9           & 48          & 0.9998     & 0.9987     & 0.9992      & 0.995           \\ \hline
\multicolumn{2}{|r|}{+OOD}                            & 37,783      & 8           & 69          & 0.9998     & 0.9982     & 0.9990      & 0.995           \\ \hline
\multicolumn{1}{|l|}{S5}             & 99,740         & 97,144      & 14          & 143         & 0.9999     & 0.9985     & 0.9992      & 0.995           \\ \hline
\multicolumn{2}{|r|}{+OOD}                            & 97,144      & 12          & 143         & 0.9999     & 0.9985     & 0.9992      & 0.995           \\ \hline
\multicolumn{1}{|l|}{S6}             & 778            & 609         & 16          & 169         & 0.9744     & 0.7823     & 0.8679      & 0.9211          \\ \hline
\multicolumn{2}{|r|}{+OOD}                            & 593         & 13          & 185         & 0.9785     & 0.7618     & 0.8567      & 0.8899          \\ \hline
\multicolumn{1}{|l|}{S7}             & 69,238         & 67,470      & 14          & 99          & 0.9998     & 0.9985     & 0.9992      & 0.995           \\ \hline
\multicolumn{2}{|r|}{+OOD}                            & 67,460      & 11          & 109         & 0.9998     & 0.9984     & 0.9991      & 0.995           \\ \hline
\multicolumn{1}{|l|}{S8}             & 69,288         & 67,479      & 9           & 91          & 0.9999     & 0.9987     & 0.9993      & 0.995           \\ \hline
\multicolumn{2}{|r|}{+OOD}                            & 67,468      & 9           & 102         & 0.9999     & 0.9985     & 0.9992      & 0.995           \\ \hline
\end{tabular}
\end{table}

\subsubsection{ML Verification Results \textbf{[Z]}} \label{amlas:z}
\begin{figure*}
        \centering
        \begin{subfigure}[b]{0.475\textwidth}
            \centering
            \includegraphics[width=\textwidth]{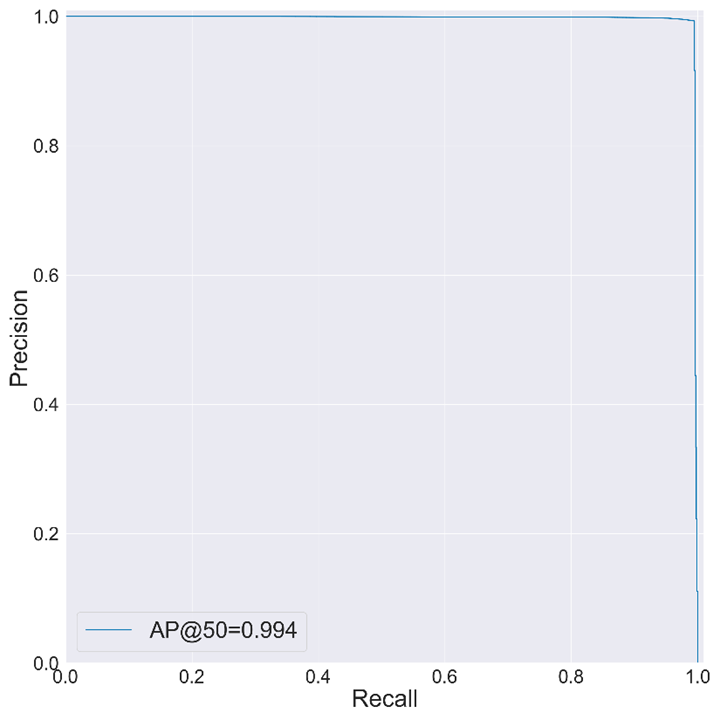}
            \caption[GreatExp A]%
            {{\small P--R curve.}}    
            \label{fig:verification_data_a}
        \end{subfigure}
        \hfill
        \begin{subfigure}[b]{0.475\textwidth}  
            \centering 
            \includegraphics[width=\textwidth]{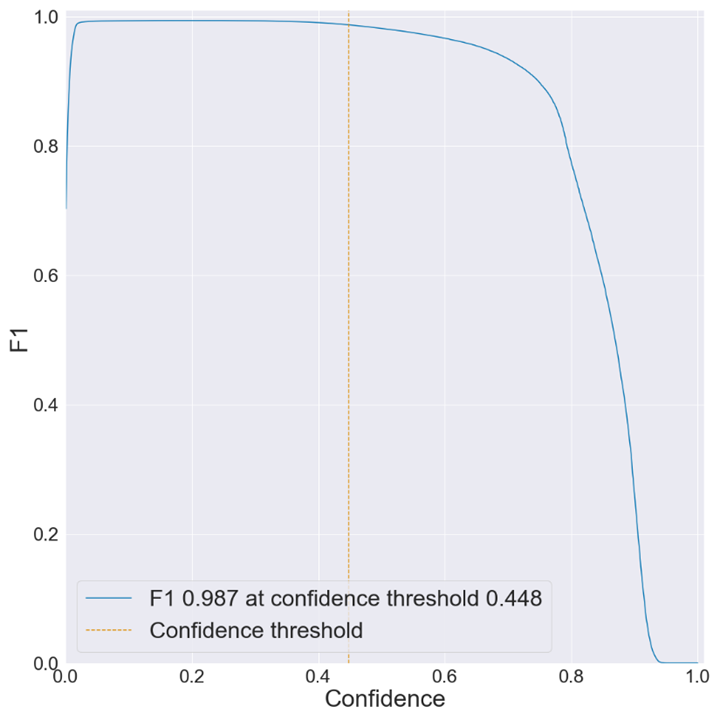}
            \caption[GreatExp B]%
            {{\small F1 vs. Conf.}}    
            \label{fig:verification_data_b}
        \end{subfigure}
        \vskip\baselineskip
        \begin{subfigure}[b]{0.475\textwidth}   
            \centering 
            \includegraphics[width=\textwidth]{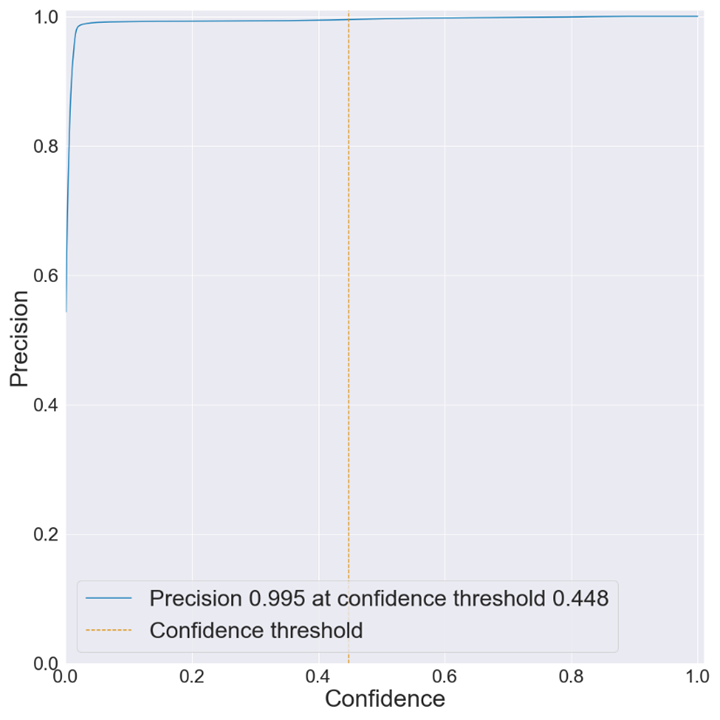}
            \caption[GreatExp C]%
            {{\small P vs. Conf.}}    
            \label{fig:verification_data_c}
        \end{subfigure}
        \hfill
        \begin{subfigure}[b]{0.475\textwidth}   
            \centering 
            \includegraphics[width=\textwidth]{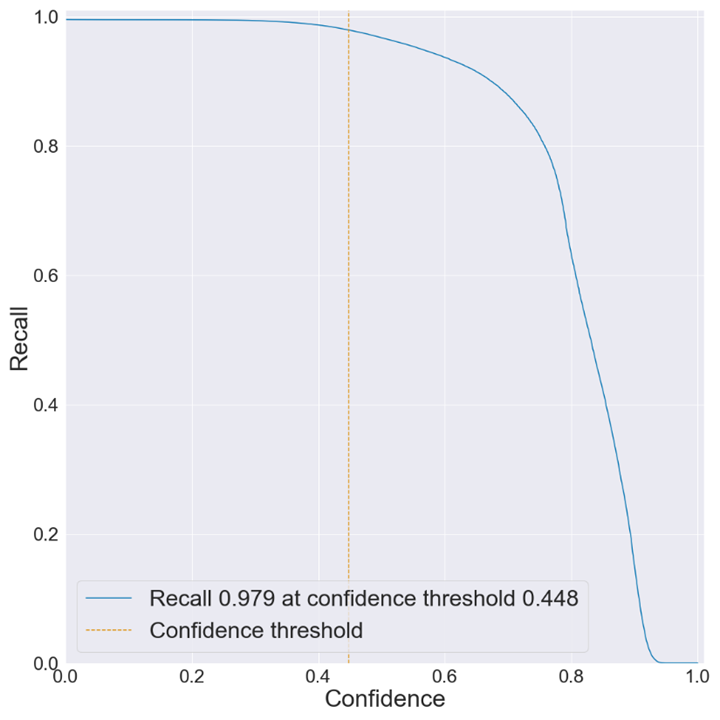}
            \caption[GreatExp D]%
            {{\small R vs. Conf.\\}}    
            \label{fig:verification_data_d}
        \end{subfigure}
        \caption[]
        {\small Evaluation of the ML model on the verification data at IoU=0.5.} 
        \label{fig:verification_data}
\end{figure*}

This section reports the key findings from conducting the independent ML model testing, i.e., the Verification Log in the AMLAS terminology. These results provide independent evidence that the ML model satisfies the ML safety requirements (see Section~\ref{amlas:h}) on the verification data. The total number of images in the verification data is 208,884 (202,712 pedestrians (97.0\%) and 6,172 non-pedestrians (3.0\%)). Analogous to Section~\ref{amlas:x}, Figure~\ref{fig:verification_data} depicts four subfigures representing IoU = 0.5: P vs R (cf.~\ref{fig:verification_data_a}), F1 vs. Conf (cf.~\ref{fig:verification_data_b}), P vs. Conf (cf.~\ref{fig:verification_data_c}), and R vs. Conf (cf.~\ref{fig:verification_data_d}). We observe that the appearance of the four subfigures closely resembles the corresponding plots for the internal test data (cf. Figure~\ref{fig:internal_test_data}).

\begin{figure}
\centering
\includegraphics[width=0.7\textwidth]{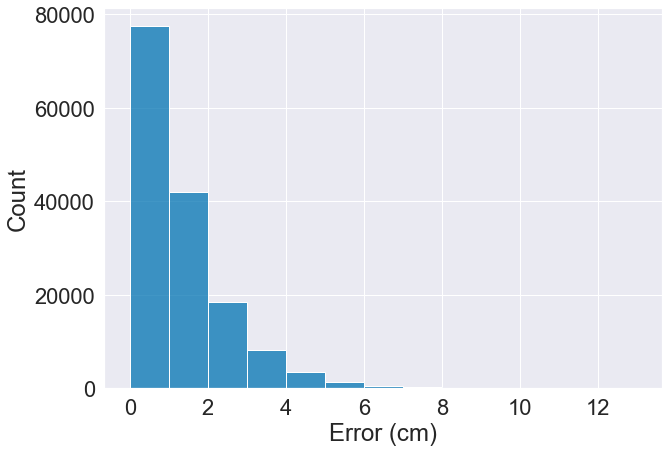}
\caption{Distribution of position errors for the verification data.}
\label{fig:verification_distances}
\end{figure}

Table~\ref{tab:verification} shows the output from the ML model using the Conf threshold 0.448 on the verification data. The table is organized into 1) all distances from the ego car, 2) within 80 m, and 3) within 50 m, respectively. The table also shows the effect of adding OOD detection using the autoencoder, i.e., the number of FPs is decreased just as for the internal test data. Table~\ref{tab:verification_reqts} demonstrates how the ML model satisfies the performance requirements on the verification data. Similar to the results for the internal test data, the FPPI (0.21\%) is too high to satisfy \textbf{SYS-PER-REQ3} without additional OOD detection, i.e., we observed 330 FPs (roughly an equal share of pyramids and children). The rightmost column in Table~\ref{tab:verification_reqts} shows how the FPPI decreased to 0.015\% with the autoencoder. All basic shapes were rejected, instead children at a long distance with too low IoU scores dominate the FPs. We acknowledge that it is hard for the YOLOv5 to achieve a high IoU for the few pixels representing a child almost 80~m away. However, commencing emergency braking in such cases is an appropriate action -- a child detected with a low IoU is not an example of the ghost braking hazard described in Section~\ref{amlas:e}. \textbf{SYS-PER-REQ4} is satisfied as the fraction of rolling windows with more than a single FN is 2.3\%. Figure~\ref{fig:verification_distances} shows the distribution of position errors. The median error is 1.0~cm, the 99\% percentile is 5.4~cm, and the largest observed error is 12.8~cm. Consequently, we show that \textbf{SYS-PER-REQ5} is satisfied for the verification data.

\begin{table}[]
\caption{ML model accuracy on the verification data at the Conf threshold 0.448. The three rows show results for all distances, within 80 m, and within 50 m, respectively. Every second row show results for the ML model followed by OOD detection using the autoencoder.}
\label{tab:verification}
\begin{tabular}{|lc|c|c|c|c|c|c|c|}
\hline
\multicolumn{1}{|l|}{\textbf{Distance}} & \textbf{Total} & \textbf{TP} & \textbf{FP} & \textbf{FN} & \textbf{P} & \textbf{R} & \textbf{F1} & \textbf{AP@0.5} \\ \hline
\multicolumn{1}{|l|}{All}               & 208,884        & 198,457     & 990         & 4,255       & 0.9950     & 0.9790     & 0.9870      & 0.9942          \\ \hline
\multicolumn{2}{|c|}{+OOD}                               & 195,695     & 533         & 7,017       & 0.9973     & 0.9654     & 0.9811      & 0.9878          \\ \hline
\multicolumn{1}{|l|}{$\le$ 80 m}        & 158,066        & 151,976     & 330         & 210         & 0.9978     & 0.9986     & 0.9982      & 0.9945          \\ \hline
\multicolumn{2}{|c|}{+OOD}                               & 149,214     & 23          & 2,972       & 0.9998     & 0.9905     & 0.9901      & 0.988           \\ \hline
\multicolumn{1}{|l|}{$\le$ 50 m}        & 92,592         & 86,847      & 178         & 165         & 0.9980     & 0.9981     & 0.9980      & 0.9949          \\ \hline
\multicolumn{2}{|c|}{+OOD}                               & 84,085      & 21          & 2,972       & 0.9998     & 0.9805     & 0.9901      & 0.988           \\ \hline
\end{tabular}
\end{table}

\begin{table}[]
\caption{ML model satisfaction of the performance requirements on the verification data at the Conf threshold 0.448. R1--R4 = SYS-PER-REQ1--4. The rightmost column show results for the YOLOv5 model followed by OOD detection using the autoencoder.}
\label{tab:verification_reqts}
\begin{tabular}{|l|c|c|c|}
\hline
\multicolumn{1}{|c|}{\textbf{Req.}} & \textbf{Expected}                                                                                                             & \textbf{Observed (Model)}          & \textbf{Observed (Model+OOD)}      \\ \hline
R1                                  & \begin{tabular}[c]{@{}c@{}}TP rate $\geq$ 93\%\\ for $\le$ 80 m\end{tabular}                                                  & $\frac{151,976}{158,066} = 96.1\%$ & $\frac{149,214}{158,066} = 94.4\%$ \\ \hline
R2                                  & \begin{tabular}[c]{@{}c@{}}FN rate $\leq$ 7\%\\ for $\le$ 50 m\end{tabular}                                                   & $\frac{165}{92,592} = 0.18\%$      & $\frac{2,927}{92,592} = 3.2\%$     \\ \hline
R3                                  & \begin{tabular}[c]{@{}c@{}}FPPI $\leq$ 0.1\%\\ for $\le$ 80 m\end{tabular}                                                    & $\frac{330}{158,066} = 0.21\%$     & $\frac{23}{158,066} = 0.015\%$     \\ \hline
R4                                  & \begin{tabular}[c]{@{}c@{}}$\leq$ 3\% of rolling windows\\  contain $\ge$ 2 misses in \\ 5 frames for $\le$ 80 m\end{tabular} & $\frac{201}{152,395} = 0.13\%$     & $\frac{3,499}{152,090} = 2.3\%$    \\ \hline
\end{tabular}
\end{table}

Table~\ref{tab:verification_slices} presents the output of the ML model on the nine slices of the verification data defined in Section~\ref{amlas:aa}. In relation to the robustness requirements, we notice that there the accuracy is slightly lower for S9 (children). This finding is related to the size requirement \textbf{SYS-ROB-REQ3}. Table~\ref{tab:children_analysis} contains an in-depth analysis of children at different distances with OOD detection. We confirm that most FPs occur outside of the ODD, i.e., 507 out of 512 FPs occur for children more than 80~m from ego car. In extension, we show that the performance requirements are still satisfied for the most troublesome slice of data as follows: 

\begin{itemize}
    \item TP rate children $\leq 80 m$: $\frac{50,402}{50,696} = 99.4\%$
    \item FN rate children $\leq 50 m$: $\frac{249}{30,731} = 0.81\%$ 
    \item FPPI children $\leq 80 m$: $\frac{5}{52,463} = 0.0099\%$
\end{itemize}

\begin{table}[]
\caption{Detailed analysis for children. The rows show results results for the ML model followed by OOD detection using the autoencoder for four distance ranges. The bottom row in italic font is outside the ODD.}
\label{tab:children_analysis}
\begin{tabular}{|lc|c|c|c|c|c|c|c|}
\hline
\multicolumn{1}{|l|}{\textbf{Distance}} & \textbf{Total} & \textbf{TP} & \textbf{FP} & \textbf{FN} & \textbf{P} & \textbf{R} & \textbf{F1} & \textbf{AP@0.5} \\ \hline
\multicolumn{1}{|l|}{All}             & 69,301        & 63,360      & 512         & 4,174       & 0.992      & 0.9382     & 0.9643      & 0.9877          \\ \hline
\multicolumn{1}{|l|}{$\le$ 80 m}        & 50,696        & 50,402     & 5         & 294         & 0.9999     & 0.9942     & 0.9971       & 0.995          \\ \hline
\multicolumn{1}{|l|}{$\le$ 50 m}        & 30,731         & 28,715      & 3         & 249         & 0.9999     & 0.9914     & 0.9956      & 0.995          \\ \hline \hline
\multicolumn{1}{|l|}{\textit{$\gt$ 80 m}}        & \textit{16,838}         & \textit{12,803}      & \textit{507}         & \textit{4,035}         & \textit{0.9619}     & \textit{0.7604}     & \textit{0.8493}      & \textit{0.942}          \\ \hline
\end{tabular}
\end{table}

The independent verification concludes that all requirements are met, based on the same argumentation as for the internal test results. The complete verification report is available on GitHub.

\begin{table}[]
\caption{ML model accuracy on nine slices of the verification data. S1=All data, S2=close distance, S3=far distance, S4=running pedestrians, S5=walking pedestrians, S6=occluded pedestrians, S7=males, S8=females, and S9=children. Every second rows show results for the ML model followed by OOD detection using the autoencoder.}
\label{tab:verification_slices}
\begin{tabular}{|lc|c|c|c|c|c|c|c|}
\hline
\multicolumn{1}{|l|}{\textbf{Slice}} & \textbf{Total} & \textbf{TP} & \textbf{FP} & \textbf{FN} & \textbf{P} & \textbf{R} & \textbf{F1} & \textbf{AP@0.5} \\ \hline
\multicolumn{1}{|l|}{S1}             & 208,884        & 198,457     & 990         & 4,255       & 0.995      & 0.979      & 0.987       & 0.9942          \\ \hline
\multicolumn{2}{|r|}{+OOD}                            & 195,695     & 533         & 7,017       & 0.9998     & 0.9616     & 0.9803      & 0.9878          \\ \hline
\multicolumn{1}{|l|}{S2}             & 92,028         & 86,691      & 22          & 165         & 0.9997     & 0.9981     & 0.9989      & 0.995           \\ \hline
\multicolumn{2}{|r|}{+OOD}                            & 83,929      & 21          & 2,927       & 0.9997     & 0.9663     & 0.9827      & 0.9761          \\ \hline
\multicolumn{1}{|l|}{S3}             & 65,330         & 65,285      & 2           & 45          & 1          & 0.9993     & 0.9996      & 0.995           \\ \hline
\multicolumn{2}{|r|}{+OOD}                            & 65,285      & 2           & 45          & 1          & 0.9993     & 0.9996      & 0.995           \\ \hline
\multicolumn{1}{|l|}{S4}             & 58,267         & 56,130      & 110         & 716         & 0.998      & 0.9874     & 0.9927      & 0.9949          \\ \hline
\multicolumn{2}{|r|}{+OOD}                            & 54,964      & 110         & 1,882       & 0.998      & 0.9669     & 0.9822      & 0.9818          \\ \hline
\multicolumn{1}{|l|}{S5}             & 149,617        & 142,328     & 424         & 3,538       & 0.997      & 0.9757     & 0.9863      & 0.9949          \\ \hline
\multicolumn{2}{|r|}{+OOD}                            & 140,732     & 423         & 5,134       & 0.997      & 0.9648     & 0.9806      & 0.9882          \\ \hline
\multicolumn{1}{|l|}{S6}             & 1,031          & 866         & 22          & 165         & 0.9752     & 0.84       & 0.9026      & 0.9289          \\ \hline
\multicolumn{2}{|r|}{+OOD}                            & 805         & 21          & 226         & 0.9746     & 0.7808     & 0.867       & 0.8783          \\ \hline
\multicolumn{1}{|l|}{S7}             & 69,292         & 67,555      & 15          & 54          & 0.9998     & 0.9992     & 0.9995      & 0.995           \\ \hline
\multicolumn{2}{|r|}{+OOD}                            & 65,009      & 14          & 2,600       & 0.9998     & 0.9616     & 0.9803      & 0.9741          \\ \hline
\multicolumn{1}{|l|}{S8}             & 69,291         & 67,495      & 7           & 74          & 0.9999     & 0.9989     & 0.9994      & 0.995           \\ \hline
\multicolumn{2}{|r|}{+OOD}                            & 67,482      & 7           & 87          & 0.9999     & 0.9987     & 0.9993      & 0.995           \\ \hline
\multicolumn{1}{|l|}{S9}             & 69,301         & 63,408      & 512         & 4,126       & 0.992      & 0.9389     & 0.9647      & 0.9879          \\ \hline
\multicolumn{2}{|r|}{+OOD}                            & 63,205      & 512         & 4,329       & 0.992      & 0.9359     & 0.9631      & 0.9871          \\ \hline
\end{tabular}
\end{table}

\subsection{Results from System Testing \textbf{[FF]}} \label{amlas:ff}
This section presents an overview of the results from testing SMIRK in ESI Pro-SiVIC, which corresponds to the Integration Testing Results in AMLAS. As explained in Section~\ref{sec:system_TCs}, we measure seven metrics for each test case execution, i.e., MinDist, TimeTrig, DistTrig, TimeBrake, DistBrake, Coll, and CollSpeed.

Table~\ref{tab:system_test_results} presents the results from executing the test cases representing operational scenarios with pedestrians, i.e., TC-OS-[1--25]. From the left, the columns show 1) test case ID, 2) the minimum distance between ego car and the pedestrian during the scenario, 3) the difference between TimeTrig and TimeBrake, 4) the difference between DistTrig and DistBrake, 5) whether there was a collision, 6) the speed of ego car at the collision, and 7) the initial speed of ego car. We note that 2) and 3) are 0 for all 25 test cases, showing that the pedestrian is always detected at the first possible frame when TTC $\le$ 4s, which means that SMIRK commenced emergency braking in all cases. Moreover, we see that SMIRK successfully avoids collisions in all but two test cases. In TC-OS-5, the pedestrian starts 20~m from ego car and runs towards it while it drives at 16 m/s -- SMIRK brakes but barely reduces the speed. In TC-OS-9, the pedestrian starts only 15~m from ego car but SMIRK significantly reduces the speed by emergency braking.

\begin{figure}
\centering
\includegraphics[width=0.7\textwidth]{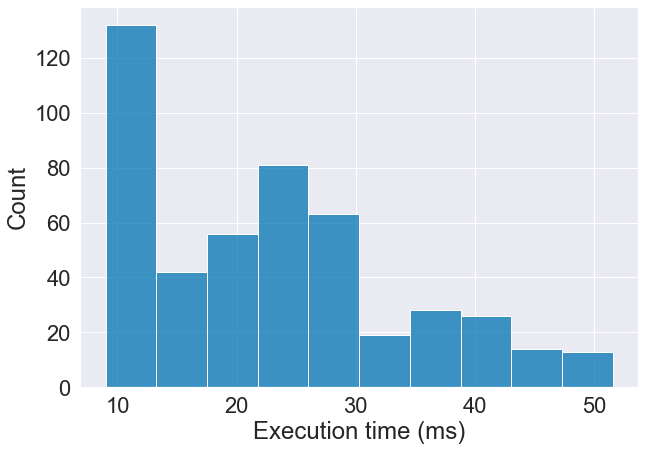}
\caption{Distribution of inference speeds during system testing.}
\label{fig:inference_speeds}
\end{figure}

The remaining system test cases corresponding to non-pedestrian operational scenarios (TC-OS-[26--38]) and all test cases with jitter (TC-RAND-[1--38]) were also executed with successful test verdicts. All scenarios with basic shapes on collision course were rejected by the safety cage architecture, i.e., SMIRK did never commence any ghost braking. In a virtual conclusion of test meeting, the first three authors concluded that TC-RBT-1 and TC-RBT-2 had passed successfully. Finally, Figure~\ref{fig:inference_speeds} shows the distribution of inference speeds during the system testing. The median inference time is 22.0~ms and the longest inference time observed is 51.6~ms. Based on these results we conclude that TC-RBT-3 passed successfully and thus provide evidence that \textbf{SYS-PER-REQ6} is satisfied. The complete system test report is available on GitHub.

\begin{table}[]
\caption{Test results and metrics collected during execution of TC-OS-[1-25].}
\label{tab:system_test_results}
\resizebox{0.9\textwidth}{!}{
\begin{tabular}{lrrrlrr}
\toprule
ID & MinDist(m) &  $\Delta$ Time(s) &  $\Delta$ Distance(m) &  Collision &  Collision Speed(m/s) &  Initial Speed(m/s) \\
\midrule
TC-OS-1   &       18 &                   0.0 &                       0.0 &      False &              - &           3.47 \\
TC-OS-2   &       21 &                   0.0 &                       0.0 &      False &              - &           5.78 \\
TC-OS-3   &       12 &                   0.0 &                       0.0 &      False &              - &          14.00 \\
TC-OS-4   &        3 &                   0.0 &                       0.0 &      False &              - &          18.90 \\
TC-OS-5   &        0 &                   0.0 &                       0.0 &       True &        15.99 &          16.00 \\
TC-OS-6   &       18 &                   0.0 &                       0.0 &      False &              - &           3.73 \\
TC-OS-7   &       34 &                   0.0 &                       0.0 &      False &              - &           5.80 \\
TC-OS-8   &       28 &                   0.0 &                       0.0 &      False &              - &          10.00 \\
TC-OS-9   &        0 &                   0.0 &                       0.0 &       True &         3.11 &          11.00 \\
TC-OS-10   &       18 &                   0.0 &                       0.0 &      False &              - &           3.14 \\
TC-OS-11   &       33 &                   0.0 &                       0.0 &      False &              - &           5.66 \\
TC-OS-12   &       27 &                   0.0 &                       0.0 &      False &              - &          10.22 \\
TC-OS-13   &       16 &                   0.0 &                       0.0 &      False &              - &          18.60 \\
TC-OS-14   &       19 &                   0.0 &                       0.0 &      False &              - &           3.68 \\
TC-OS-15   &        1 &                   0.0 &                       0.0 &      False &              - &          15.00 \\
TC-OS-16   &       18 &                   0.0 &                       0.0 &      False &              - &           3.46 \\
TC-OS-17   &       32 &                   0.0 &                       0.0 &      False &              - &           9.00 \\
TC-OS-18   &       17 &                   0.0 &                       0.0 &      False &              - &           5.52 \\
TC-OS-19   &       25 &                   0.0 &                       0.0 &      False &              - &           4.71 \\
TC-OS-20   &       29 &                   0.0 &                       0.0 &      False &              - &          10.30 \\
TC-OS-21   &       28 &                   0.0 &                       0.0 &      False &              - &          19.00 \\
TC-OS-22   &        4 &                   0.0 &                       0.0 &      False &              - &           6.00 \\
TC-OS-23   &       16 &                   0.0 &                       0.0 &      False &              - &           3.46 \\
TC-OS-24   &       30 &                   0.0 &                       0.0 &      False &              - &           4.81 \\
TC-OS-25   &       36 &                   0.0 &                       0.0 &      False &              - &          12.89 \\
\bottomrule
\end{tabular}
}
\end{table}

\subsection{Erroneous Behaviour Log \textbf{[DD]}} \label{amlas:dd}
As prescribed by AMLAS, the characteristics of erroneous outputs shall be predicted and documented. This section presents the key observations from internal testing of the ML model, independent verification activities, and system testing in ESI Pro-SiVIC. The findings can be used to design appropriate responses by other vehicular systems in the SMIRK context.

Tables~\ref{tab:internal_slices} and \ref{tab:verification_slices} show that the AP@0.5 are lower for occluded pedestrians (S6). As occlusion is an acknowledged challenge for object detection, which we previously have studied for automotive pedestrian detection \citep{henriksson2021understanding}, this is an expected result. Table~\ref{tab:verification_slices} also reveals that the number of FPs and FNs for the S9 slice (children) is relatively high, resulting in slightly lower AP@0.5. Table~\ref{tab:children_analysis} shows that the problem with children is primarily far away, explained by the few pixels available for the object detection at long distances. While the SMIRK fulfils the robustness requirements within the ODD, we recognize this perception issue in the erroneous behavior log.

\begin{figure}
\centering
\includegraphics[width=1\textwidth]{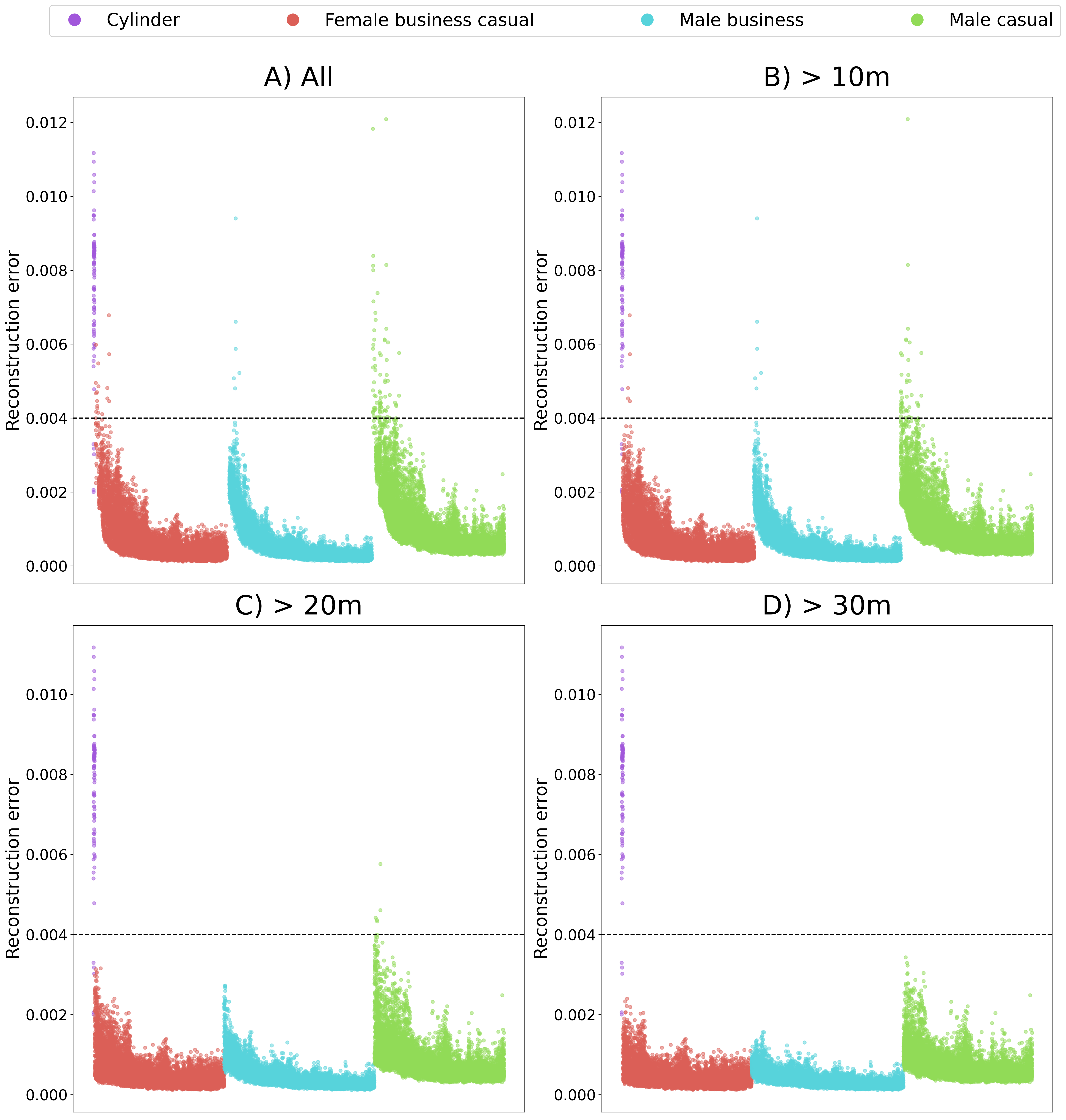}
\caption{Reconstruction errors for different objects on the validation subset of the development data at different distances from ego car (purple=cylinder, red=female business casual, blue=male business, green=male casual). The dashed lines show the threshold for rejecting objects. In SMIRK, we use alternative B) in the safety cage.}
\label{fig:vae_distances}
\end{figure}

During the iterative SMIRK development (cf. E) in Figure~\ref{fig:smile}), it became evident that OOD detection using the autoencoder was inadequate at close range. Figure~\ref{fig:vae_distances} shows reconstruction errors (on the y-axis) for all objects in the validation subset of the development data at A) all distances, B) $\gt$ 10 m, C) $\gt$ 20 m, and D) $\gt$ 30 m. The visualization clearly shows that the autoencoder cannot convincingly distinguish the cylinders from the pedestrians at all distances (in subplot A), different objects appear above the threshold, but the OOD detection is more accurate when objects at close distance are excluded (subplot D) displays high accuracy). Based on validation of the four distances, comparing the consequences of the trade-off between safety cage availability and accuracy, the design decision for SMIRK's autoencoder is to only perform OOD detection for objects that are at least 10 m away. We explain the less accurate behaviour at close range by limited training data, a vast majority of images contain pedestrians at a larger distance -- which is reasonable since the SMIRK ODD is limited to rural country roads.

\section{Lessons Learned and Practical Advice} \label{sec:lessons}
This section shares the most valuable lessons learned during our project. We organize the section into the perspectives of AI engineering and industry-academia collaboration.

\subsection{AI Engineering in the Safety Context}

SOTIF and AMLAS are compatible and complementary. Our experience in this R\&D project is that the processes are feasible to combine during systems development and safety engineering. We expected this based on reading process documentation before embarking on this project, and our hands-on experience confirmed the compatibility in our case under study. As presented in Figures~\ref{fig:sotif} and~\ref{fig:amlas}, both processes are iterative which is key to AI engineering. In our project, we first used SOTIF (including a formal HARA) to iterate toward an initial SRS. From this point, subsequent development iterations that adhered to both the cycles of SOTIF and AMLAS followed. Moreover, SOTIF and AMLAS complement each other as SOTIF maintains a systems perspective (sensors, ML algorithms, and actuators, cf. Section~\ref{sec:prod_func}) whereas AMLAS provides an artifact-oriented focus on safety evidence for activities related to ML. Finally, as stressed by industry partners, we found that we could apply both SOTIF and AMLAS after initial prototyping (cf. Figure~\ref{fig:smile}) and still harness insights from previous \textit{ad hoc} work. While safety rarely can be added on top of a complex system, we found that prototyping accelerated subsequent safety engineering for SMIRK.

Using a simulator to create data sets limits the validity of the negative examples. On one hand, our data generation scripts enable substantial freedom and cheap access to data. On the other hand, there is barely any variation in the scenarios (apart from clouds moving on the skydome) as would be the case for naturalistic data. As anything that is not a pedestrian in our data is a \textit{de facto} negative example (see rationale for \textbf{DAT-BAL-REQ3}), and nothing ever appears in our simulated scenarios unless we add it in our scripts, the diversity of our negative examples is very limited. Our approach to negative examples in the development data, referred to as ``background images'' in Section~\ref{sec:data_split}, involved including the outlier example Cylinder [N5]. From experiments on the validation subset of the development data, we found that adding frames with cylinders representing negative examples was essential to let the model distinguish between pedestrians and basic shapes. For ML components designed for use in the real world, trained on outcomes from real data collection campaigns, the natural variation of the negative examples would be completely different. When working with synthetic data from simulators, how to specify data requirements on negative examples remains an open question. 

Evaluation of object detection models is non-trivial. We spent substantial time to align the understanding within the project and we believe other development endeavors will need to do the same. In particular, we observed that the definition of TP, FP, TN, and FN based on IoU (explained in Section~\ref{amlas:aa}) is difficult to grasp for novices. The fact that FPs appear due to low IoU scores despite parts of a pedestrian indeed is detected is often counter-intuitive, i.e., ``how can a detected pedestrian ever be a FP?'' To align the development team, organizations should ensure that the true meaning of those KPIs are communicated as part of internal training. In the same vein, FP rate is not a valid metric (as TNs do not exist) whereas FN rate is used in \textbf{SYS-PER-REQ2} -- again internal training is important to align the understanding. What intuitively is perceived as a FP on the system level is not the same as a FP on the ML model level. To make the distinction clear, we restrict the use of FPs to the model level and refer to incorrect braking on the system level as ``ghost braking.''

ML model selection post learning involves fundamental decisions. Model selection is an essential activity in ML. When training ML models over several epochs, the best performing model given some criterion shall be kept. Also, when training alternative models with alternative architectures or hyperparameter settings, there must be a way to select the best candidate. How to tailor a fitness function to quantitatively measure what ``best'' involves is a delicate engineering effort with inevitable tradeoffs. The default fitness function in YOLOv5 puts 10\% of the weight at AP@0.5 and 90\% at Mean AP for a range of ten IoU values between 0.5 to 0.95. It would be possible to further customize the fitness function to also cover fairness aspects, i.e., to already during model selection value models that fulfill various quality aspects. There is no upper limit to the possible complexity, as this could encompass gender, size, ODD aspects etc. 
For SMIRK, however, we decided to do the opposite, i.e., to prioritize simplicity to gain interpretability by using a simpler metric. As explained in Section~\ref{amlas:aa}, our fitness function solely uses AP@0.5. Future work could also explore sets of complementary fitness functions and evaluate approaches for multi-criteria optimization for ML model selection \citep{koch2015efficient,ali2017accurate}.

OOD scores can be measured for different parts of an image. What pixels to send to the autoencoder is another important design decision. Initially, we used the entire image as input to the autoencoder, which showed promising results in detecting major changes in the environmental conditions, e.g., leaving the ODD due to nightfall or heavy fog. However, it quickly became evident that this input generated too small differences in the autoencoder's reconstruction error between inliers and outliers, i.e., it was not a feasible approach to reject basic shapes. We find this to be in line with how the ``curse of dimensionality'' affects unsupervised anomaly detection in general~\citep{zimek2012survey} -- the anomalies we try to find are dwarfed by the background information. Instead, we decided to focus on squares (a good shape for the autoencoder) containing pixels close to the bounding box of the detected object, and tried three solutions: 1) extracting a square centered on the middle pixel, 2) extracting the entire bounding box and padding with gray pixels to make it a square, and 3) stretching the contents of the bounding box to fit a rectangle  matching the average aspect ratio of pedestrians in the development set. The third approach was the most successful in our study, and is now used in SMIRK. Future OOD architectures will likely combine different selection of the input images.

The fidelity of the radar signatures in the simulator matters. While it is easy for a human to tell how realistic the visual appearance of objects are in ESI Pro-SiVIC, assessing the appropriateness of its radar signature model requires a sensor expert. In SMIRK, we attached the same radar signature to all pedestrians, i.e., the one provided for human bodies in the object catalog. For all basic shapes, on the other hand, we attach the same simplistic spherical radar signature. Designing customized signatures is beyond the scope of our project, thus we acknowledge this limitation as a threat to validity. It is possible that system testing results would have been different if more elaborate radar signatures were used.

\subsection{Reflections on Industry-Academia Collaboration}
Engineering research is ideally done in projects that involve partners from both industry and academia. While it might sound easy to accomplish, the software and systems engineering research community acknowledges that it is hard. Numerous papers address challenges and best practices in industry-academia collaboration. A systematic review by \citet{garousi2016challenges} identified 10 challenge themes (e.g., different time horizons, contrasting reward systems, limited practical relevance, and limited resources) and 17 best practice themes (e.g., select real-world problems, work in an agile fashion, organize regular meetings, and identify industry champions). In this section, we share three main reflections related to industry-academia collaboration around safety-critical ML-based demonstrator development that complement previously reported perspectives.

Collaborating in a safety-critical context is sensitive. The research relation between industry and academia in Sweden is recognized in the (empirical) software engineering community as particularly good. The relations have developed over decades, and we have successfully conducted several research projects guided by the best practices from the literature. However, developing a publicly available demonstrator with an accompanying safety case was a new experience. We found that industry partners are highly reluctant to put their names anywhere in anyway that could suggest any form of liability. Legal discussions can completely stall research projects. For SMIRK, the only reasonable way forward was to 1) largely remove the traceability of individual partner's contributions, and 2) add explicit disclaimers that SMIRK is only intended for simulators and that all users assume all responsibility and risk etc. We accept that this compromise threatens the validity of our work.

Preparing for long-term maintenance of an OSS demonstrator is difficult. Initiating development of a demonstrator system in a public repository under an OSS license is no problem. However, preparing for long-term maintenance of the system is a different story. Research funding is typically project-based and when projects conclude, it might be hard to motivate maintenance efforts -- even if there are active users. Long-term support of OSS tends to depend on individual champions, but, this is rarely sustainable. We acknowledge the challenge and support SMIRK's longevity by 1) hosting the source code in a GitHub repository managed by the independent non-profit institute RISE, 2) publishing careful contribution guidelines including a branching model, and 3) explicitly stating the responsible unit in the RISE line organization. Even more importantly, before we initiated the development, we aligned the goals of SMIRK with RISE's long-term research roadmap and project portfolio. Time will tell whether our efforts are sufficient to ensure long-term maintenance.

Finding long-term hosting of large data sets is hard. Fully open ML-based projects must go beyond sharing its source code by also making corresponding data sets available. For us, this turned out to be more difficult than expected. First, GitHub is not a feasible choice for data hosting as they recommend repositories to remain small, i.e., less than 5~GB is strongly recommended. Also, GitHub does not support sophisticated data versioning but must be combined with third party solutions such as DVC\footnote{\url{https://dvc.org/}}. Second, no project partners volunteered to assure long-term hosting of the 185~GB SMIRK data set. Data hosting requires appropriate solutions to accommodate access control, backup, bandwidth etc. Moreover, even with appropriate solutions in place, storing data is not free -- and when the research project is over, someone must keep paying. Our solution, which involved negotiations with long lead-times, was to reach out to a national non-profit AI ecosystem. Luckily, AI~Sweden agreed to host the SMIRK data set as part of their Data Factory initiative.

\section{Limitations and Threats to Validity} \label{sec:threats}
This section discusses the primary limitations of our work and the most important threats to validity. The overall goal of engineering research is to produce general design knowledge rather than to solve the problems of unique instances. Consequently, we critically discuss our research with respect to the specialized technological rule: ``To \textit{develop a safety case} for \textit{ML-based perception in ADAS} apply \textit{AMLAS}.'' As proposed by \citet{engstrom2020software}, we discuss the design knowledge resulting from our engineering research with respect to rigor, relevance, and novelty. Our discussion of rigor further addresses categories of validity in qualitative research as presented by \citet{maxwell1992understanding}, 

\subsection{Rigor}
Rigor refers to the degree to which the research is conducted in a thorough, rigorous, and systematic manner. This includes factors such as the use of appropriate research methods, the careful design of the study, and the robustness of the data analysis. Presenting engineering research constructs explicitly supports communicating the research contributions to readers and peer researchers. In this work, we express both a design problem and how we seek to specialize a technological rule from ML in autonomous systems to ML in ADAS.

We primarily support the rigor of our engineering research by applying the engineering frameworks SOTIF and AMLAS. We claim that we properly adapted the frameworks to the SMIRK development context and Section~\ref{sec:bg} shares our preunderstanding, which allows others to assess our process interpretations. We highlight that we have been in contact with developers of both frameworks. The initiator of SOTIF is part of our research collaboration network in Sweden and RISE Research Institutes of Sweden is involved in the standardization process. Regarding AMLAS, to support a valid implementation  in our R\&D project, we invited the developers to give a workshop with plenty of interactivity. 

The scope of our specialized technological rule, and the underlying engineering research, involves the perspectives of automotive stakeholders. \textit{Interpretive validity} addresses the researchers' ability to interpret the perspectives of the R\&D participants. There is a risk that the views of the authors influenced the reported lessons learned and practical advice without properly reflecting all perspectives involved. We mitigate this researcher bias through \textit{member checking}, i.e., letting participants validate our final outcome.

Our primary approach to mitigate threats to \textit{evaluation validity} is through \textit{transparency}.
We carefully describe design decisions for SMIRK and the work involved in the six AMLAS phases. Thanks to the complete safety case argumentation in Appendix~\ref{app:amlas}, we stress that the traceability from arguments to the design and implementation supports external assessments. The description of our safety evidence can be traced on GitHub at the level of commits. Finally, as \textit{descriptive validity} also involves issues of omission, we claim that our reported ML safety case is complete. The transparency allows others to scrutinize our claim.

\subsection{Relevance}
Relevance refers to the degree to which the research addresses a relevant and important problem. Involved factors include the validity of the design problem, the potential impact of the study findings on the field, the relevance for practitioners and researchers, and the generalizability.

We support relevance by anchoring the engineering research in a real problem instance. Many companies are currently facing challenges related to safety assurance for ML-based automotive perception systems. Working with the concrete challenge of safety assurance for ML in SMIRK helped connecting all stakeholders. Threats to relevance have been mitigated in the SMILE~III project through \textit{prolonged involvement}, i.e., the long-term relations that evolved during the study. The first SMILE project started already in 2016 and several partners have worked jointly on safety cage architectures since 2018. What we report does not represent individual interviews but joint work and regular project meetings.

Generalizabiliy is the condition of extending the findings to other contexts. The rich description of the systems and safety engineering that led to the open SMIRK ADAS~\citep{socha2022smirk} supports \textit{analytical generalization} to other contexts. Furthermore, just like AMLAS is applicable to safety argumentation in different domains, we believe that the safety case provided in Appendix~\ref{app:amlas} can inspire similar projects beyond the automotive sector -- especially if computer vision is involved. Still, we acknowledge three limitations that threaten the generalizability.

First, we developed SMIRK for the simulator ESI~Pro-SiVIC. We have not systematically analyzed how different aspects of the safety case generalize to an ML-based ADAS intended for deployment in a real-world vehicle. However, we mitigated this threat through \textit{prolonged industry involvement}. All SMILE projects have received an equal share of public and industry funding, which assured that the R\&D activities entailed were considered practically relevant to industry. We believe that the SMIRK requirements would generalize to real-world systems, whereas the two AMLAS stages data management and model deployment would change the most. Future work should investigate this in detail. However, we argue that our work with synthetic images is relevant for real-world systems due to the growing interest in combining synthetic and natural images~\citep{poucin2021boosting}.

Second, SMIRK is a single independent ADAS designed for a minimalisic ODD. The main strategy to mitigate threats related to the minimalism was again \textit{prolonged involvement}, i.e., our industry partners found our safety case relevant, thus also others in the community are likely to share this view. It is less clear how the independence of the SMIRK ADAS influences the generalizability of our contributions as a modern car can be considered a system-of-systems~\citep{pelliccione2020beyond}. We leave it to future work to assess how the safety case would change if additional ADAS were involved in the same safety argumentation.

Third, Python is dynamically typed and not an ideal choice for development of safety-critical applications. We chose Python to get easy access to numerous state-of-the-art ML libraries. Also, as it is the dominating language in the research community others can more easily build on our work. A real-world in-vehicle implementation would lead to another language choice, e.g., adhering to MISRA C~\citep{motor2012misra}, a widely accepted set of software development guidelines for using the C programming language in safety-critical systems.

\subsection{Novelty}
Novelty refers to the newness or originality of an idea, approach, or method. The novelty of our contributions primarily originate in two concepts useful in engineering research. First, \textit{combinatorial creativity} involves the combination of existing ideas, concepts, or elements to create something new and novel. Among other things, our work combines research on requirements engineering and software testing in the development of a safety case for ML. 

Second, \textit{research synthesis} is a method of systematically combining the results of multiple studies. There are several approaches to synthesizing results, e.g., statistical meta-analysis as popular in medicine, less formal narrative methods, and meta-ethnography that relies on interpretation to preserve the social contexts in which the original findings emerged. Closer to our work, \citet{denyer2008developing} discuss how design-oriented
research synthesis can address fragmentation and increase the chances of industrial adoption.

There is an ever-increasing number of software engineering publications and the research field is diverse and fragmented. Moreover, the practical relevance of the research has been frequently questioned~\citep{garousi2020practical}. We argue that our approach to design-oriented synthesis of fragmented requirements on ML-based perception systems, safety cage architectures, DNNs and ML testing --- identified in various software engineering subfields and connected using the AMLAS framework --- represents a holistic case that has been missing. Collecting these pieces in a joint publication increases the chances for industrial impact.

Finally, as reported by \citet{engstrom2020software}, expressing technological rules clearly and at a carefully selected level of abstraction helps in communicating the novelty. To the best of our knowledge, SMIRK was the first completely transparent ML-based OSS ADAS. On the same note, related to our specialized technological rule, we believe this publication is the first complete transparent safety case for an ADAS ML component. We also believe that this is the first peer-reviewed work from a complete application of AMLAS that does not involve any of the inventors of the framework. Consequently, this also constitutes novelty related to the general technological rule presented in Section~\ref{sec:method}.

 %We have not conducted a systematic quality requirements prioritization, such as an analytical hierarchy process workshop~\citep{kassab2015applying}. Qualities were analyzed and prioritized less formally in several SMILE~III workshops.

%\textbf{Evaluation validity} refers to the researchers' ability to look critically at the results and the research itself as a way of learning and expanding understanding. \citet{maxwell1992understanding} argues that it is often less critical as ``many researchers make no claims to evaluate what they study.'' We, on the other hand, present a safety case that we evaluated thoroughly. 

\section{Conclusion and Future Work} \label{sec:conc}
Safe ML is going to be fundamental when increasing the level of vehicle automation. Several automotive standardization initiatives are ongoing to allow safety certification for ML in road vehicles, e.g., ISO 21448 SOTIF. However, standards provide high-level requirements that must be operationalized in each development context. Unfortunately, there is a lack of publicly available ML-based automotive demonstrator systems that can be used to study safety case development. We set out to remedy this lack through engineering research in an industry-academia collaboration in Sweden. The design knowledge provided through our engineering research relates to the technological rule: ``To \textit{develop a safety case} for \textit{ML-based perception in ADAS} apply \textit{AMLAS}.

We present a safety argumentation for SMIRK, a PAEB designed for operation in the industry-grade simulator ESI Pro-SiVIC \citep{socha2022smirk}, available on GitHub under an OSS license \citep{github_smirk}. SMIRK uses a radar sensor for object detection and an ML-based component relying on a DNN for pedestrian recognition. Originating in SMIRK's minimalistic ODD, we present a complete safety case for its ML-based component by following the AMLAS framework \citep{amlas2021}. To the best of our knowledge, this work constitutes the first complete application of AMLAS independent from its authors. Guided by AMLAS, we argue that we demonstrate how to provide sufficient evidence that ML in SMIRK is safe given its ODD. We conclude that even for a very restricted ODD, the size of the ML safety case is considerable, i.e., there are many aspects of the AI engineering that must be clearly explained. As the devil is in the detail, we carefully report all steps -- and we recommend future research studies to follow suit.

We report several lessons learned related to AI engineering in the SOTIF/AMLAS context. First, using a simulator to create synthetic data sets for ML training particularly limits the validity of the negative examples. Second, the complexity of object detection evaluations necessitates internal training within the project team. Third, composing the fitness function used for model selection is a delicate engineering activity that forces explicit tradeoff decisions. Fourth, what parts of an image to send to a autoencoder for OOD detection is an open question -- for SMIRK, we stretch the content of bounding boxes to a larger square. Finally, we report three reflections related to industry-academia collaboration around safety-critical ML-based demonstrator development.

Thanks to the complete safety case, SMIRK can be used as a starting point for several avenues of future research. First, the SMIRK MVP enables studies on efficient and effective approaches to conduct safety assurance for ODD extension \citep{weissensteiner2021virtual}. In this context, SMIRK could be used as a platform to study dynamic safety cases \citep{denney2015dynamic}, i.e., updating the safety case as the system evolves, and reuse of safety evidence for new operational contexts \citep{de2019amass}. Second, SMIRK could be used as a realistic test benchmark for automotive ML testing, including search-based techniques for test case generation. The testing community has largely worked on offline testing of single frames, but we know that this is insufficient \citep{haq2021can}. Also, we recommend comparative studies involving real-world testing in controlled environments, as discrepancies do exist between simulations and the physical world \citep{stocco2022mind}. Third, we recommend the community to port SMIRK to other simulators beyond ESI Pro-SiVIC. As we investigated in previous work, running highly similar test scenarios in different simulators can lead to considerably different results \citep{borg2021digital} -- further exploring this phenomenon using SMIRK would be a valuable research direction. Finally, while SOTIF explicitly excludes antagonistic attacks, there are good reasons to consider both safety and cybersecurity jointly in automotive systems through co-engineering \citep{amorim2017systematic}. We plan to use SMIRK as a starting point in future studies on adversarial ML attacks.

\section*{Acknowledgment}
Thanks go to ESI Group for supporting us with technical details along the way, especially Erik Abenius and Fran\c cois-Xavier Jegeden. We also thank AI Sweden for agreeing to host the SMIRK data set as part of their Data Factory initiative. This work was carried out within the SMILE~III project financed by Vinnova, FFI, Fordonsstrategisk forskning och innovation under the grant number 2019-05871 and partially supported by the Wallenberg AI, Autonomous Systems and Software Program (WASP) funded by Knut and Alice Wallenberg Foundation.

\section*{Conflict of Interest Statement}
The authors have no conflicts of interest to declare.

\section*{Data Availability Statement}
All data related to this study is publicy available. The SMIRK source code is publicly available on GitHub under a GPL-3.0 license: \url{https://github.com/RI-SE/smirk/}. This repository also contains all safety evidence and instructions to allow reproduction of our results. The data set used for development of SMIRK's ML component is available at AI~Sweden: \url{https://www.ai.se/en/data-factory/datasets/data-factory-datasets/smirk-dataset}.

%\begin{appendices}

%\section{Section title of first appendix}\label{secA1}

%An appendix contains supplementary information that is not an essential part of the text itself but which may be helpful in providing a more comprehensive understanding of the research problem or it is information that is too cumbersome to be included in the body of the paper.

%\end{appendices}

\bibliography{smirk}

\begin{thebibliography}{83}
\providecommand{\natexlab}[1]{#1}
\providecommand{\url}[1]{{#1}}
\providecommand{\urlprefix}{URL }
\providecommand{\doi}[1]{\url{https://doi.org/#1}}
\providecommand{\eprint}[2][]{\url{#2}}
 \bibcommenthead

\bibitem[{van Aken(2004)}]{aken2004management}
van Aken JE (2004) Management research based on the paradigm of the design
  sciences: the quest for field-tested and grounded technological rules.
  Journal of management studies 41(2):219--246

\bibitem[{Ali et~al(2017)Ali, Lee, and Chung}]{ali2017accurate}
Ali R, Lee S, Chung TC (2017) Accurate multi-criteria decision making
  methodology for recommending machine learning algorithm. Expert Systems with
  Applications 71:257--278

\bibitem[{Amorim et~al(2017)Amorim, Martin, Ma, Schmittner, Schneider, Macher,
  Winkler, Krammer, and Kreiner}]{amorim2017systematic}
Amorim T, Martin H, Ma Z, et~al (2017) Systematic pattern approach for safety
  and security co-engineering in the automotive domain. In: Proc. of the Int'l.
  Conf. on Computer Safety, Reliability, and Security, pp 329--342

\bibitem[{An and Cho(2015)}]{an2015variational}
An J, Cho S (2015) Variational autoencoder based anomaly detection using
  reconstruction probability. Special Lecture on IE 2(1):1--18

\bibitem[{Arrieta et~al(2020)Arrieta, D{\'\i}az-Rodr{\'\i}guez, Del~Ser,
  Bennetot, Tabik, Barbado, Garc{\'\i}a, Gil-L{\'o}pez, Molina, Benjamins
  et~al}]{arrieta2020explainable}
Arrieta AB, D{\'\i}az-Rodr{\'\i}guez N, Del~Ser J, et~al (2020) Explainable
  artificial intelligence {(XAI)}: Concepts, taxonomies, opportunities and
  challenges toward responsible {AI}. Information Fusion 58:82--115

\bibitem[{Ashmore et~al(2021)Ashmore, Calinescu, and
  Paterson}]{ashmore2021assuring}
Ashmore R, Calinescu R, Paterson C (2021) Assuring the machine learning
  lifecycle: Desiderata, methods, and challenges. ACM Computing Surveys
  54(5):1--39

\bibitem[{{Assurance Case Working Group}(2021)}]{gsn2021}
{Assurance Case Working Group} (2021) {Goal Structuring Notation Community
  Standard (Version 3)}. Tech. Rep. SCSC-141C, Safety-Critical Systems Club, UK

\bibitem[{Barr et~al(2014)Barr, Harman, McMinn, Shahbaz, and
  Yoo}]{barr2014oracle}
Barr ET, Harman M, McMinn P, et~al (2014) The oracle problem in software
  testing: A survey. IEEE Transactions on Software Engineering 41(5):507--525

\bibitem[{Ben~Abdessalem et~al(2016)Ben~Abdessalem, Nejati, Briand, and
  Stifter}]{ben_abdessalem_testing_2016}
Ben~Abdessalem R, Nejati S, Briand LC, et~al (2016) Testing advanced driver
  assistance systems using multi-objective search and neural networks. In:
  Proc. of the 31st {Int'l.} {Conf.} on {Automated} {Software} {Engineering},
  pp 63--74

\bibitem[{{Ben Abdessalem} et~al(2018{\natexlab{a}}){Ben Abdessalem}, Nejati,
  Briand, and Stifter}]{abdessalem2018testinglearnable}
{Ben Abdessalem} R, Nejati S, Briand LC, et~al (2018{\natexlab{a}}) Testing
  vision-based control systems using learnable evolutionary algorithms. In:
  Proc. of the 40th Int'l. Conf. on Software Engineering, pp 1016--1026

\bibitem[{{Ben Abdessalem} et~al(2018{\natexlab{b}}){Ben Abdessalem},
  Panichella, Nejati, Briand, and Stifter}]{abdessalem2018testing}
{Ben Abdessalem} R, Panichella A, Nejati S, et~al (2018{\natexlab{b}}) Testing
  autonomous cars for feature interaction failures using many-objective search.
  In: Proc. of the 33rd Int'l. Conf. on Automated Software Engineering, pp
  143--154

\bibitem[{Bolya et~al(2020)Bolya, Foley, Hays, and Hoffman}]{bolya2020tide}
Bolya D, Foley S, Hays J, et~al (2020) Tide: A general toolbox for identifying
  object detection errors. In: Proc. of the European Conf. on Computer Vision,
  pp 558--573

\bibitem[{Borg et~al(2019)Borg, Englund, Wnuk, Durann, Lewandowski, Gao, Tan,
  Kaijser, L{\"o}nn, and T{\"o}rnqvist}]{borg2019safely}
Borg M, Englund C, Wnuk K, et~al (2019) Safely entering the deep: A review of
  verification and validation for machine learning and a challenge elicitation
  in the automotive industry. Journal of Automotive Software Engineering
  1(1):1--19

\bibitem[{Borg et~al(2021{\natexlab{a}})Borg, {Ben Abdessalem}, Nejati,
  Jegeden, and Shin}]{borg2021digital}
Borg M, {Ben Abdessalem} R, Nejati S, et~al (2021{\natexlab{a}}) Digital twins
  are not monozygotic: Cross-replicating {ADAS} testing in two industry-grade
  automotive simulators. In: Proc. of the 14th Conf. on Software Testing,
  Verification and Validation, pp 383--393

\bibitem[{Borg et~al(2021{\natexlab{b}})Borg, Bronson, Christensson, Olsson,
  Lennartsson, Sonnsj{\"o}, Ebabi, and Karsberg}]{borg2021exploring}
Borg M, Bronson J, Christensson L, et~al (2021{\natexlab{b}}) Exploring the
  assessment list for trustworthy {AI} in the context of advanced
  driver-assistance systems. In: Prov. of the 2nd Int'l. Workshop on Ethics in
  Software Engineering Research and Practice, pp 5--12

\bibitem[{Bosch et~al(2021)Bosch, Olsson, and Crnkovic}]{bosch2021engineering}
Bosch J, Olsson HH, Crnkovic I (2021) Engineering {AI} systems: A research
  agenda. In: Artificial Intelligence Paradigms for Smart Cyber-Physical
  Systems. IGI global, p 1--19

\bibitem[{Chen et~al(2012)Chen, Babar, and Nuseibeh}]{chen2012characterizing}
Chen L, Babar MA, Nuseibeh B (2012) Characterizing architecturally significant
  requirements. IEEE Software 30(2):38--45

\bibitem[{Denney et~al(2015)Denney, Pai, and Habli}]{denney2015dynamic}
Denney E, Pai G, Habli I (2015) Dynamic safety cases for through-life safety
  assurance. In: Proc. of the 37th Int'l. Conf. on Software Engineering, pp
  587--590

\bibitem[{Denyer et~al(2008)Denyer, Tranfield, and
  Van~Aken}]{denyer2008developing}
Denyer D, Tranfield D, Van~Aken JE (2008) Developing design propositions
  through research synthesis. Organization studies 29(3):393--413

\bibitem[{Dollar et~al(2011)Dollar, Wojek, Schiele, and
  Perona}]{dollar2011pedestrian}
Dollar P, Wojek C, Schiele B, et~al (2011) Pedestrian detection: An evaluation
  of the state of the art. IEEE Transactions on Pattern Analysis and Machine
  Intelligence 34(4):743--761

\bibitem[{Ebadi et~al(2021)Ebadi, Moghadam, Borg, Gay, Fontes, and
  Socha}]{ebadi2021efficient}
Ebadi H, Moghadam MH, Borg M, et~al (2021) Efficient and effective generation
  of test cases for pedestrian detection-search-based software testing of
  {Baidu Apollo} in {SVL}. In: Proc. of the Int'l. Conf. on Artificial
  Intelligence Testing, pp 103--110

\bibitem[{Engstr{\"o}m et~al(2020)Engstr{\"o}m, Storey, Runeson, H{\"o}st, and
  Baldassarre}]{engstrom2020software}
Engstr{\"o}m E, Storey MA, Runeson P, et~al (2020) How software engineering
  research aligns with design science: a review. Empirical Software Engineering
  25(4):2630--2660

\bibitem[{Fagan(1976)}]{fagan1976design}
Fagan M (1976) Design and code inspections to reduce errors in program
  development. IBM Systems Journal 15(3):182--211

\bibitem[{Garousi et~al(2016)Garousi, Petersen, and
  Ozkan}]{garousi2016challenges}
Garousi V, Petersen K, Ozkan B (2016) Challenges and best practices in
  industry-academia collaborations in software engineering: {A} systematic
  literature review. Information and Software Technology 79:106--127

\bibitem[{Garousi et~al(2020)Garousi, Borg, and Oivo}]{garousi2020practical}
Garousi V, Borg M, Oivo M (2020) Practical relevance of software engineering
  research: synthesizing the community’s voice. Empirical Software
  Engineering 25(3):1687--1754

\bibitem[{Gauerhof et~al(2020)Gauerhof, Hawkins, Picardi, Paterson, Hagiwara,
  and Habli}]{gauerhof2020assuring}
Gauerhof L, Hawkins R, Picardi C, et~al (2020) Assuring the safety of machine
  learning for pedestrian detection at crossings. In: Proc. of the Int'l. Conf.
  on Computer Safety, Reliability, and Security, pp 197--212

\bibitem[{Haq et~al(2021{\natexlab{a}})Haq, Shin, Briand, Stifter, and
  Wang}]{haq2021automatic}
Haq FU, Shin D, Briand LC, et~al (2021{\natexlab{a}}) Automatic test suite
  generation for key-points detection dnns using many-objective search
  (experience paper). Proc. of the 30th Int'l. Symposium on Software Testing
  and Analysis, pp 91--102

\bibitem[{Haq et~al(2021{\natexlab{b}})Haq, Shin, Nejati, and
  Briand}]{haq2021can}
Haq FU, Shin D, Nejati S, et~al (2021{\natexlab{b}}) Can offline testing of
  deep neural networks replace their online testing? Empirical Software
  Engineering 26(5):1--30

\bibitem[{Hauer et~al(2019)Hauer, Schmidt, Holzm{\"u}ller, and
  Pretschner}]{hauer2019did}
Hauer F, Schmidt T, Holzm{\"u}ller B, et~al (2019) Did we test all scenarios
  for automated and autonomous driving systems? In: Proc. of the IEEE
  Intelligent Transportation Systems Conf., pp 2950--2955

\bibitem[{Hawkins et~al(2021)Hawkins, Paterson, Picardi, Jia, Calinescu, and
  Habli}]{amlas2021}
Hawkins R, Paterson C, Picardi C, et~al (2021) Guidance on the assurance of
  machine learning in autonomous systems (amlas). Tech. Rep. Version 1.1,
  Assuring Autonomy Int'l. Programme, University of York

\bibitem[{Henriksson et~al(2019)Henriksson, Berger, Borg, Tornberg, Englund,
  Sathyamoorthy, and Ursing}]{henriksson2019towards}
Henriksson J, Berger C, Borg M, et~al (2019) Towards structured evaluation of
  deep neural network supervisors. In: Proc. of the Int'l. Conf. on Artificial
  Intelligence Testing, pp 27--34

\bibitem[{Henriksson et~al(2021{\natexlab{a}})Henriksson, Berger, Borg,
  Tornberg, Sathyamoorthy, and Englund}]{henriksson2021performance}
Henriksson J, Berger C, Borg M, et~al (2021{\natexlab{a}}) Performance analysis
  of out-of-distribution detection on trained neural networks. Information and
  Software Technology 130:106,409

\bibitem[{Henriksson et~al(2021{\natexlab{b}})Henriksson, Berger, and
  Ursing}]{henriksson2021understanding}
Henriksson J, Berger C, Ursing S (2021{\natexlab{b}}) Understanding the impact
  of edge cases from occluded pedestrians for {ML} systems. In: Proc. of the
  47th Euromicro Conf. on Software Engineering and Advanced Applications, pp
  316--325

\bibitem[{{High-Level Expert Group on Artificial Intelligence}(2019)}]{eu2019}
{High-Level Expert Group on Artificial Intelligence} (2019) Ethics guidelines
  for trustworthy {AI}. Tech. rep., Directorate-General for Communications
  Networks, Content and Technology, European Commission

\bibitem[{Horkoff(2019)}]{horkoff2019non}
Horkoff J (2019) Non-functional requirements for machine learning: Challenges
  and new directions. In: Proc. of the IEEE 27th Int'l. Requirements
  Engineering Conf., pp 386--391

\bibitem[{IEEE(1998)}]{ieee_srs}
IEEE (1998) {IEEE} recommended practice for software requirements
  specifications. Tech. Rep. IEEE 830-1998, Institute of Electrical and
  Electronics Engineers

\bibitem[{Jia et~al(2022)Jia, Mcdermid, Lawton, and Habli}]{jia2022role}
Jia Y, Mcdermid JA, Lawton T, et~al (2022) The role of explainability in
  assuring safety of machine learning in healthcare. IEEE Transactions on
  Emerging Topics in Computing

\bibitem[{K{\"a}pyaho and Kauppinen(2015)}]{kapyaho2015agile}
K{\"a}pyaho M, Kauppinen M (2015) Agile requirements engineering with
  prototyping: A case study. In: Proc. of the 23rd Int'l. requirements
  engineering Conf., pp 334--343

\bibitem[{Klaise et~al(2020)Klaise, Van~Looveren, Cox, Vacanti, and
  Coca}]{klaise2020monitoring}
Klaise J, Van~Looveren A, Cox C, et~al (2020) Monitoring and explainability of
  models in production. In: Proc. of the ICML Workshop on Challenges in
  Deploying and Monitoring Machine Learning Systems

\bibitem[{Koch et~al(2015)Koch, Wagner, Emmerich, B{\"a}ck, and
  Konen}]{koch2015efficient}
Koch P, Wagner T, Emmerich MT, et~al (2015) Efficient multi-criteria
  optimization on noisy machine learning problems. Applied Soft Computing
  29:357--370

\bibitem[{Kruchten(1995)}]{kruchten19954+}
Kruchten PB (1995) The 4+1 view model of architecture. IEEE Software
  12(6):42--50

\bibitem[{Lin et~al(2014)Lin, Maire, Belongie, Hays, Perona, Ramanan,
  Doll{\'a}r, and Zitnick}]{lin2014microsoft}
Lin TY, Maire M, Belongie S, et~al (2014) Microsoft {COCO}: Common objects in
  context. In: European Conf. on Computer Vision, pp 740--755

\bibitem[{Liu et~al(2018)Liu, Qi, Qin, Shi, and Jia}]{liu2018path}
Liu S, Qi L, Qin H, et~al (2018) Path aggregation network for instance
  segmentation. In: Proc. of the IEEE Conf. on Computer Vision and Pattern
  Recognition, pp 8759--8768

\bibitem[{Masuda(2017)}]{masuda2017software}
Masuda S (2017) Software testing design techniques used in automated vehicle
  simulations. In: Proc. of the Int'l. Conf. on Software Testing, Verification
  and Validation Workshops, pp 300--303

\bibitem[{Maxwell(1992)}]{maxwell1992understanding}
Maxwell J (1992) Understanding and validity in qualitative research. Harvard
  educational review 62(3):279--301

\bibitem[{Mohseni et~al(2020)Mohseni, Pitale, Singh, and
  Wang}]{mohseni2020practical}
Mohseni S, Pitale M, Singh V, et~al (2020) Practical solutions for machine
  learning safety in autonomous vehicles. In: Proc. of the Artificial
  Intelligence Safety (SafeAI) Workshop at AAAI 2020,
  \urlprefix\url{http://ceur-ws.org/Vol-2560/}

\bibitem[{{Motor Industry Software Reliability Association}
  et~al(2012)}]{motor2012misra}
{Motor Industry Software Reliability Association}, et~al (2012) {MISRA-C}
  guidelines for the use of the {C} language in critical systems

\bibitem[{Panichella et~al(2015)Panichella, Kifetew, and
  Tonella}]{panichella2015reformulating}
Panichella A, Kifetew FM, Tonella P (2015) Reformulating branch coverage as a
  many-objective optimization problem. In: Proc. of the 8th Int'l. Conf. on
  Software Testing, Verification and Validation, pp 1--10

\bibitem[{Pei et~al(2017)Pei, Cao, Yang, and Jana}]{pei2017deepxplore}
Pei K, Cao Y, Yang J, et~al (2017) Deepxplore: {A}utomated whitebox testing of
  deep learning systems. In: Proc. of the 26th Symposium on Operating Systems
  Principles, pp 1--18

\bibitem[{Pelliccione et~al(2020)Pelliccione, Knauss, {\AA}gren, Heldal,
  Bergenhem, Vinel, and Brunneg{\aa}rd}]{pelliccione2020beyond}
Pelliccione P, Knauss E, {\AA}gren SM, et~al (2020) Beyond connected cars: A
  systems of systems perspective. Science of Computer Programming 191:102,414

\bibitem[{Petersson et~al(2004)Petersson, Thelin, Runeson, and
  Wohlin}]{petersson2004capture}
Petersson H, Thelin T, Runeson P, et~al (2004) Capture-recapture in software
  inspections after 10 years research: Theory, evaluation and application.
  Journal of Systems and Software 72(2):249--264

\bibitem[{Picardi et~al(2020)Picardi, Paterson, Hawkins, Calinescu, and
  Habli}]{picardi2020assurance}
Picardi C, Paterson C, Hawkins RD, et~al (2020) Assurance argument patterns and
  processes for machine learning in safety-related systems. In: Proc. of the
  Workshop on Artificial Intelligence Safety, pp 23--30

\bibitem[{Poucin et~al(2021)Poucin, Kraus, and Simon}]{poucin2021boosting}
Poucin F, Kraus A, Simon M (2021) Boosting instance segmentation with synthetic
  data: A study to overcome the limits of real world data sets. In: Proc. of
  the IEEE/CVF Int'l. Conf. on Computer Vision, pp 945--953

\bibitem[{Preschern et~al(2015)Preschern, Kajtazovic, and
  Kreiner}]{preschern2015building}
Preschern C, Kajtazovic N, Kreiner C (2015) Building a safety architecture
  pattern system. In: Proc. of the 18th European Conf. on Pattern Languages of
  Program, pp 1--55

\bibitem[{Rajput(2020)}]{rajput2020}
Rajput M (2020) {YOLO V5 -- Explained and Demystified}.
  \url{https://towardsai.net/p/computer-vision/yolo-v5\%E2\%80\%8A-\%E2\%80\%8Aexplained-and-demystified}

\bibitem[{Ralph et~al(2020)Ralph, {bin Ali}, Baltes, Bianculli, Diaz, Dittrich,
  Ernst, Felderer, Feldt, Filieri et~al}]{ralph2020empirical}
Ralph P, {bin Ali} N, Baltes S, et~al (2020) Empirical standards for software
  engineering research. arXiv preprint arXiv:201003525

\bibitem[{Redmon et~al(2016)Redmon, Divvala, Girshick, and
  Farhadi}]{redmon2016you}
Redmon J, Divvala S, Girshick R, et~al (2016) You only look once: Unified,
  real-time object detection. In: Proc. of the IEEE Conf. on Computer Vision
  and Pattern Recognition, pp 779--788

\bibitem[{Riccio et~al(2020)Riccio, Jahangirova, Stocco, Humbatova, Weiss, and
  Tonella}]{riccio2020testing}
Riccio V, Jahangirova G, Stocco A, et~al (2020) Testing machine learning based
  systems: A systematic mapping. Empirical Software Engineering
  25(6):5193--5254

\bibitem[{{RISE Research Institutes of Sweden}(2022)}]{github_smirk}
{RISE Research Institutes of Sweden} (2022) {SMIRK} {GitHub} repository.
  \urlprefix\url{https://github.com/RI-SE/smirk/}

\bibitem[{Runeson et~al(2020)Runeson, Engstr{\"o}m, and
  Storey}]{runeson2020design}
Runeson P, Engstr{\"o}m E, Storey MA (2020) The design science paradigm as a
  frame for empirical software engineering. In: Contemporary Empirical Methods
  in Software Engineering. Springer, p 127--147

\bibitem[{Salay et~al(2018)Salay, Queiroz, and Czarnecki}]{salay2018analysis}
Salay R, Queiroz R, Czarnecki K (2018) An analysis of {ISO 26262}: Machine
  learning and safety in automotive software

\bibitem[{Schwalbe and Schels(2020)}]{schwalbe2020survey}
Schwalbe G, Schels M (2020) A survey on methods for the safety assurance of
  machine learning based systems. In: Proc. of the 10th European Congress on
  Embedded Real Time Software and Systems

\bibitem[{Schwalbe et~al(2020)Schwalbe, Knie, S{\"a}mann, Dobberphul, Gauerhof,
  Raafatnia, and Rocco}]{schwalbe2020structuring}
Schwalbe G, Knie B, S{\"a}mann T, et~al (2020) Structuring the safety
  argumentation for deep neural network based perception in automotive
  applications. In: Proc. of the Int'l. Conf. on Computer Safety, Reliability,
  and Security, Springer, pp 383--394

\bibitem[{Schyllander(2014)}]{msb2014}
Schyllander J (2014) Fotgängarolyckor - statistik och analys. Tech. Rep.
  MSB744, Swedish Civil Contingencies Agency,
  \urlprefix\url{https://rib.msb.se/filer/pdf/27438.pdf}

\bibitem[{Socha et~al(2022)Socha, Borg, and Henriksson}]{socha2022smirk}
Socha K, Borg M, Henriksson J (2022) {SMIRK}: A machine learning-based
  pedestrian automatic emergency braking system with a complete safety case.
  Software Impacts 13:100,352

\bibitem[{Song et~al(2022)Song, Borg, Engström, Ardö, and
  Rico}]{song2022exploring}
Song Q, Borg M, Engström E, et~al (2022) Exploring {ML} testing in practice:
  Lessons learned from an interactive rapid review with axis communications.
  In: Proc. of the 1st Int'l. Conf. on AI Engineering -- Software Engineering
  for AI

\bibitem[{Stocco et~al(2022)Stocco, Pulfer, and Tonella}]{stocco2022mind}
Stocco A, Pulfer B, Tonella P (2022) Mind the gap! {A} study on the
  transferability of virtual vs physical-world testing of autonomous driving
  systems. IEEE Transactions on Software Engineering

\bibitem[{Tambon et~al(2022)Tambon, Laberge, An, Nikanjam, Mindom, Pequignot,
  Khomh, Antoniol, Merlo, and Laviolette}]{tambon2021certify}
Tambon F, Laberge G, An L, et~al (2022) How to certify machine learning based
  safety-critical systems? {A} systematic literature review. Automated Software
  Engineering 29(38)

\bibitem[{Tao et~al(2019)Tao, Li, Wotawa, Felbinger, and
  Nica}]{tao2019industrial}
Tao J, Li Y, Wotawa F, et~al (2019) On the industrial application of
  combinatorial testing for autonomous driving functions. In: Proc. of the
  Int'l. Conf. on Software Testing, Verification and Validation Workshops, pp
  234--240

\bibitem[{Thorn et~al(2018)Thorn, Kimmel, Chaka, Hamilton
  et~al}]{thorn2018framework}
Thorn E, Kimmel SC, Chaka M, et~al (2018) A framework for automated driving
  system testable cases and scenarios. Tech. rep., US Department of
  Transportation. National Highway Traffic Safety Administration.

\bibitem[{Tian et~al(2018)Tian, Pei, Jana, and Ray}]{tian2018deeptest}
Tian Y, Pei K, Jana S, et~al (2018) Deeptest: {A}utomated testing of
  deep-neural-network-driven autonomous cars. In: Proc. of the 40th Int'l.
  Conf. on Software Engineering, pp 303--314

\bibitem[{Tsilionis et~al(2021)Tsilionis, Wautelet, Faut, and
  Heng}]{tsilionis2021unifying}
Tsilionis K, Wautelet Y, Faut C, et~al (2021) Unifying behavior driven
  development templates. In: Proc. of the 29th Int'l. Requirements Engineering
  Conf., pp 454--455

\bibitem[{de~la Vara et~al(2019)de~la Vara, Ruiz, Gallina, Blondelle,
  Ala{\~n}a, Herrero, Warg, Skoglund, and Bramberger}]{de2019amass}
de~la Vara JL, Ruiz A, Gallina B, et~al (2019) The {AMASS} approach for
  assurance and certification of critical systems. In: Embedded World 2019

\bibitem[{Weissensteiner et~al(2021)Weissensteiner, Stettinger, Rumetshofer,
  and Watzenig}]{weissensteiner2021virtual}
Weissensteiner P, Stettinger G, Rumetshofer J, et~al (2021) Virtual validation
  of an automated lane-keeping system with an extended operational design
  domain. Electronics 11(1):72

\bibitem[{Wiegers(2008)}]{wiegers_srs}
Wiegers K (2008) Karl {W}iegers' software requirements specification ({SRS})
  template. Tech. rep., Process Impact,
  \urlprefix\url{https://www.modernanalyst.com/Resources/Templates/tabid/146/ID/497/Karl-Wiegers-Software-Requirements-Specification-SRS-Template.aspx}

\bibitem[{Wieringa(2014)}]{wieringa2014design}
Wieringa RJ (2014) Design science methodology for information systems and
  software engineering. Springer

\bibitem[{Willers et~al(2020)Willers, Sudholt, Raafatnia, and
  Abrecht}]{willers2020safety}
Willers O, Sudholt S, Raafatnia S, et~al (2020) Safety concerns and mitigation
  approaches regarding the use of deep learning in safety-critical perception
  tasks. In: Proc. of the Int'l. Conf. on Computer Safety, Reliability, and
  Security, pp 336--350

\bibitem[{Wozniak et~al(2020)Wozniak, C{\^a}rlan, Acar-Celik, and
  Putzer}]{wozniak2020safety}
Wozniak E, C{\^a}rlan C, Acar-Celik E, et~al (2020) A safety case pattern for
  systems with machine learning components. In: Proc. of the Int'l. Conf. on
  Computer Safety, Reliability, and Security, pp 370--382

\bibitem[{Wu and Nevatia(2008)}]{wu2008optimizing}
Wu B, Nevatia R (2008) Optimizing discrimination-efficiency tradeoff in
  integrating heterogeneous local features for object detection. In: Proc. of
  the IEEE Conf. on Computer Vision and Pattern Recognition, pp 1--8

\bibitem[{Wu and Kelly(2004)}]{wu2004safety}
Wu W, Kelly T (2004) Safety tactics for software architecture design. In: Proc.
  of the 28th Annual Int'l. Computer Software and Applications Conf., pp
  368--375

\bibitem[{Zablocki et~al(2022)Zablocki, Ben-Younes, P{\'e}rez, and
  Cord}]{zablocki2022explainability}
Zablocki {\'E}, Ben-Younes H, P{\'e}rez P, et~al (2022) Explainability of deep
  vision-based autonomous driving systems: {R}eview and challenges. Int'l
  Journal of Computer Vision (130):2425–--2452

\bibitem[{Zhang et~al(2018)Zhang, Zhang, Zhang, Liu, and
  Khurshid}]{zhang2018deeproad}
Zhang M, Zhang Y, Zhang L, et~al (2018) Deeproad: {GAN}-based metamorphic
  testing and input validation framework for autonomous driving systems. In:
  Proc. of the 33rd Int'l. Conf. on Automated Software Engineering, pp 132--142

\bibitem[{Zimek et~al(2012)Zimek, Schubert, and Kriegel}]{zimek2012survey}
Zimek A, Schubert E, Kriegel HP (2012) A survey on unsupervised outlier
  detection in high-dimensional numerical data. Statistical Analysis and Data
  Mining: The ASA Data Science Journal 5(5):363--387

\end{thebibliography}

\newpage

\begin{appendices}

\section{AMLAS Safety Argumentation for Machine Learning in SMIRK} \label{app:amlas}
This section describes the complete SMIRK safety argumentation organized by the six AMLAS stages \citep{amlas2021}. For each step, we present an argument pattern using GSN notation \citep{gsn2021} and present the final argument in a text box.

\subsection{Stage 1: Machine Learning Assurance Scoping} \label{amlas:g}
\begin{figure}
\centering
\includegraphics[width=1\textwidth]{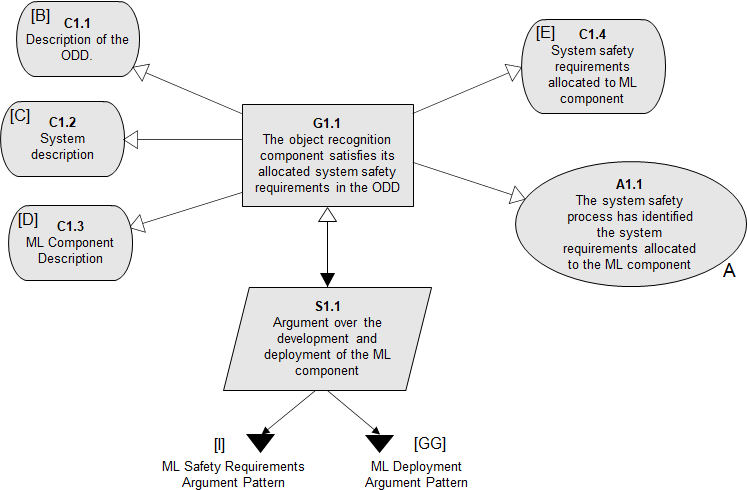}
\caption{ML Assurance Scoping Argument Pattern \textbf{[F]}.}
\label{fig:pattern_ml_assurance_scoping}
\label{amlas:f}
\end{figure}

Figure~\ref{fig:pattern_ml_assurance_scoping} shows the overall ML assurance scoping argument pattern for SMIRK. The pattern, as well as all subsequent patterns in this paper, follows the examples provided in AMLAS, but adapts it to the specific SMIRK case. Furthermore, we provide evidence that supports our arguments. 

The top claim, i.e., the starting point for the safety argument for the ML-based component, is that the system safety requirements that have been allocated to the pedestrian recognition component are satisfied in the ODD (G1.1). The safety claim for the pedestrian recognition component is made within the context of the information that was used to establish the safety requirements allocation, i.e., the system description (\textbf{[C]}), the ODD (\textbf{[B]}), and the ML component description (\textbf{[D]}). The allocated system safety requirements (\textbf{[E]}) are provided as context. An explicit assumption is made that the allocated safety requirements have been correctly defined (A1.1), as this is part of the overall system safety process (FuSa and SOTIF) preceding AMLAS. Our claim to the validity of this assumption is presented in relation to the HARA described in \textbf{[E]}. As stated in AMLAS, ``the primary aim of the ML Safety Assurance Scoping argument is to explain and justify the essential relationship between, on the one hand, the system-level safety requirements and associated hazards and risks, and on the other hand, the ML-specific safety requirements and associated ML performance and failure conditions.''

The ML safety claim is supported by an argument split into two parts. First, the development of the ML component is considered with an argument that starts with the elicitation of the ML safety requirements. Second, the deployment of the ML component is addressed with a corresponding argument.

\begin{center}
\begin{framed}

\noindent \textbf{ML Safety Assurance Scoping Argument \textbf{[G]}}\\
SMIRK instantiates the ML safety assurance scoping argument through the artifacts listed in the Table~\ref{tab:amlas_index}. The set of artifacts constitutes the safety case for SMIRK's ML-based pedestrian recognition component.

\end{framed}
\end{center}

\subsection{Stage 2: Machine Learning Requirements Assurance} \label{amlas:k}
\begin{figure}
\centering
\includegraphics[width=1\textwidth]{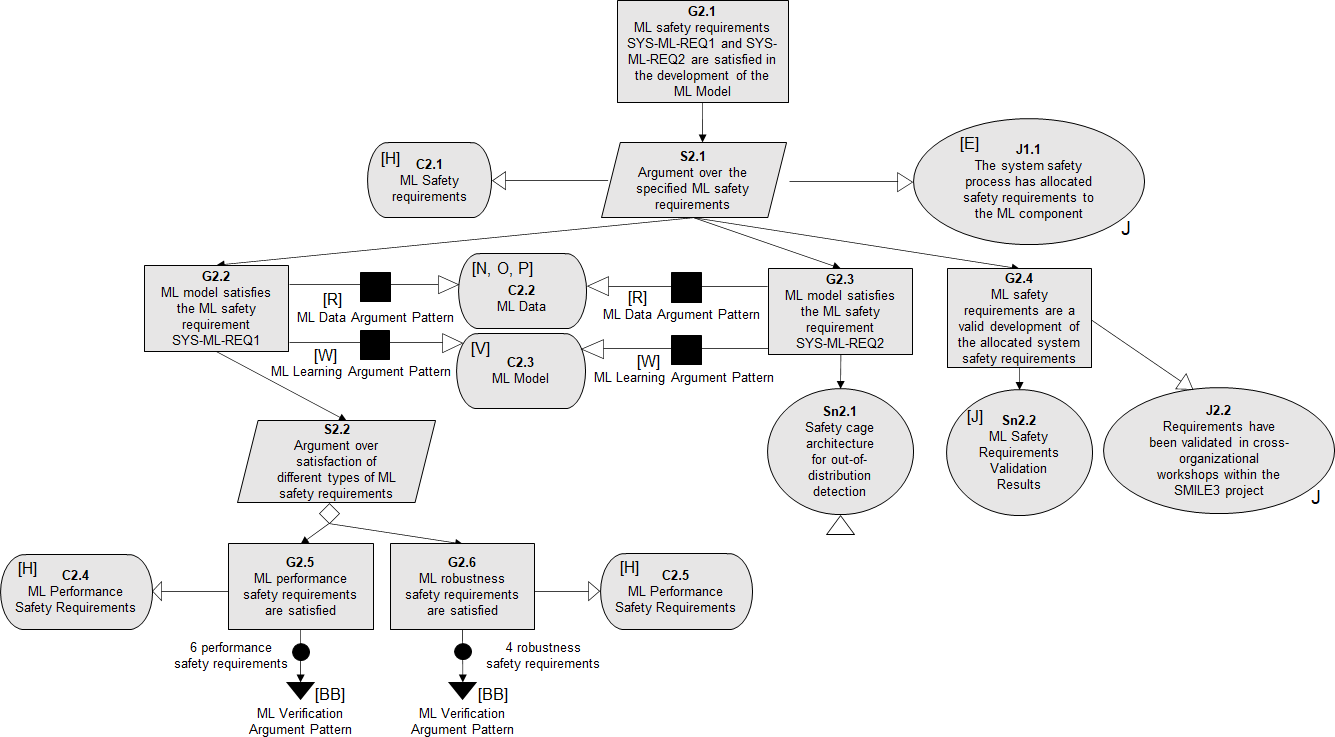}
\caption{ML Safety Requirements Argument Pattern \textbf{[I]}.}
\label{fig:pattern_ml_safety_reqts}
\label{amlas:i}
\end{figure}

Figure~\ref{fig:pattern_ml_safety_reqts} shows the ML Safety Requirements Argument Pattern \textbf{[I]}. The top claim is that the system safety requirements that have been allocated to the ML component are satisfied by the model that is developed (G2.1). This is demonstrated through considering explicit ML safety requirements defined for the ML model \textbf{[H]}. The argument approach is a refinement strategy translating the allocated safety requirements into two concrete ML safety requirements (S2.1) provided as context (C2.1). Justification J2.1 explains how we allocated safety requirements to the ML component as part of the system safety process, including the HARA.

Strategy S2.1 is refined into two subclaims about the validity of the ML safety requirements corresponding to missed pedestrians and ghost braking, respectively. Furthermore, a third subclaim concerns the satisfaction of those requirements. G2.2 focuses on the ML safety requirement \textbf{SYS-ML-REQ1}, i.e., that the nominal functionality of the pedestrian recognition component shall be satisfactory. G2.2 is considered in the context of the ML data (C2.2) and the ML model (C2.3), which in turn are supported by the ML Data Argument Pattern \textbf{[R]} and the ML Learning Argument Pattern \textbf{[W]}. The argumentation strategy (S2.2) builds on two subclaims related to two types of safety requirements with respect to safety-related outputs, i.e., performance requirements (G2.5 in context of C2.4) and robustness requirements (G2.6 in context of C2.5). The satisfaction of both G2.5 and G2.6 are addressed by the ML Verification Argument Pattern \textbf{[BB]}. 

Subclaim G2.3 focuses on the ML safety requirement \textbf{SYS-ML-REQ2}, i.e., that the pedestrian recognition component shall reject input that does not resemble the training data to avoid ghost braking. G2.3 is again considered in the context of the ML data (C2.2) and the ML model (C2.3). For SMIRK, the solution is the safety cage architecture (Sn2.1) developed in the SMILE research program \citep{henriksson2021performance}, described in Section~\ref{sec:safetycage}.

Subclaim G2.4 states that the ML safety requirements are a valid development of the allocated system safety requirements. The justification (J2.2) is that the requirements have been validated in cross-organizational workshops within the SMILE~III research project. We provide evidence through ML Safety Requirements Validation Results \textbf{[J]} originating in a Fagan inspection (Sn2.2).

\begin{center}
\begin{framed}
\noindent \textbf{ML Safety Requirements Argument \textbf{[K]}}\\
SMIRK instantiates the ML safety requirements argument through a subset of the artifacts listed in Table~\ref{tab:amlas_index}, i.e., ML Safety Requirements Argument Pattern \textbf{[I]}, as well as: Safety Requirements Allocated to ML Component \textbf{[E]}, ML Safety Requirements \textbf{[H]}, and ML Safety Requirements Validation Results \textbf{[J]}.
\end{framed}
\end{center}

\subsection{Stage 3: Data Management Assurance} \label{amlas:t}
\begin{figure}
\centering
\includegraphics[width=1\textwidth]{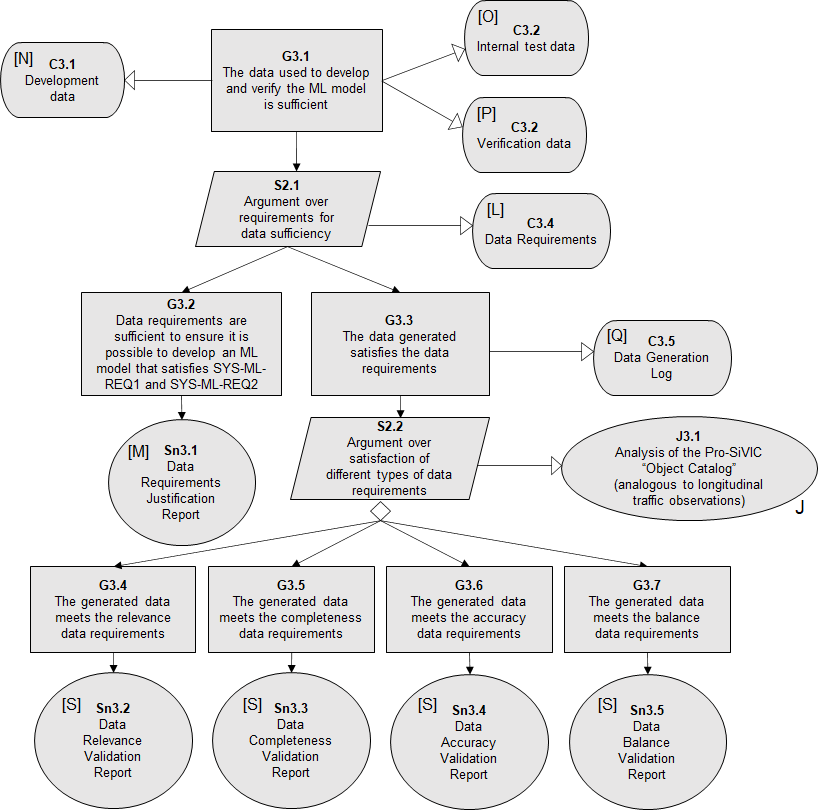}
\caption{ML Data Argument Pattern \textbf{[R]}.}
\label{fig:pattern_ml_data}
\label{amlas:r}
\end{figure}

Figure~\ref{fig:pattern_ml_data} shows the ML Data Argument Pattern \textbf{[R]}. The top claim is that the data used during the development and verification of the ML model is sufficient (G3.1). This claim is made for all three data sets: development data \textbf{[N]}, internal test data \textbf{[O]}, and verification data \textbf{[P]}. The argumentation strategy (S2.1) involves how the sufficiency of these data sets is demonstrated given the Data Requirements \textbf{[L]}. The strategy is supported by arguing over subclaims demonstrating sufficiency of the Data Requirements (G3.2) and that the Data Requirements are satisfied (G3.3). Claim G3.2 is supported by evidence in the form of a data requirements justification report \textbf{[M]}. As stated in AMLAS, ``It is not possible to claim that the data alone can guarantee that the ML safety requirements will be satisfied, however the data used must be sufficient to enable the model that is developed to do so.''

Claim G3.3 states that the generated data satisfies the data requirements in context of the decisions made during data collection. The details of the data collection, along with rationales, are recorded in the Data Generation Log \textbf{[Q]}. The argumentation strategy (S2.2) uses refinement mapping to the assurance-related desiderata of the data requirements. The refinement of the desiderata into concrete data requirements for the pedestrian recognition component of SMIRK, given the ODD, is justified by an analysis of the expected traffic agents and objects that can appear in ESI Pro-SiVIC. For each subclaim corresponding to a desideratum, i.e., relevance (G3.4), completeness (G3.5), accuracy (G3.6), and balance (G3.7), there is evidence in a matching section in the ML Data Validation Report \textbf{[S]}.

\begin{center}
\begin{framed}
\noindent \textbf{ML Data Argument \textbf{[T]}}\\
SMIRK instantiates the ML Data Argument through a subset of the artifacts listed in Table~\ref{tab:amlas_index}, i.e., the ML Data Argument Pattern \textbf{[R]}, as well as: ML Safety Requirements \textbf{[H]}, Data Requirements \textbf{[L]}, Data Requirements Justification Report \textbf{[M]}, Development Data \textbf{[N]}, Internal Test Data \textbf{[O]}, Verification Data \textbf{[P]}, Data Generation Log \textbf{[Q]}, and ML Data Validation Results \textbf{[S]}.
\end{framed}
\end{center}

\subsection{Stage 4: Model Learning Assurance} \label{amlas:y}
\begin{figure}
\centering
\includegraphics[width=1\textwidth]{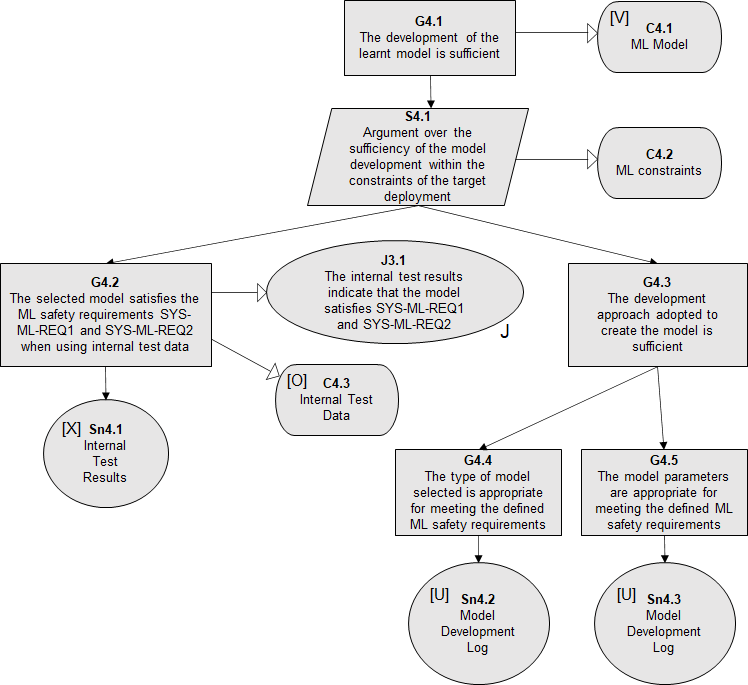}
\caption{ML Learning Argument Pattern \textbf{[W]}.}
\label{fig:pattern_ml_learning}
\label{amlas:w}
\end{figure}

Figure~\ref{fig:pattern_ml_learning} shows the ML Learning Argument Pattern \textbf{[W]}. The top claim (G4.1) is that the development of the learned model \textbf{[V]} is sufficient. The strategy is to argue over the internal testing of the model and that the ML development was appropriate (S4.1) in context of creating a valid model that meets practical constraints such as real-time performance and cost (C4.2). Subclaim (G4.2) is that the ML model satisfies the ML safety requirements when using the internal test data \textbf{[O]}. We justify that the internal test results indicate that the ML model satisfies the ML safety requirements (J3.1) by presenting evidence from the internal test results \textbf{[X]}.

Subclaim G4.3 addresses the approach that was used when developing the model. This claim is in turn supported by two additional subclaims regarding the type of model selected and the model parameters selected, respectively. First, G4.4 claims that the type of model is appropriate for the specified ML safety requirements and the other model constraints. ML development processes, including transfer learning, are highly iterative thus rationales for development decisions must be recorded. Second, G4.5 claims that the parameters of the ML model are appropriately selected to tune performance toward the object detection task within the specified ODD. Rationales for all relevant decisions in G4.4 and G4.5 are recorded in the model development log \textbf{[U]}.

\begin{center}
\begin{framed}
\noindent \textbf{ML Learning Argument \textbf{[T]}}\\
SMIRK instantiates the ML Learning Argument through a subset of the artifacts listed in Table~\ref{tab:amlas_index}, i.e., the ML Learning Argument Pattern \textbf{[W]}, as well as: ML Safety Requirements \textbf{[H]}, Development Data \textbf{[N]}, and Internal Test Data \textbf{[O]}.
\end{framed}
\end{center}

\subsection{Stage 5: Model Verification Assurance} \label{amlas:cc}
\begin{figure}
\centering
\includegraphics[width=1\textwidth]{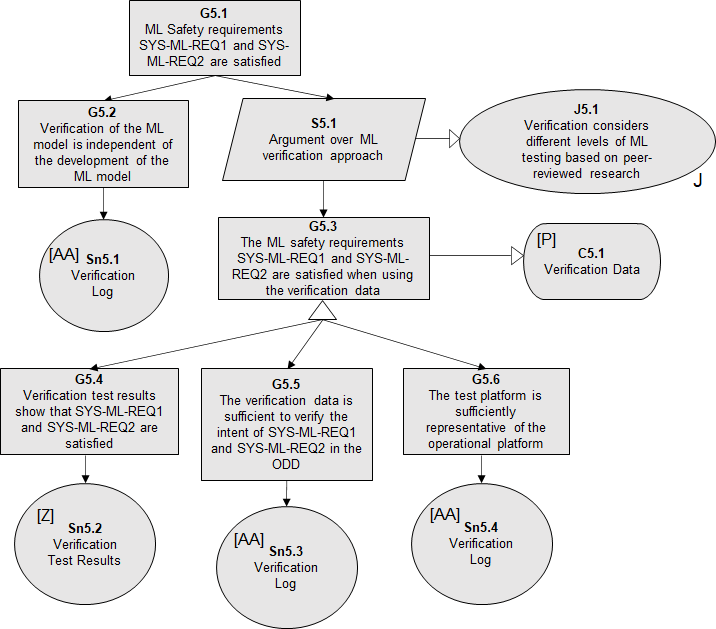}
\caption{ML Verification Argument Pattern \textbf{[BB]}.}
\label{fig:pattern_ml_verification}
\label{amlas:bb}
\end{figure}

Figure~\ref{fig:pattern_ml_verification} shows the ML Verification Argument Pattern \textbf{[BB]}. The top claim (G5.1) corresponds to the bottom claim in the safety requirements argument pattern \textbf{[I]}, i.e., that all ML safety requirements are satisfied. The argumentation builds on a subclaim and an argumentation strategy. First, subclaim G5.2 is that the verification of the ML model is independent of its development. The verification log \textbf{[AA]} specifies how this has been achieved for SMIRK (Sn5.1). Second, the strategy S5.1 argues that test-based verification is an appropriate approach to generate evidence that the ML safety requirements are met. The justification (J5.1) is that the SMIRK test strategy follows the proposed organization in peer-reviewed literature on ML testing, which is a better fit than using less mature formal methods for ML models as complex as YOLOv5.

Following the test-based verification approach, the subclaim G5.3 argues that the ML model satisfies the ML safety requirements when the verification data (C5.1) is applied. The testing claim is supported by three subclaims. First, G5.4 argues that the test results demonstrate that the ML safety requirements are satisfied, for which Verification Test Results \textbf{[Z]} are presented as evidence. Second, G5.5 argues that the Verification Data \textbf{[P]} is sufficient to verify the intent of the ML safety requirements in the ODD. Third, G5.6 argues that the test platform is representative of the operational platform. Evidence for both G5.5 and G5.6 is presented in the Verification Log \textbf{[AA]}.

\begin{center}
\begin{framed}
\noindent \textbf{ML Verification Argument \textbf{[CC]}}\\
SMIRK instantiates the ML Verification Argument through a subset of the artifacts listed in Table~\ref{tab:amlas_index}, i.e., the ML Verification Argument Pattern \textbf{[W]}, as well as: ML Safety Requirements \textbf{[H]}, Verification Data \textbf{[P]}, and the ML Model \textbf{[V]}.
\end{framed}
\end{center}

\subsection{Stage 6: Model Deployment Assurance} \label{amlas:hh}
\begin{figure}
\centering
\includegraphics[width=1\textwidth]{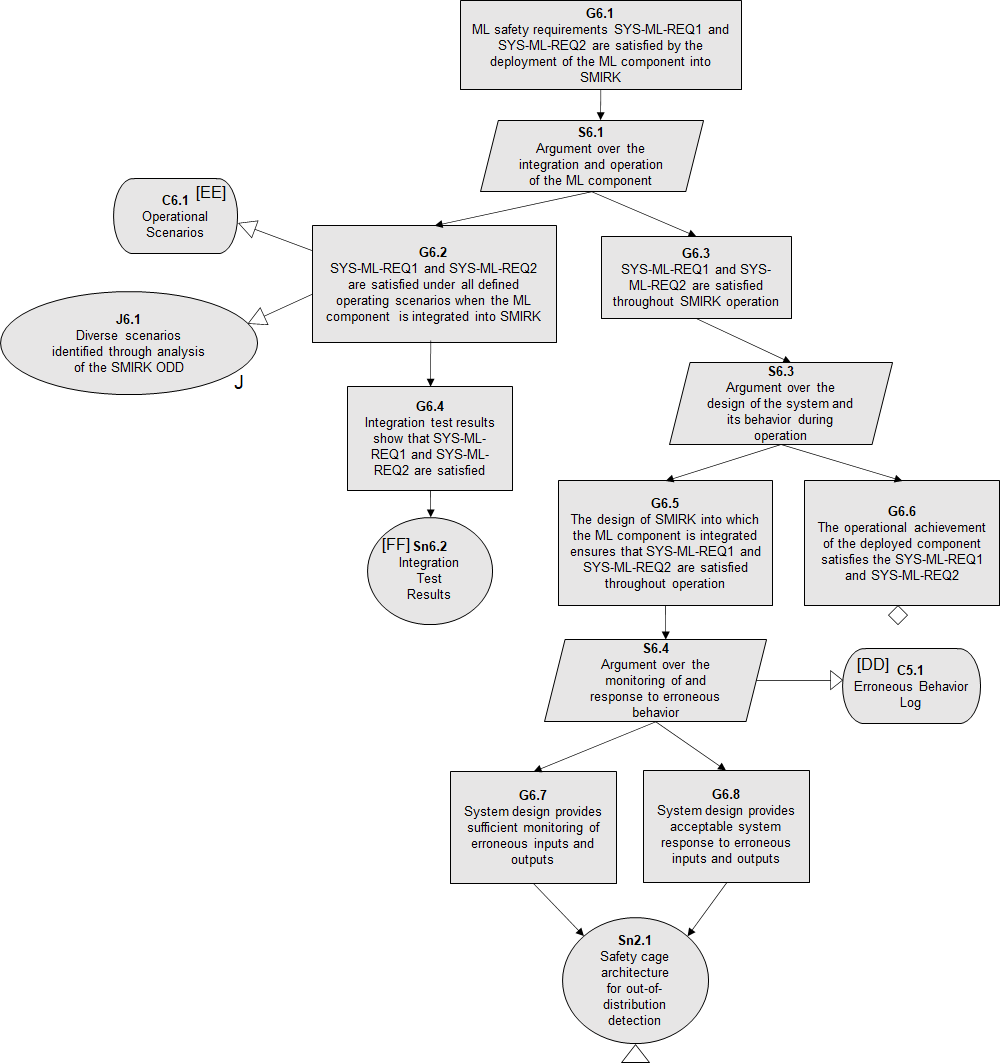}
\caption{ML Deployment Argument Pattern \textbf{[GG]}.}
\label{fig:pattern_ml_deployment}
\label{amlas:gg}
\end{figure}

Figure~\ref{fig:pattern_ml_deployment} shows the ML Verification Argument Pattern \textbf{[GG]}. The top claim (G6.1) is that the ML safety requirements \textbf{SYS-ML-REQ1} and \textbf{SYS-ML-REQ2} are satisfied when deployed to the ego car in which SMIRK operates. The argumentation strategy S6.1 is two-fold. First, subclaim G6.2 is that the ML safety requirements are satisfied under all defined operating scenarios when the ML component is integrated into SMIRK in the context (C6.1) of the specified operational scenarios \textbf{[EE]}. Justification J6.1 explains that the scenarios were identified through an analysis of the SMIRK ODD. G6.2 has another subclaim (G6.4), arguing that the integration test results \textbf{[FF]} show that \textbf{SYS-ML-REQ1} and \textbf{SYS-ML-REQ2} are satisfied.

Second, subclaim G6.3 argues that \textbf{SYS-ML-REQ1} and \textbf{SYS-ML-REQ2} continue to be satisfied during the operation of SMIRK. The supporting argumentation strategy (S6.3) relates to the design of SMIRK and is again two-fold. First, subclaim G6.6 argues that the operational achievement of the deployed component satisfies the ML safety requirements. Second, subclaim G6.5 argues that the design of SMIRK into which the ML component is integrated ensures that \textbf{SYS-ML-REQ1} and \textbf{SYS-ML-REQ2} are satisfied throughout operation. The corresponding argumentation strategy (S6.4) is based on demonstrating that the design is robust by taking into account identified erroneous behavior in the context (C5.1) of the Erroneous Behavior Log \textbf{[DD]}. More specifically, the argumentation entails that predicted erroneous behavior will not result in the violation of the ML safety requirements. This is supported by two subclaims, i.e., that the system design provides sufficient monitoring of erroneous inputs and outputs (G6.7) and that the system design provides acceptable response to erroneous inputs and outputs (G6.8). Both G6.7 and G6.8 are addressed by the safety cage architecture that monitors input through OOD detection using an autoencoder that rejects anomalies accordingly. The acceptable system response is to avoid emergency braking and instead let the human driver control ego car.

\begin{center}
\begin{framed}
\noindent \textbf{ML Verification Argument \textbf{[HH]}}\\
SMIRK instantiates the ML Deployment Argument through a subset of the artifacts listed in Table~\ref{tab:amlas_index}, i.e., the ML Deployment Argument Pattern \textbf{[GG]}, as well as: System Safety Requirements \textbf{\textbf{[A]}}, Environment Description \textbf{[B]}, System Description \textbf{[C]}, ML Model \textbf{[V]}, Erroneous Behaviour Log \textbf{[DD]}, Operational Scenarios \textbf{[EE]}, and Integration Testing Results \textbf{[FF]}.
\end{framed}
\end{center}

\end{appendices}

\end{document}